\documentclass{aa}

  \usepackage[latin1]{inputenc}
  \usepackage{graphicx}
  \usepackage{tabularx}
  \usepackage{natbib}
  \usepackage[normalem]{ulem}
  \usepackage{amsmath,amssymb}

  \newcommand{\pc}{\;{\rm pc}}
  \newcommand{\kpc}{\;{\rm kpc}}
  \newcommand{\Mpc}{\;{\rm Mpc}}
  \newcommand{\mic}{\;\mu m}

  \newcommand{\zsol}{\;Z_\odot}
  \newcommand{\upfov}{''{\rm pixel}^{-1}}

  \newcommand{\ddiff}{{\, \rm d}}
  \newcommand{\papi}{this study}

  \newcommand{\ngc}[1]{NGC$\;$#1}
  \newcommand{\M}[1]{M$\;$#1}
  \newcommand{\mw}{Milky Way}
  \newcommand{\iizw}{II$\;$Zw$\;$40}
  \newcommand{\all}{\ngc{1569}, \ngc{1140} and \iizw}
  \newcommand{\smcn}{SMC$\;$N$\;$66}
  \newcommand{\smcb}{SMC$\;$B\#1}
  \newcommand{\sbs}{SBS$\;$0335-052}
  \newcommand{\xxxdor}{30$\;$Dor}

  \newcommand{\hii}{H$\,${\sc ii}}

  \newcommand{\neiii}{[Ne$\,${\sc iii}]}
  \newcommand{\neii}{[Ne$\,${\sc ii}]}
  \newcommand{\siv}{[S$\,${\sc iv}]}
  \newcommand{\ariii}{[Ar$\,${\sc iii}]}
  \newcommand{\arii}{[Ar$\,${\sc ii}]}
  \newcommand{\neiiiline}{\neiii$\lambda 15.56\mic$}
  \newcommand{\neiiline}{\neii$\lambda 12.81\mic$}
  \newcommand{\sivline}{\siv$\lambda 10.51\mic$}
  \newcommand{\ariiiline}{\ariii$\lambda 8.99\mic$}
  \newcommand{\ariiline}{\arii$\lambda 6.98\mic$}

  \graphicspath{{figures/}}

\begin{document}


\title{ISM Properties in Low-Metallicity Environments} 
\subtitle{I. Mid-Infrared Spectra of Dwarf Galaxies\thanks{Based on
       observations with ISO, an ESA project with instruments funded by ESA 
       Member States (especially the PI countries: France, Germany, the 
       Netherlands and the United Kingdom) and with the participation of ISAS 
       and NASA.}}
\titlerunning{Mid-Infrared spectra of dwarf galaxies}

\author{Suzanne~C.~Madden\inst{1} \and
        Fr\'ed\'eric~Galliano\inst{1,2} \and
        Anthony~P.~Jones\inst{3} \and 
        Marc~Sauvage\inst{1}}
\authorrunning{S. Madden et al.}
\institute{Service d'Astrophysique, CEA/Saclay, l'Orme des Merisiers,
           91191 Gif sur Yvette, France \and
           Observational Cosmology Lab., Code 665, 
           NASA Goddard Space Flight Center, 
           Greenbelt MD 20771 USA \and
           Institut d'Astrophysique Spatiale (CNRS),
           Universite de Paris XI, 91405 Orsay, France}

\abstract{
We present new ISOCAM mid-infrared spectra of three starbursting nearby dwarf 
galaxies, \all\ and the \xxxdor\ region of the LMC and explore the properties 
of the ISM in low-metallicity environments, also using additional sources from
the literature.  
We analyse the various components of the ISM probed by the mid-infrared 
observations and compare them with other Galactic and extragalactic objects. 
The MIR spectra of the low-metallicity starburst sources are dominated by the 
\neiiiline\ and \sivline\ lines, as well as a steeply rising dust continuum. 
PAH bands are generaly faint, both locally and averaged over the full galaxy, 
in stark contrast to dustier starburst galaxies, where the PAH features are 
very prominant and even dominate on global scales. 
The hardness of the modeled interstellar radiation fields for the dwarf 
galaxies increases as the presence of PAH band emission becomes less 
pronounced. 
The \neiii/\neii\ ratios averaged over the full galaxy are strikingly high, 
often $>10$. 
Thus, the hard radiation fields are pronounced and pervasive. 
We find a prominent correlation between the PAHs/VSGs and the \neiii/\neii\ 
ratios for a wide range of objects, including the low metallicity galaxies as 
well as Galactic \hii\ regions and other metal-rich galaxies.  
This effect is consistent with the hardness of the interstellar radiation 
field playing a major role in the destruction of PAHs in the low metallicity 
ISM. 
We see a PAHs/VSGs and metallicity correlation, also found by 
\citet{engelbracht+05} for a larger survey. 
Combined effects of metallicity and radiation field seem to be playing 
important roles in the observed behavior of PAHs in the low metallicity 
systems.
}
\date{Received / Accepted}

\offprints{smadden@cea.fr}

\maketitle


\section{Introduction}

Understanding the interplay between star formation and the interstellar medium
(ISM) in low-metallicity environments, perhaps characteristic of primordial
galaxies, can be approached through observations of the local universe dwarf 
galaxies. 
Much of what we have gleaned about the state of the ISM and star formation 
properties in dwarf galaxies has been derived historically, from optical 
photons. 
We have access to a vast Infrared Space 
Observatory (ISO) mid-infrared (MIR) database, and more recently, Spitzer 
observations, from which new thoughts on the 
nature of the ISM of dwarf galaxies are emerging. 
For example, the amount of dust that is actually present in dwarf galaxies is 
being called into question in several cases 
\citep{hunt+01,vacca+02,plante+02,lisenfeld+02,galliano+03,galliano+05}, 
as a result of the modeling of the dust emission spectral energy distribution 
(SED). 
The myth that the effects of dust can be ignored or treated lightly in dwarf 
galaxies is on the way to banishment, and opens our eyes to the possibility 
of optically hidden, embedded star formation activity in these low 
metallicity systems.

Until recently, our knowledge of the dust emission properties of dwarf galaxies
was dependent primarily on 4 IRAS bands 
\citep[e.g.][]{hunter+89,melisse+94,sauvage+94}. 
ISO provided the sensitivity to begin the exploration of the details of the 
MIR wavelength regime, where a wide variety of physical components of the ISM 
can be traced. 
The MIR camera on board ISO, ISOCAM \citep{cam}, with its spectro-imaging 
capability, has proven to be
invaluable in probing the various components of the MIR regime of a number of
low-metallicity dwarf galaxies. 
ISOCAM first demonstrated the widely-varying MIR characteristics of low 
metallicity starburst galaxies \citep{madden00}, which differ remarkably from 
those of the more metal-rich starburst galaxies 
\citep[e.g.][]{rigopoulou+99,laurent+00,sturm+00,sturm+02,forster+03}. 
The improved sensitivity of the Spitzer telescope will provide a means to 
delve even deeper into the detailed MIR properties of dwarf galaxies 
\citep{engelbracht+05}.
The availability of the details of the MIR wavelength regime launched the 
study of the MIR to millimetre (mm) SEDs of dwarf galaxies, from which the 
nature of the dust properties in galaxies can be more accurately modeled. 
The spectral properties of the MIR SED are 
particularly important factors in constraining dust models for the MIR to 
millimetre (mm) SEDs of dwarf galaxies  
\citep{plante+02,takeuchi+03,galliano+03,vanzi+04,galliano+05}.  

Some of the brightest dwarf galaxies have also been valiantly observed from 
the ground at MIR wavelengths \citep [e.g.][]{rieke+72,frogel+82,roche+91,dudley99,vacca+02,plante+02,vanzi+04}. 
In this paper, we examine the MIR spectra of a sample of low metallicity 
galaxies, 
using ISOCAM spectra from $\lambda=5\mic$ to $16\mic$.

Great progress has been made in understanding the MIR interstellar gas and dust
characteristics through recent studies of Galactic regions and galaxies with 
metal-rich ISM \citep[e.g.][]{helou+00,vigroux+01,roussel+01b,roussel+01a,dale+02,lutz+96,sturm+00}, and
we can now explore the effects of low-metallicity on the
dust and gas properties. 
The MIR regime provides unique advantages
for characterising the physical properties of the dust and gas in
galaxies, particularly due to the fact that it contains a wide variety of ISM 
diagnostics that are not heavily affected by extinction. 
There is a
factor of 10 to 50 less extinction over MIR wavelengths compared to optical
wavelengths \citep{mathis90}. 
Nebular emission lines from neon, argon and sulfur are
present in the ISOCAM range, displaying unique new global properties
of star formation activity, which can have some bearing on the
low-metallicity environment, directly or indirectly.  
An important
aspect in studying dwarf galaxies in the MIR wavelength range is to
discern and understand the variations in the spectral characteristics of the 
dust and gas
emission between the low-metallicity and more metal-rich ISM,
perhaps establishing useful tracers for possible primordial conditions
in more distant galaxies.

For this study we present new MIR ISOCAM spectra for the 3 dwarf
galaxies, \all, whose full infrared (IR) SEDs have been succesfully modeled by
\citet{galliano+03,galliano+05}, as well as the MIR spectra toward \xxxdor\ in
the LMC. 
Additionally, we add to our study other low metallicity sources which have 
previously been presented: \ngc{5253} \citep{crowther+99}; the
$1/41\zsol$ galaxy, \sbs\ \citep{thuan+99}; a quiescent molecular cloud in the
SMC, \smcb\ \citep{reach+00} and N66 in the SMC \citep{contursi+00}. 
All of these galaxies possess massive young star clusters and/or super star 
clusters (SSCs), which represent unusual and relatively rare modes of star
formation with stellar surface densities orders of magnitude in excess
of normal \hii\ regions and OB associations, some
of which can be deeply embedded, concealing the precise nature of the star
formation. 
The combination
of the relatively high angular resolution and high sensitivity in the
MIR using ISOCAM, provides a unique opportunity to study the impact of
the environment of SSCs on the nature of the  dust.  
As SSCs have strong stellar winds from the densely packed,
rapidly-evolving stars, the presence of dust, as well as gas
in the immediate vicinity may not immediately be obvious
and may be hindering our view of star formation.

The 7 low metallicity sources discussed in this article range in metallicity 
from $Z=1/41\zsol$ to $1/2\zsol$, and provide a variety of spatial scales, 
from our neighboring Magellanic Clouds ($D\simeq 50\kpc$ to $60\kpc$), to 
that of our most distant source, \ngc{1140} ($D\simeq 23\Mpc$; 
Table~\ref{tab:sources}). 
Thus, the linear physical scales range from $1.2\pc$ to $500\pc$, given the 
short wavelength spatial resolution ($\sim 5''$) of ISOCAM on ISO.

The paper is organised as follows: 
Sect.~\ref{sec:obs} presents the new ISOCAM CVF observations, 
Sect.~\ref{sec:anal} is the spectral analysis, describing how the components 
of MIR spectra are modeled, isolating the MIR nebular lines, the PAH bands 
and the continua. 
Also in Sect.~\ref{sec:anal}, we compare the properties of some of the 
galaxies with the Galactic \hii\ region/molecular cloud, \M{17}. 
Further interpretation of the ionic line ratios, the PAH bands, and the 
continuum emission are presented in Sect.~\ref{sec:interp}, along with other 
sources presented here and in the literature. 
Sect.~\ref{sec:conc} ties up the paper with a summary.


\section{Observations and Data Reduction}
\label{sec:obs}

\begin{table*}
  \centering
  \caption[]{Sources used in this study. 
             The dates indicate when the observations were made by ISO; 
             the R.A. and Dec. are the coordinates of the center of the field.
             References for the distances and metallicity values are the 
             following:
                   (a)~\citet{israel88}; 
                   (b)~\citet{kobulnicky+97};
                   (c)~\citet{thuan+81};
                   (d)~\citet{brinks+88};
                   (e)~\citet{guseva+00};
                   (f)~\citet{hunter+94};
                   (g)~\citet{heckman+98};
                   (i)~\citet{calzetti97};
                   (j)~\citet{freedman+01};
                   (k)~\citet{kobulnicky+97a};
                   (l)~\citet{vandenbergh99};
                   (m)~\citet{dufour+82}.
             For comparison, the solar metallicity is 
	     $12+\log{\rm O/H}=8.83$ \citep{grevesse+98}.}
  \label{tab:sources}
\begin{tabularx}{\textwidth}{*{6}{X}l} 
\hline 
\hline
  Source       & Date     & R.A. (J2000)       & Dec. (J2000)             
    & Distance   & $12+\log({\rm O/H})$ & MIR References      \\
\hline
  \ngc{1569}   & Mar 1998 & $04^h30^m49\fs1$   & $64\degr50\arcmin52.8\arcsec$
    & $2.2\Mpc^{(a)}$  & $8.19^{(b)}$   & \papi               \\
  \iizw        & Oct 1997 & $05^h55^m42\fs7$   & $3\degr23\arcmin29.5\arcsec$ 
    & $10\Mpc^{(c,d)}$ & $8.09^{(e)}$   & \papi               \\
  \ngc{1140}   & Feb 1998 & $2^h54^m33\fs5$    & $-10\degr01\arcmin44.0\arcsec$
    & $23\Mpc^{(f,g,h)}$& $8.44^{(i)}$  & \papi               \\
  \ngc{5253}   & Jan 1997 & $13^h39^m56\fs0$   & $-31\degr38\arcmin29.0\arcsec$
    & $3.25\Mpc^{(j)}$ & $8.16^{(k)}$   & \citet{crowther+99} \\
  \smcn        & Sep 1996 & $0^h59^m02\fs0$    & $-72\degr10\arcmin36.0\arcsec$
    & $60\kpc^{(l)}$   & $8.04^{(m)}$         & \citet{contursi+00} \\
  \smcb        & Jul 1996 & $0^h45^m33\fs0$    & $-73\degr18\arcmin46.0\arcsec$
    & $60\kpc^{(l)}$   & $8.04^{(m)}$         & \citet{reach+00}    \\
  LMC~\xxxdor  & Oct 1997 & $5^h38^m34\fs0$    & $-69\degr05\arcmin57.0\arcsec$
    & $50\kpc^{(l)}$   & $8.37^{(m)}$         & \papi               \\   \hline
\end{tabularx}
\end{table*}

The CVF spectro-images were made with ISOCAM \citep{cam} on board the ISO 
satellite \citep{kessler+96}, using a $32\times 32$ detector array with
a sampling of $6\upfov$ for \ngc{1569} and \ngc{1140}, and $3\upfov$ for 
\iizw, giving a camera field of view of $192''\times192''$ and 
$96''\times96''$, respectively.
The CVF performed spectral imaging from $\lambda=5\mic$ to $16.5\mic$ with one
pointing of two CVFs, from $\lambda=5$ to $9.5\mic$ and from $\lambda=9.0$ to
$16.5\mic$, with a spectral resolution of $\lambda/\Delta\lambda=35$ to 51 
across the full spectra.

With the pixel field of view (PFOV) used, the camera observed beyond the 
extended emission of the entire galaxies so that the foreground emission 
could be obtained directly from the maps. 
The total integration time for each of these observations was 
$1^{\rm h}30^{\rm m}$.

\begin{figure*}[htbp]
\centering
  \begin{tabular}{cc}
    \includegraphics[width=0.43\textwidth]{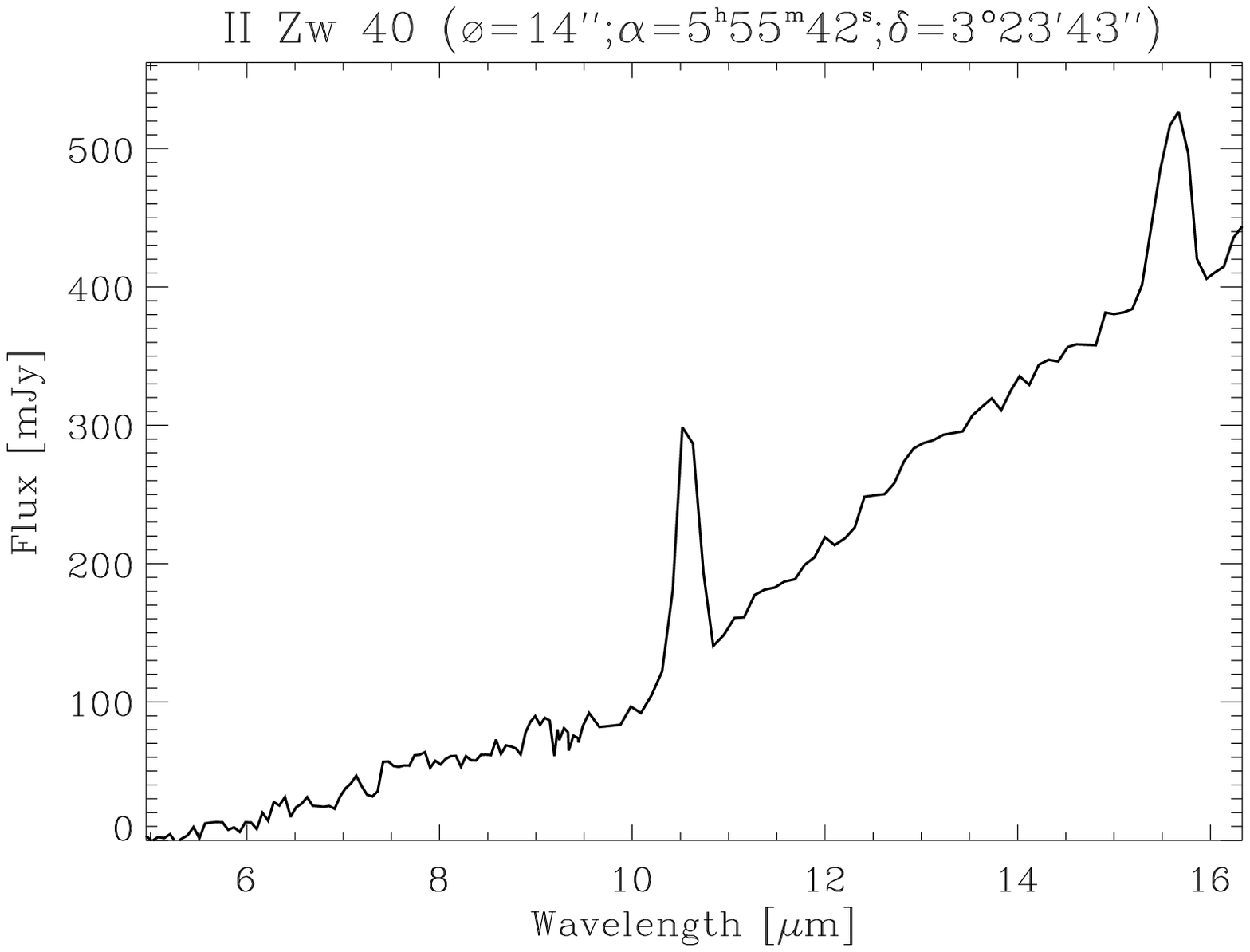}      &
	\includegraphics[width=0.43\textwidth]{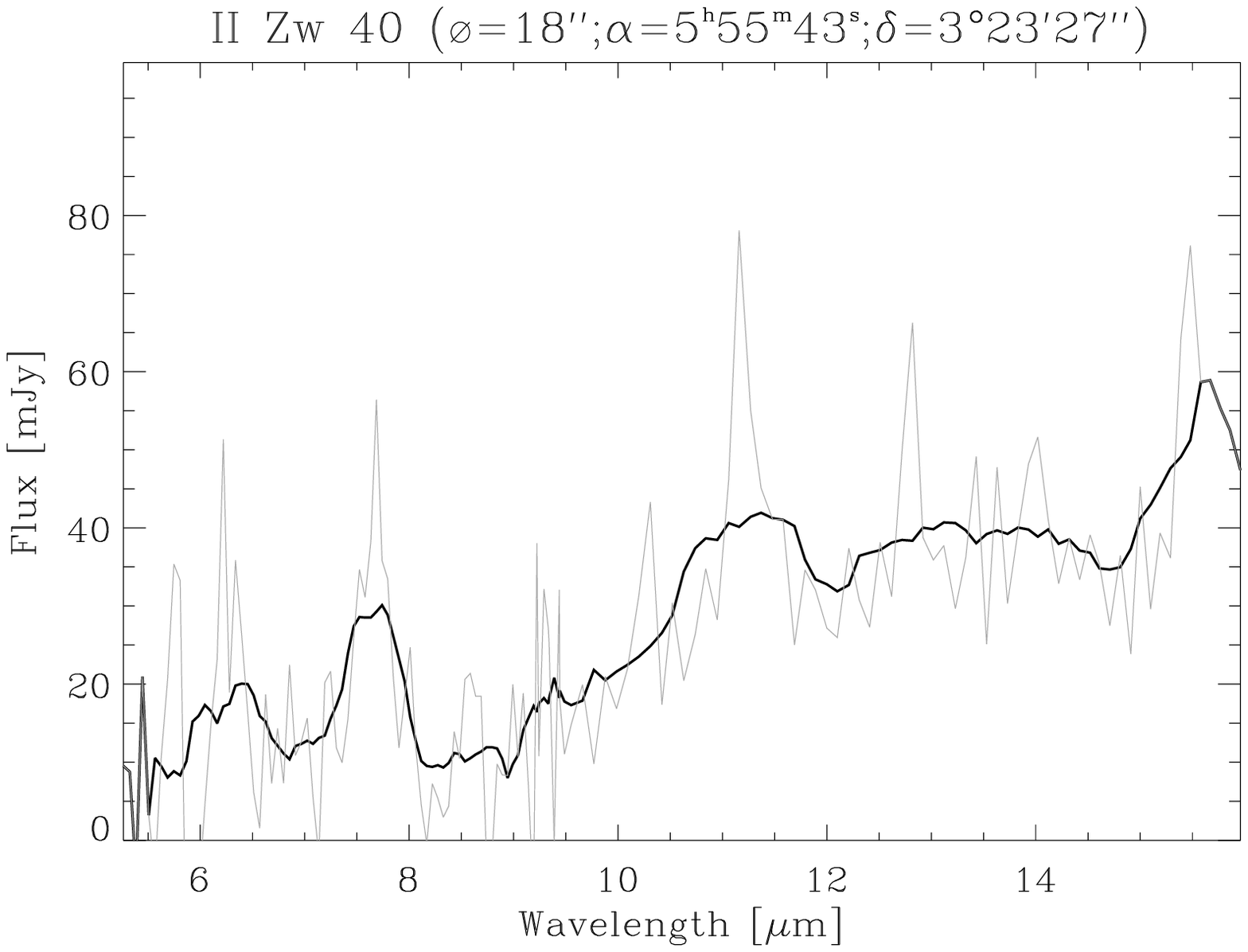}      \\
	\includegraphics[width=0.43\textwidth]{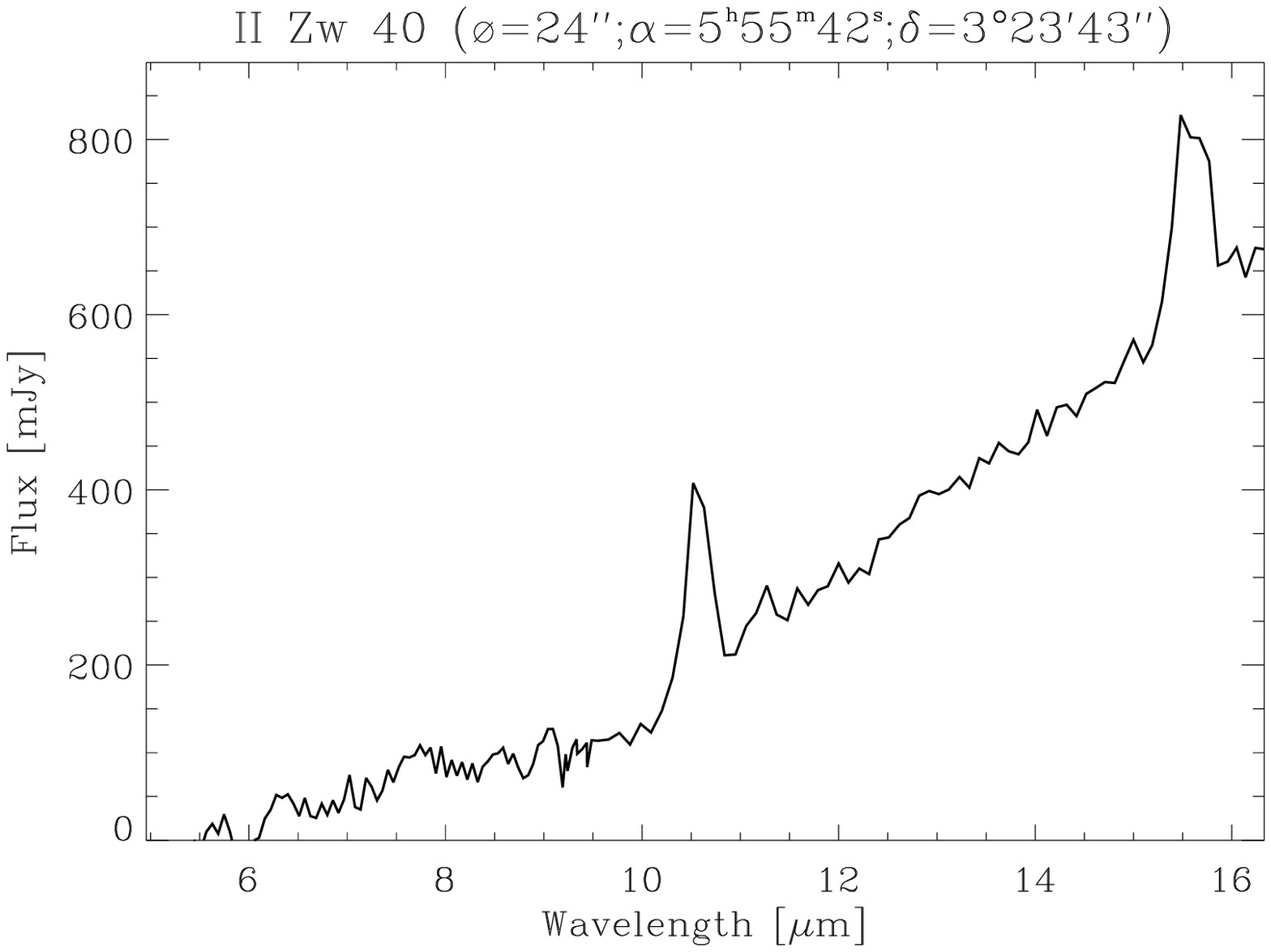}      &
    \includegraphics[width=0.43\textwidth]{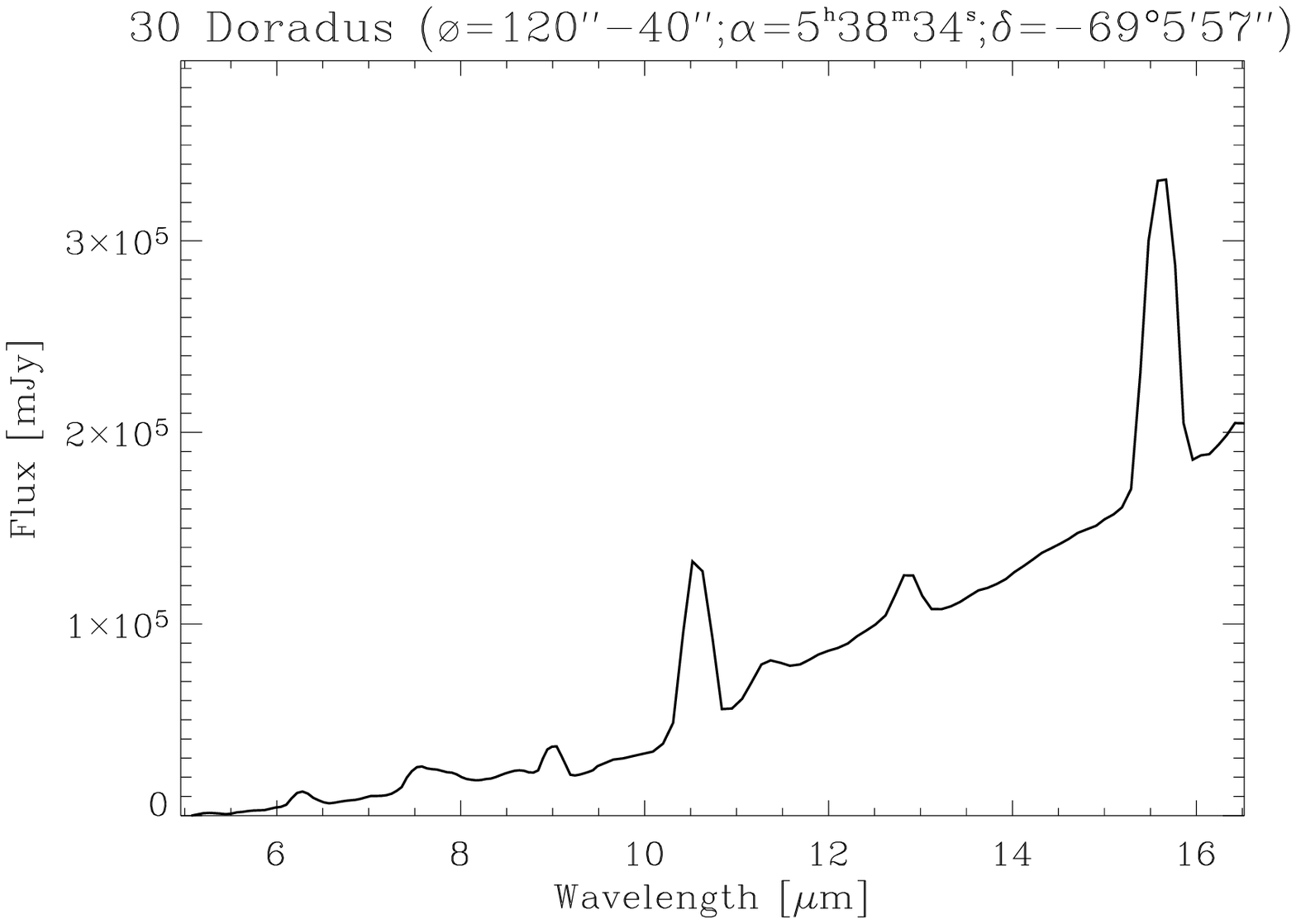}      \\	
    \includegraphics[width=0.43\textwidth]{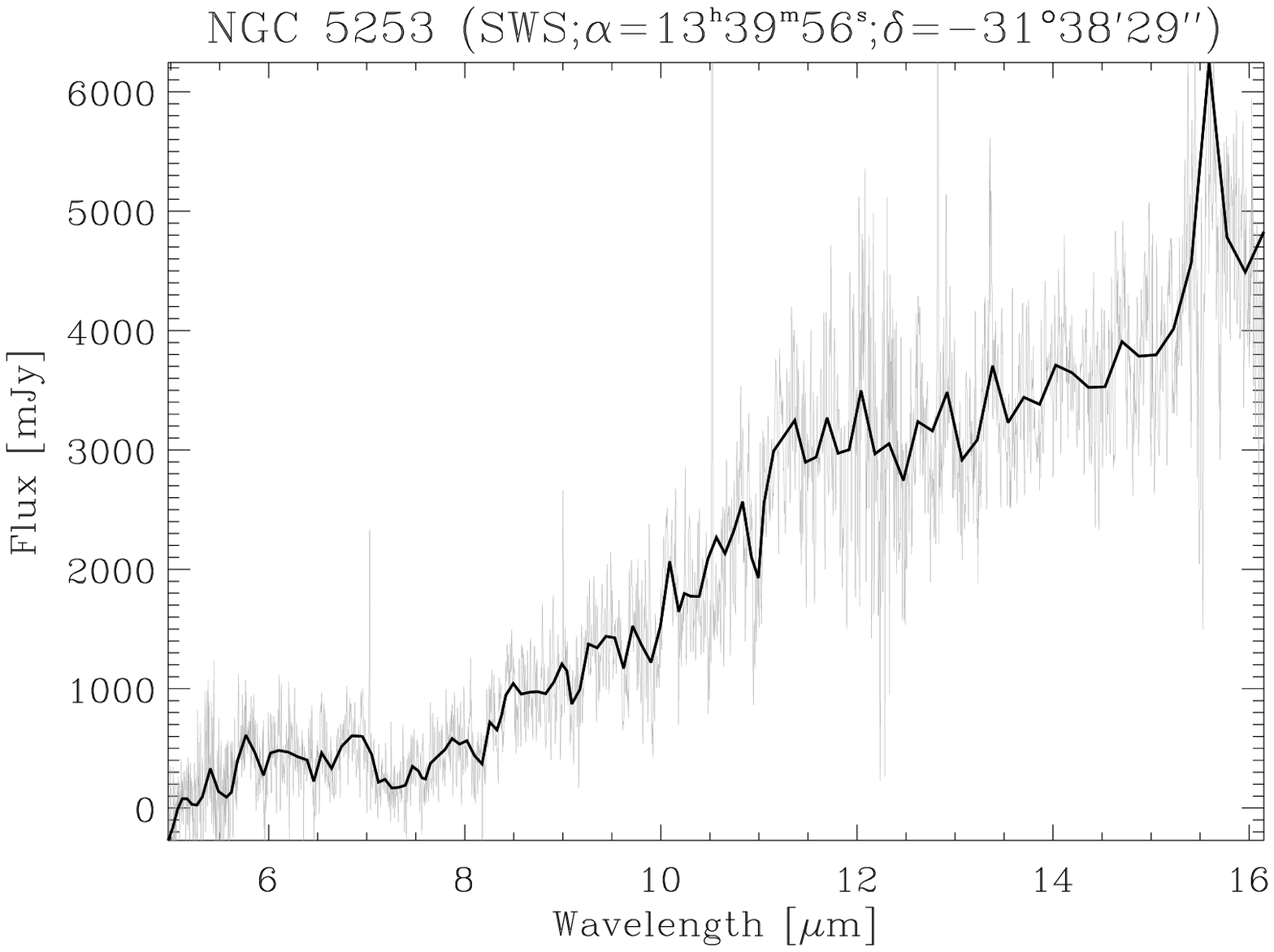}          &
	\includegraphics[width=0.43\textwidth]{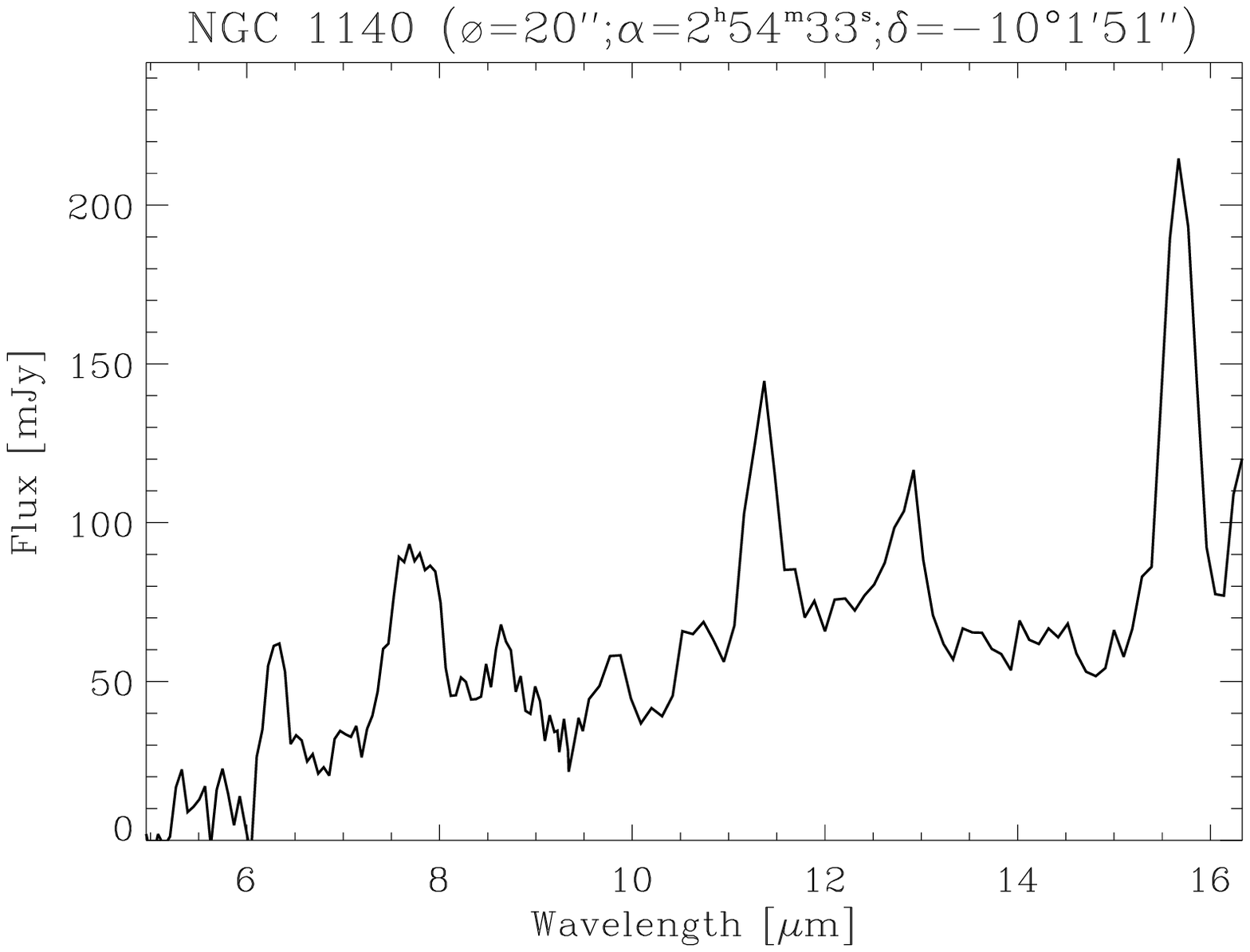}     \\
    \includegraphics[width=0.43\textwidth]{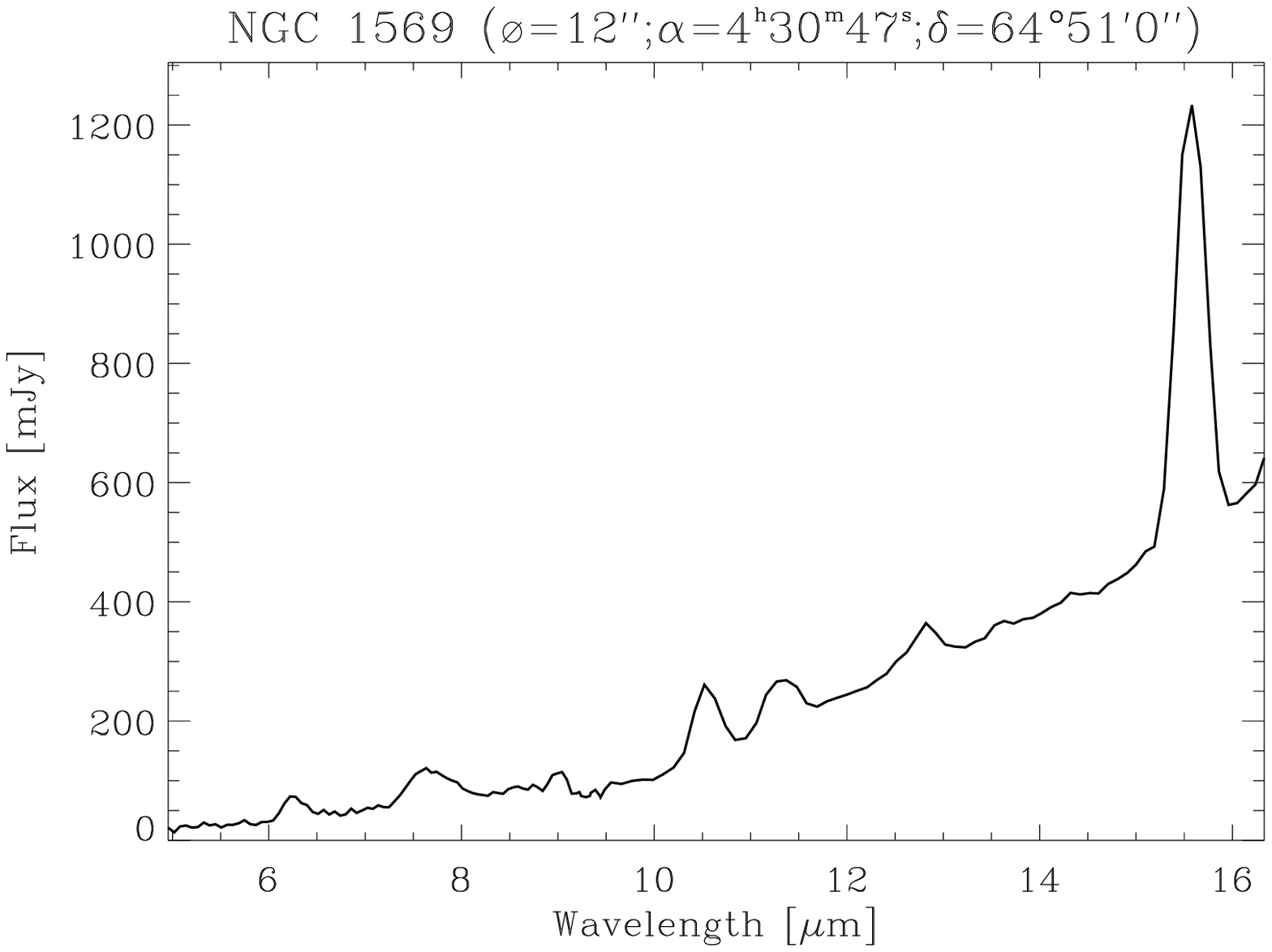}     &
    \includegraphics[width=0.43\textwidth]{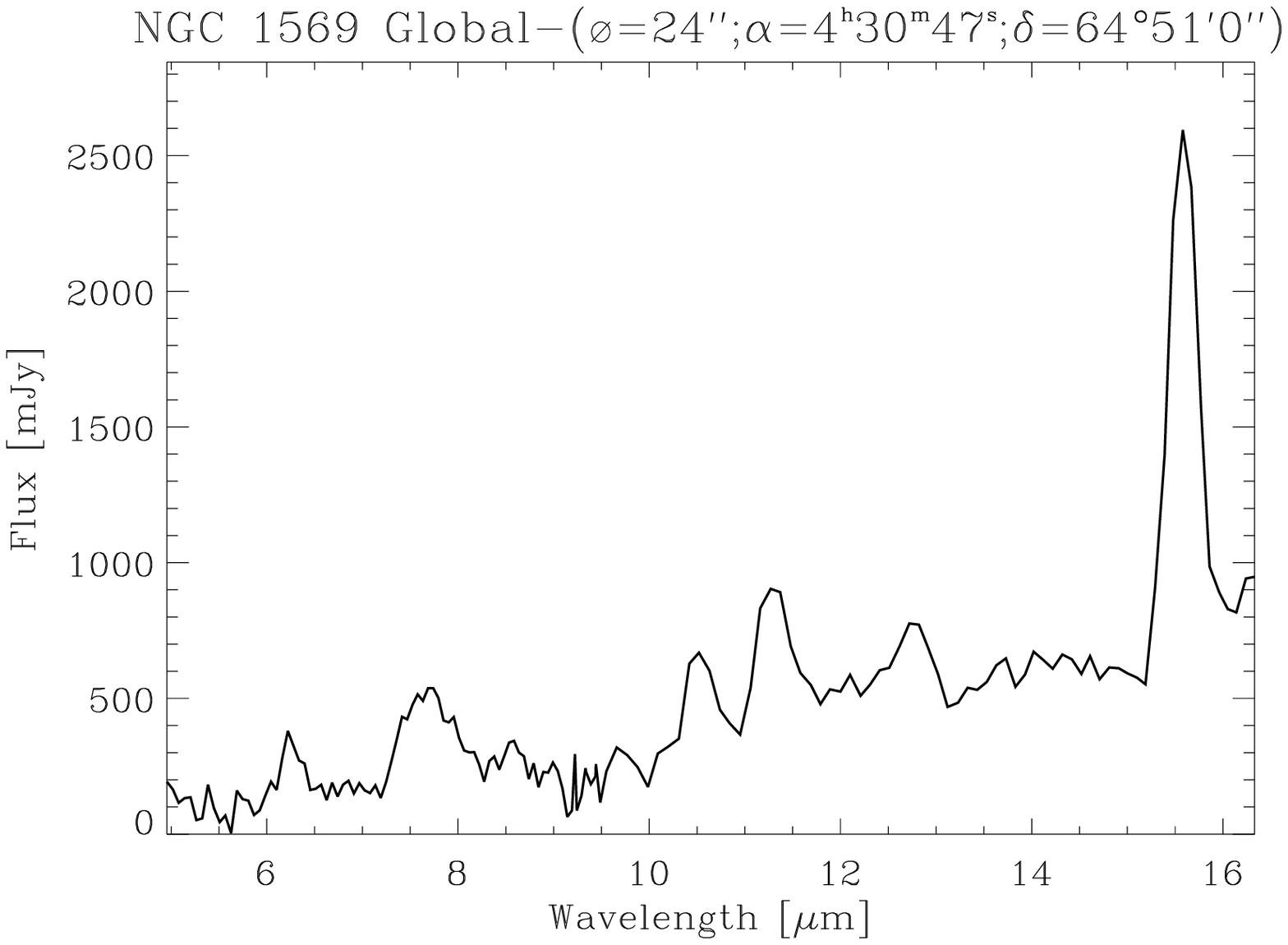}    \\
   \end{tabular}
  \caption{ISOCAM MIR spectra of dwarf galaxies. 
	   The position of the central pointing as well as the diameter 
	   of the circular aperture are indicated next to the name of the 
           source, above each spectrum. 
           The spectrum of \iizw\ in the upper right panel of is taken 
            toward the merger tail. 
           Both the smoothed (black) and original (grey) profiles are 
           displayed (see Sect.~\ref{sec:iizw40} and 
           Fig.~\ref{fig:iizw40_tail}). 
           Two profiles are displayed for \ngc{5253}: the original SWS 
           spectrum is the grey profile \citep{crowther+99} and the black 
           profile is the original spectrum smoothed to the spectral 
           resolution of ISOCAM. 
           The spectrum of the peak 12\arcsec\ region of \ngc{1569} (lower 
           left) is shown in contrast with the integrated emission of the
	   ``extended'' region of \ngc{1569} (lower right): the full galaxy 
           with the central 24\arcsec\ removed.}
\label{fig:spec1}		
\end{figure*}

\begin{figure*}[htbp]
\centering
  \begin{tabular}{cc}
	\includegraphics[width=0.43\textwidth]{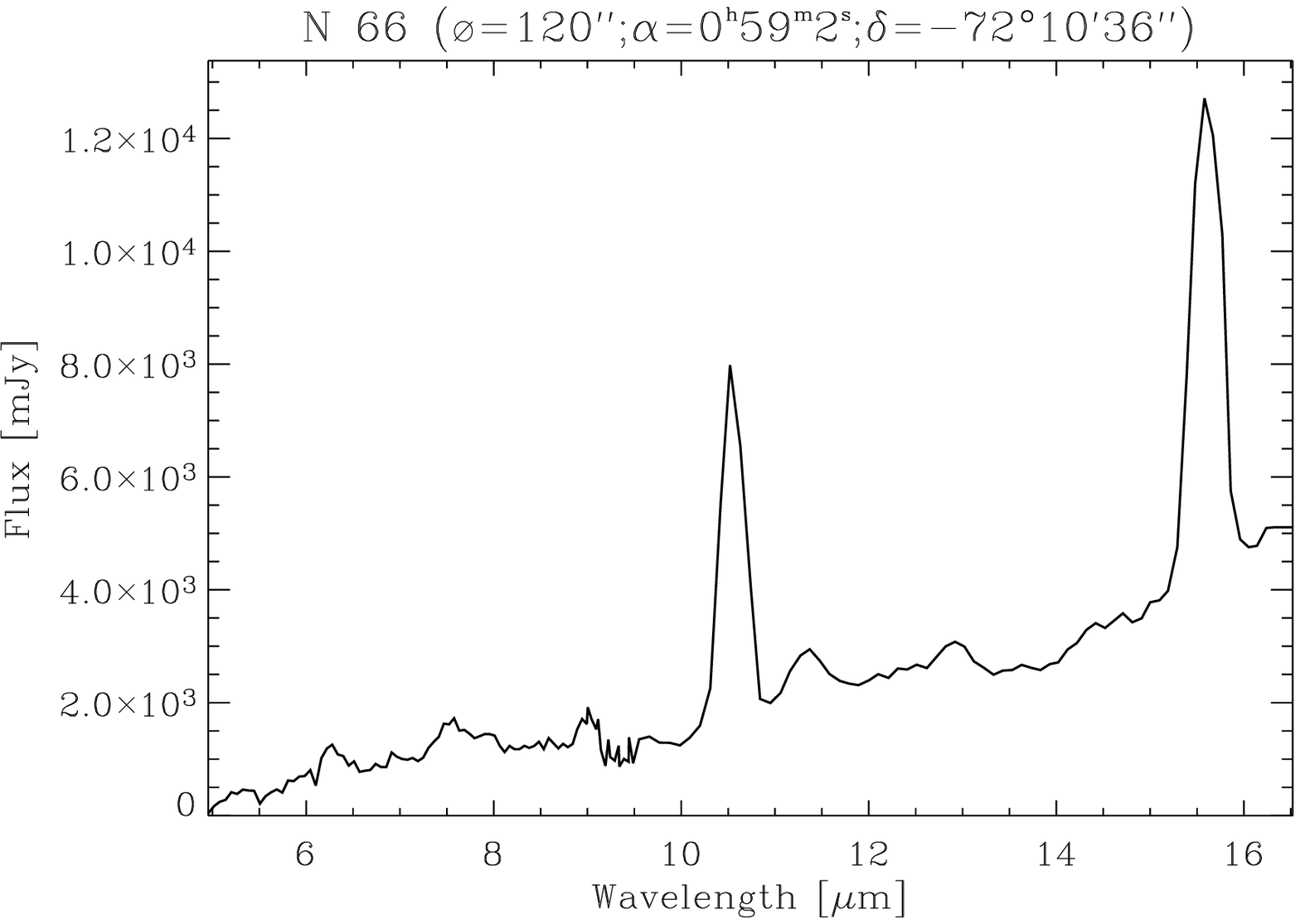}        &
	\includegraphics[width=0.43\textwidth]{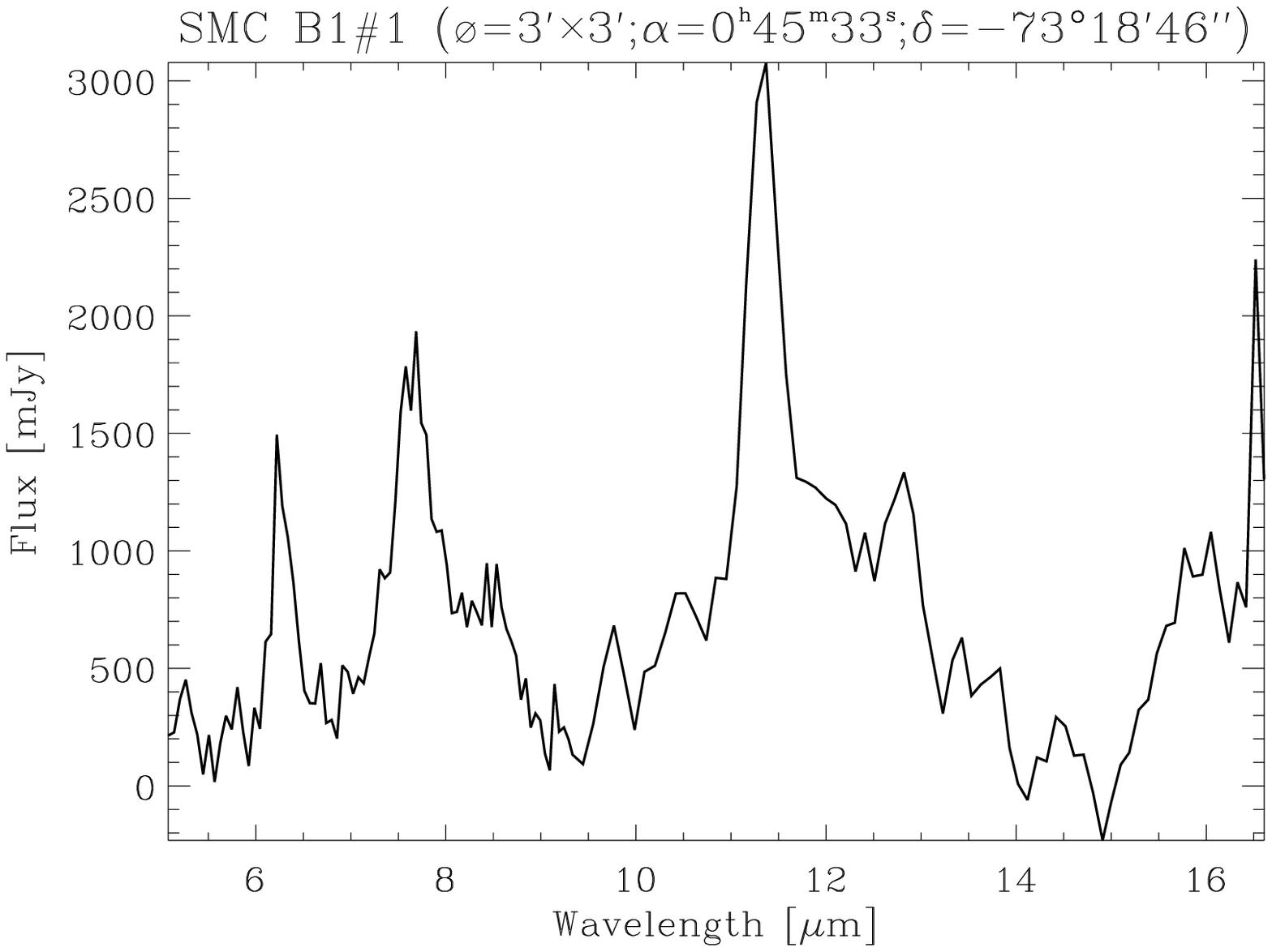}            \\
    \includegraphics[width=0.43\textwidth]{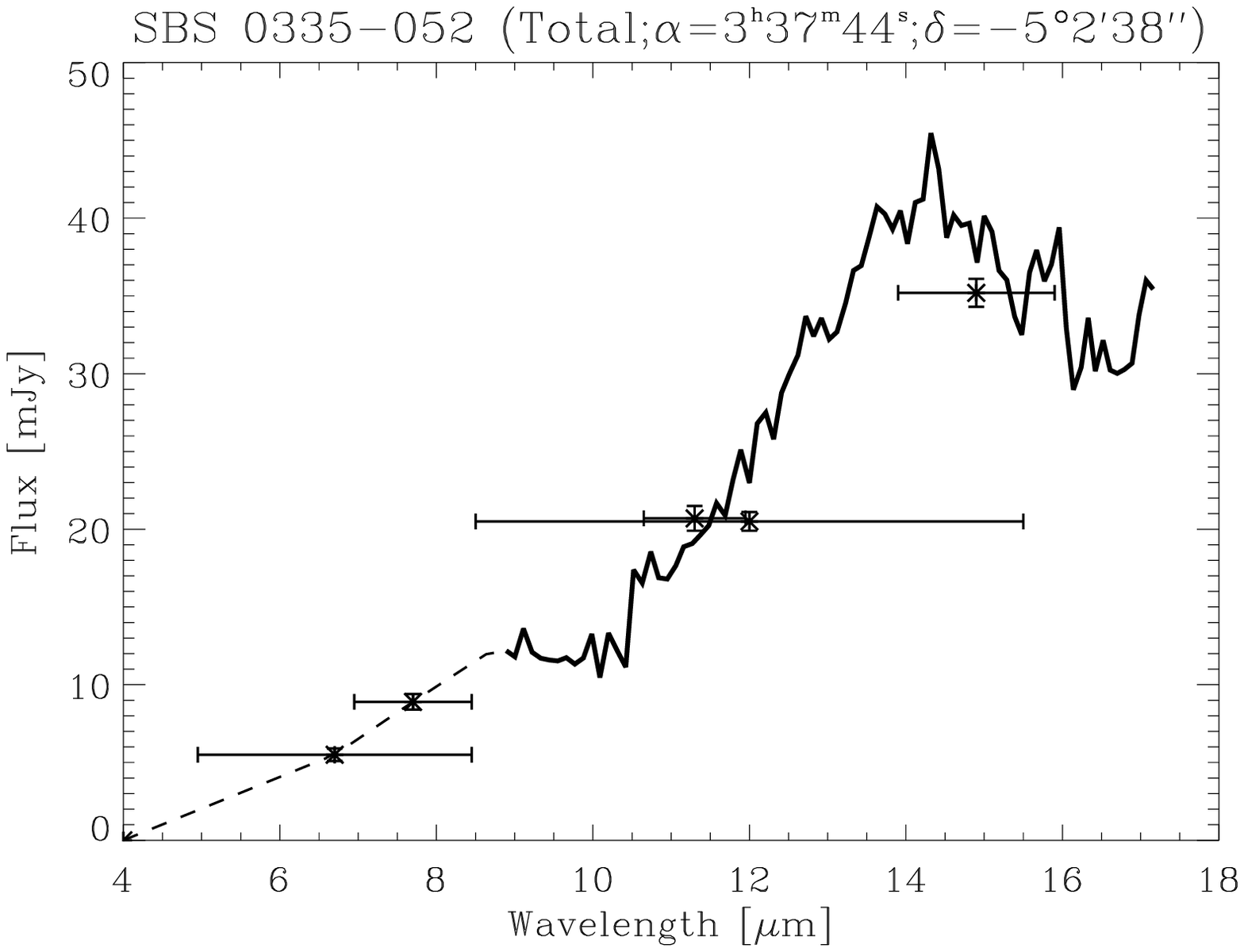}            &
    \includegraphics[width=0.43\textwidth]{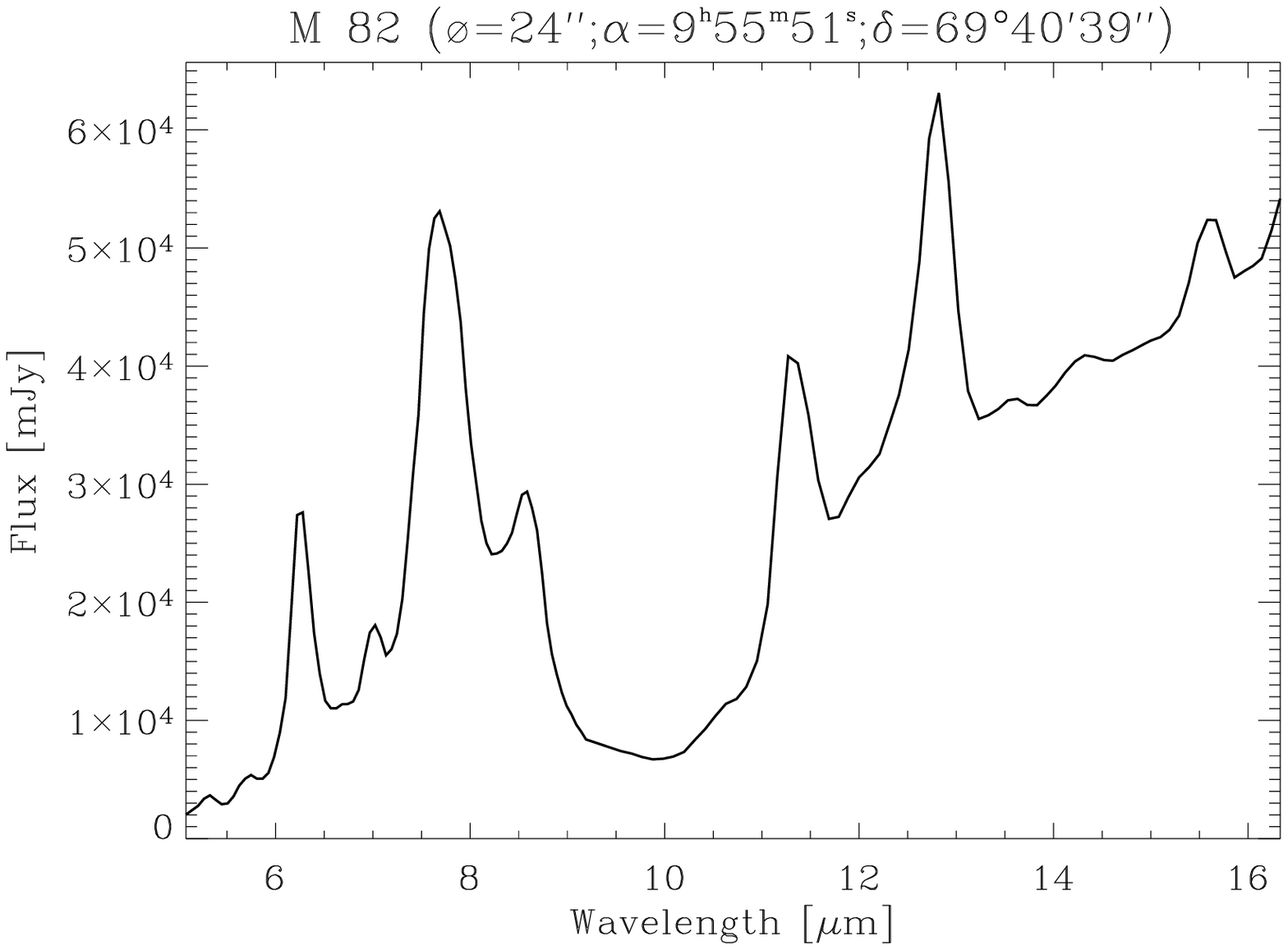}         \\
  \end{tabular}
  \caption{ISOCAM MIR spectra from the SMC: \smcn\ and \smcb\ and the very low
	   metallicity galaxy, \sbs\ \citep{thuan+99}. 
           For comparison with a metal rich starburst galaxy, we also include 
           the global ISOCAM CVF spectra of \M{82}. 
           The position of the central pointing as well as the diameter 
	   of the circular aperture are indicated next to the name of the 
           source, above each spectrum. 
           For comparison with a metal rich starburst galaxy, we also include 
           the ISOCAM CVF spectra of the central region of \M{82} in 
           Fig.~\ref{fig:spec2}. 
           The spectral features identified in the spectra are discussed in 
           Sect.~\ref{sec:components}.}
  \label{fig:spec2}
\end{figure*}

We reduced the ISOCAM data of \all\ and LMC~\xxxdor\ as well as the already 
published \smcn. 
\smcb\ is taken directly from \citet{reach+00}. 
For the data treatment we used and adapted the CIR 
\citep[CAM Interactive Reduction; Version: JAN01;][]{chanialphd} package 
incorporating private IDL routines. 
Great care was taken to examine the data between each step of the processing 
in order to verify the data reduction and to look for evidence of artifacts. 
Table~\ref{tab:sources} lists the low metallicity sources used in this study 
along with some source parameters.
We proceeded in the following way:
\\
1) \emph{Dark current subtraction}: 
The subtraction of the dark currents is performed using a model which predicts
the time evolution for each row of the detector \citep{darkCAM}, taking into 
account drifts along each orbit and each revolution. 
The correction implemented in CIR is a second order correction avoiding 
negative values.
\\
2) \emph{Cosmic ray impacts}: 
We masked the glitches using multi-resolution median filtering 
\citep{starck+99} on each block of data after slicing the cube. 
This method works well for common glitches except for faders and dippers
\citep{glitchCAM}. 
In addition, we perform manual deglitching after the transient correction 
(see below), examining the temporal cut for each pixel and masking the bad 
pixels. 
Thus, we were able to remove the remaining glitches which were not found by 
the automatic method. 
Moreover, we removed the slowly decreasing tails after each glitch. 
These were not always masked by the algorithm. 
We also examined the pixels in the vicinity of all of the strong glitches.
\\
3) \emph{Transient effects}: 
ISOCAM is subject to systematic memory effects due to the very long time 
needed for the signal to stabilize. 
We corrected this using the Fouks-Schubert method \citep{fsCAM} which 
provides good photometric accuracy without any fitting. 
We noticed that the corners and the edges of the field were not sufficiently 
illuminated, thus we mask the borders of the detector to allow proper 
flat-fielding and background subtraction.
\\
4) \emph{Flat-fielding}: 
The galaxies were observed with a single pointing.
Therefore, we could not use redundancy to build a flat-field for the entire
detector. 
We computed a hybrid flat-field image placing a mask on the source and
computing a flat field outside this mask from the median of the temporal cut 
for each pixel. 
For the pixels which were on-source, the flat-field response was set to the 
corresponding calibration flat-field.
\\
5) \emph{Flux conversion}: 
The conversion from Analog Digital Units to mJy/pixel was performed using the 
standard in-flight calibration data base.
\\
6) \emph{Sky subtraction}: 
The sky contributions can, in principle, contain contributions from our 
Galaxy, background galaxies, and Galactic zodiacal light which is usually the 
dominating source. 
To remove the sky contribution, the source was masked and, for a given 
wavelength, the median of the pixels which are off-source were subtracted from
each pixel. 
A $\sigma_{\rm zodiacal}$ was determined for the calculation of the total
uncertainties (below in paragraph~8), where 
$\sigma_{\rm zodiacal} = \sigma_{\rm {bkg}}/\sqrt{N_{bkg}}$. 
$N_{bkg}$ is the number of pixels used to measure the sky background and 
$\sigma_{\rm {bkg}}$ is the $1 \sigma$ uncertainty in the  background 
determination.
\\
7) \emph{Filtering}:  
We apply a multiresolution filtering method from the MR/1 package 
\citep{starckbook}. 
Each image corresponding to a given wavelength was filtered with a threshold 
of $3\sigma$. 
The final spectra are shown in Fig.~\ref{fig:spec1} and Fig.~\ref{fig:spec2}.
\\
8) \emph{Evaluation of uncertainties}: 
To estimate the uncertainty $\Delta F_\nu (\lambda)$ on the net flux 
$F_\nu (\lambda)$ integrated in a circular aperture $\Theta$, we quantify 
the various contributions induced by the data reduction steps and the 
remaining errors. 
We propagate the statistical fluctuations for each pixel $(i,j;\lambda)$ at 
a given wavelength $\lambda$, $\sigma_{\rm RMS}(i,j;\lambda)$ along each 
step of the processing. 
The total uncertainty due to these fluctuations on $F_\nu (\lambda)$ in 
$\Theta$ is $2 \sigma_{RMS}$ or $\Delta F_\nu^{\rm RMS} (\lambda) = 
2\times \sqrt{ \sum_{(i,j)\in\Theta} \sigma_{\rm RMS}^2(i,j;\lambda) }$.
We estimate the error due to the sky subtraction (described above in 
paragraph~6) using the standard deviation, $\sigma_{\rm zodiacal}$ of the 
distribution of the points used to compute the median.
This error is 
$\Delta F_\nu^{\rm zodiacal} (\lambda) = 2\times \sqrt{ \sum_{(i,j)\in\Theta}
\sigma_{\rm zodiacal}^2 } = \sqrt{ N_\Theta \sigma_{\rm zodiacal}^2 }$ 
where $N_\Theta$ is the number of pixels inside the aperture $\Theta$. 
The uncertainty due to remaining memory effects was estimated by considering 
the amplitude of the variations of each pixel up to 
$3\sigma_{\rm RMS}$, $\delta F_{3\sigma}(i,j;\lambda)$ within a block of data. 
We also take into account the time needed for the signal to stabilize using 
a factor, $f_{\rm stabilization}$, which is equal to 1 if the signal is 
assumed to be stabilised and equals $(120\, s)/\Delta t$ if not stabilised 
\citep{roussel+01a}. 
The uncertainty due to transients is, finally, 
$\Delta F_\nu^{\rm transients} (\lambda) =
\sqrt{\sum_{(i,j)\in\Theta} \left( \delta F_{3\sigma}(i,j;\lambda) 
\times f_{\rm stabilization}(\lambda) \right)^2 }$. 
The total uncertainty on the net flux is the sum of these different 
contributions plus $5\,\%$ of the flux due to absolute calibration errors and 
$5\;\%$ due to variations along each orbit \cite{camhandbook}:
\begin{eqnarray}
  \Delta F_\nu (\lambda) & = & \left[
                               (\Delta F_\nu^{\rm RMS} (\lambda))^2
                               + 2\times (0.05\times F_\nu (\lambda))^2
                               \right. \nonumber \\
                         & + & \left.
                               (\Delta F_\nu^{\rm zodiacal}(\lambda))^2
                               + (\Delta F_\nu^{\rm transients} (\lambda))^2
                               \right]^{1/2} .
\end{eqnarray}
Table~\ref{tab:errCAM} gives an example of the level of the uncertainties for
each step, and total, for \ngc{1569}.
\begin{table}[htbp]
  \begin{center}
  \caption{Example of the various uncertainties quantified for the global 
           CVF of \ngc{1569}.}
  \begin{tabularx}{\linewidth}{l*{4}{X}}
    \hline
    \hline
          &  RMS  &  Transients
      &  Zodiacal &  Total \\
    \hline
      Average                & $13\;\%$ & $22\;\%$
      & $3\;\%$      & $27\;\%$  \\
      $\lambda = 16 \mic$ & $13\;\%$ & $9\;\%$
      & $2\;\%$      & $17\;\%$  \\
      $\lambda = 4 \mic$  & $66\;\%$ & $31\;\%$
      & $13\;\%$     & $75\;\%$  \\
    \hline
  \end{tabularx}
  \label{tab:errCAM}
  \end{center}
\end{table}
\\
9) \emph{Comparison with IRAS $12\mic$}:
To verify the photometry of our newly-presented sources, we integrate our CVF
specrta over the $12\mic$ IRAS band and measure the flux of the resulting 
image in the same aperture. 
For \ngc{1569}, we find a flux of $(880\pm 200)$~mJy using our ISOCAM spectra. 
The IRAS $12\mic$ flux given by \citet{melisse+94} is 508~mJy and is flagged 
by the authors to be uncertain. 
The IRAS flux given by \citet{hunter+89} is 
$F_{\rm 12\mic} = (880\pm 100)$~mJy which is in excellent agreement with 
the ISOCAM spectrum. 
For \iizw, we find a flux of $(290\pm 60)$~mJy integrated in the $12\mic$ 
IRAS band and the IRAS flux given by \citet{melisse+94} is 
$F_{\rm 12\mic} = 415$~mJy. 
This is higher than our measured flux value
and may indicate a substantial extended very low level $12\mic$ flux. 
\citet{beck+02} observed this galaxy with the Keck 1 Telescope and found a 
total flux of $F_{\rm 11.7\mic} = 240$~mJy. 
The ground-based narrow band measurements are consistent with our $11.7\mic$ 
flux from the ISOCAM CVF spectrum. 
For \ngc{1140}, we find a flux of $(96\pm 30)$~mJy, in agreement with the
$12\mic$ IRAS flux value from \citet{melisse+94}, who measure
$F_{\rm 12\,\mu m} = 73$~mJy.
\\

\section{Spectral analysis}
\label{sec:anal}

The MIR wavelength range covered by ISOCAM CVF ranges from
$\lambda=5$ to $16\mic$ and contains a rich assortment of tracers of various
components of gas and dust in galaxies. 
Fine structure emission lines arising from \hii\ regions, silicate features, 
aromatic hydrocarbon features, dust continuum and possibly stellar continuum, 
are all interwoven into a MIR spectrum which can be disentangled to extract 
clues of the physics of star formation. 
When observing galaxies, the telescope beam will contain a variety of 
components of the ISM and star formation regions. 
The relative contributions of each component depends on many physical 
parameters, including metallicity, star formation activity and ISM structure 
--~all conspiring together to govern the global observed emission.

Two components of dust emission can be present in MIR spectra:
\begin{itemize}
\item 
The aromatic MIR band features, the major ones being at $\lambda=6.2$, 7.7, 
8.6, 11.3, $12.6\mic$, are thought to come from large molecules or very small 
grains of carbonaceous origins, for example, PAHs 
\citep{leger+84,allamandola+85}. 
Due to their very small nature, these particles undergo thermal fluctuations 
of a few hundreds of degrees K upon single photon absorption 
\citep[e.g.][]{sellgren+90,verstraete+01}. 
From spectro-imaging of Galactic regions, the PAHs are seen to peak in 
photodissociation regions (PDRs) --~the interface envelopes between \hii\ 
regions and molecular clouds \citep{cesarskyd+96}.
\item 
Very Small Grains \citep[VSGs;]{desert+90}, wavelengths 
longward of $\lambda\simeq 10\mic$. 
The VSGs can be either in thermal equilibrium with the radiation field, or 
stochastically heated to high temperatures, depending on the radiation field 
and the sizes of the grains.
In the Galaxy, these particles range in size from $\lambda=1$~nm to 10~nm  
and are thought to be mostly stochastically heated in the diffuse ISM 
\citep{desert+90}. 
This component is seen to peak in \hii\ regions, when observed with 
sufficient spatial resolution \citep{cesarskyd+96,verstraete+01}.
\end{itemize}

Both PAHs and VSGs are heated by UV photons in the vicinity of active star
formation regions, with the PAHs observed to peak in the PDRs, and the VSGs
emitting prominently in the nebular regions 
\citep{cesarskyd+96,verstraete+96}. 
PAHs can also be excited by optical photons, as shown to be the case in 
reflection nebulae \citep{uchida+98}, and by their presence in elliptical 
galaxies \citep[e.g.][]{athey+02,xilouris+04}.

In the disks of spiral gaxies, the PAH emission bands are a very prominent 
component of the ISM, while the continuum due to the VSGs is less so 
\citep[e.g.][]{dale+00,roussel+01b,vogler+05}. 
In the nuclei of spiral galaxies, starbursts or active galactic nuclei, 
various nebular fine structure lines can also be observed in the ISO MIR 
wavelength range, notably the \neiiline, \neiiiline, \ariiiline, \ariiline\ 
and \sivline\ lines \citep
{genzel+98,thornley+00,laurent+00,rigopoulou+02,sturm+02,lutz+03,forster+03, 
vogler+05}. 

\subsection{Comparison with a Galactic \hii\ region/PDR: \M{17}}
\label{sec:m17} 

Lacking sufficient spatial resolution to isolate various physical components 
within galaxies, it is also possible to decompose a global spectra with
known template spectra representative of, for example, \hii\ regions, 
PDR regions, diffuse regions, etc. 
We use ISOCAM spectra from the Galactic region, \M{17}~SW, as one way to 
interpret the spectra of the dwarf galaxies, \ngc{1569}, \ngc{1140} and 
\iizw, which exhibit very different MIR properties.

\begin{figure*}[htbp]
  \centering
  \begin{tabular}{cc}
    \includegraphics[width=0.48\textwidth]{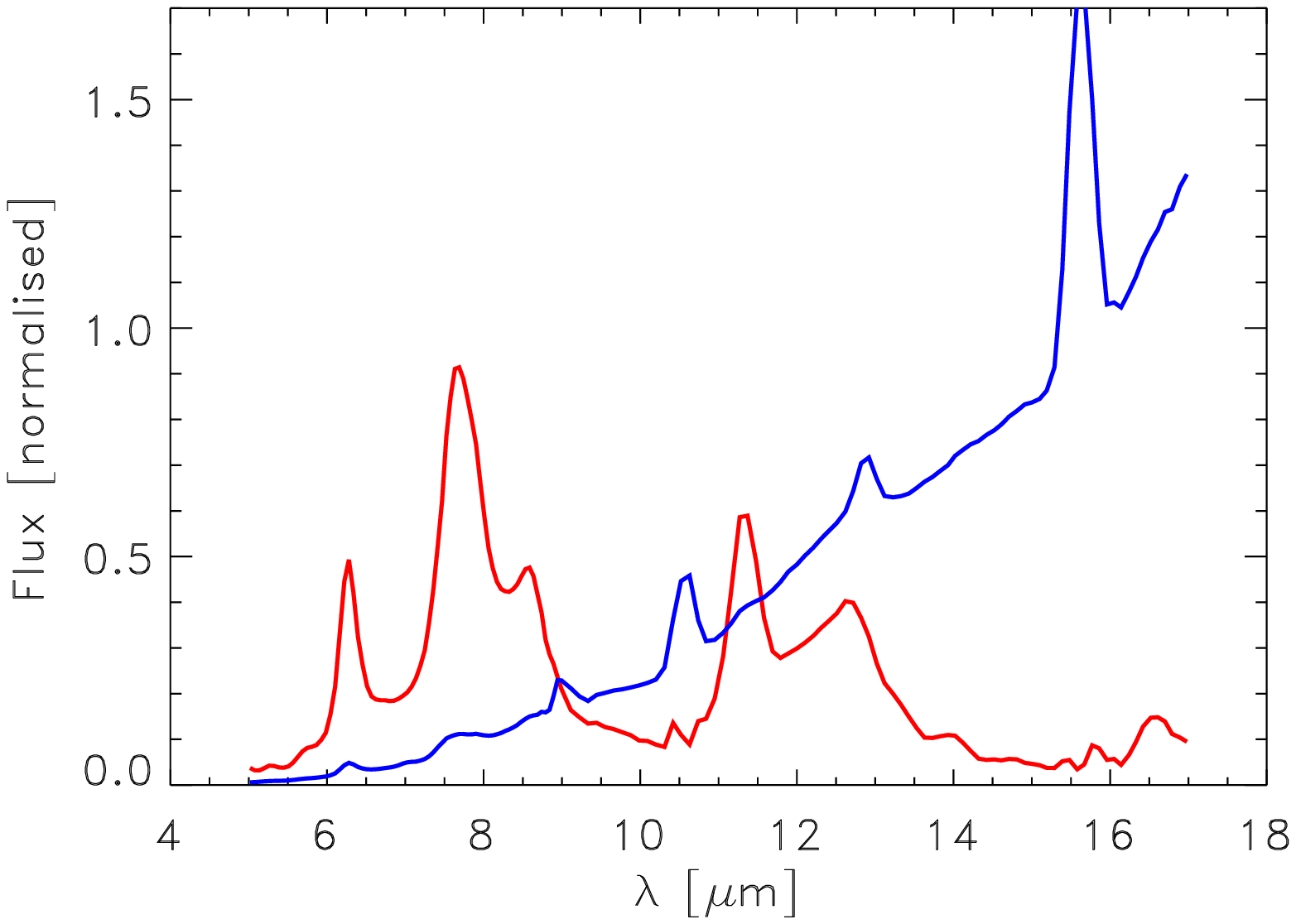} &
   \includegraphics[width=0.49\textwidth]{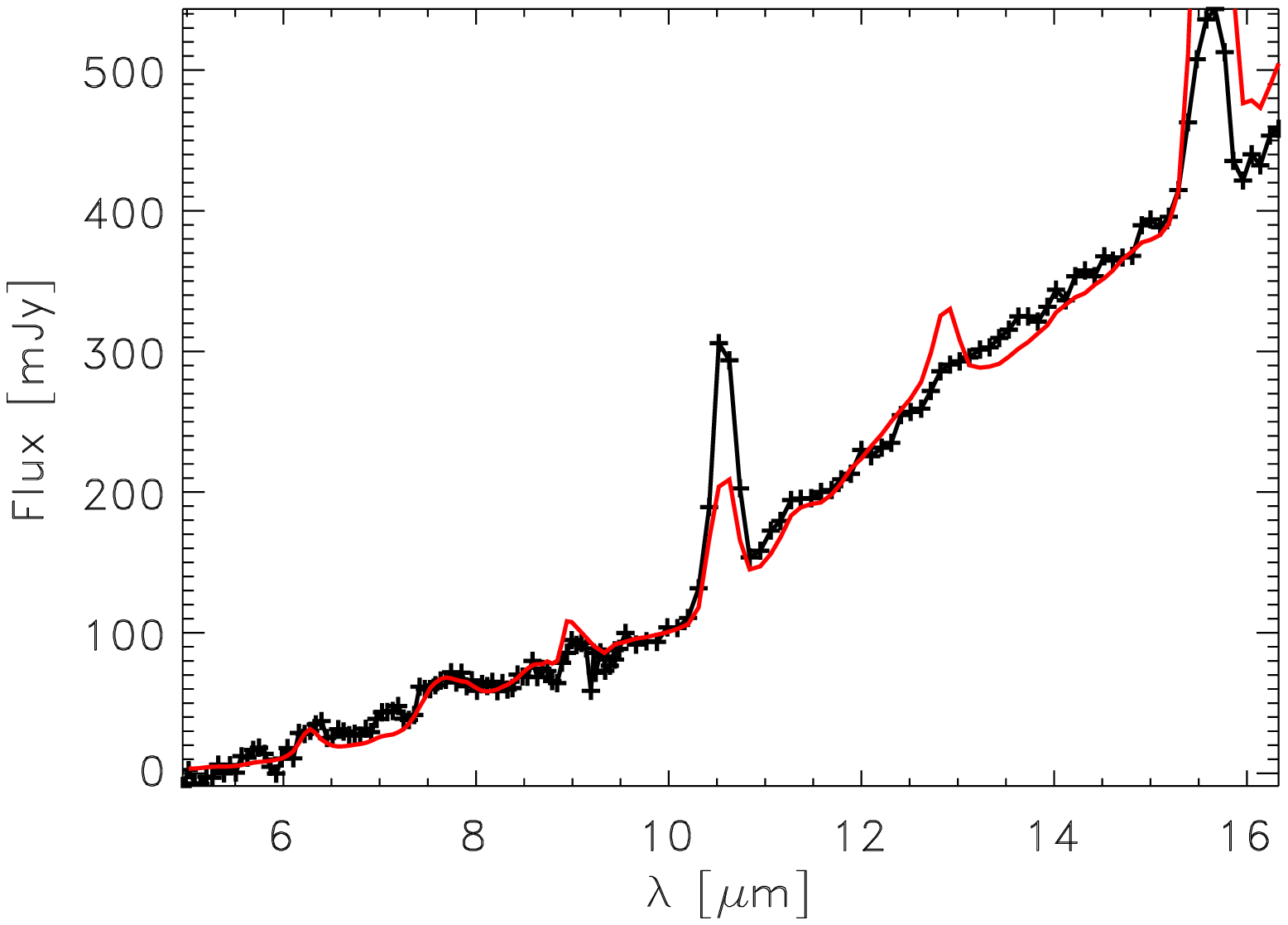}  \\
    \includegraphics[width=0.48\textwidth]{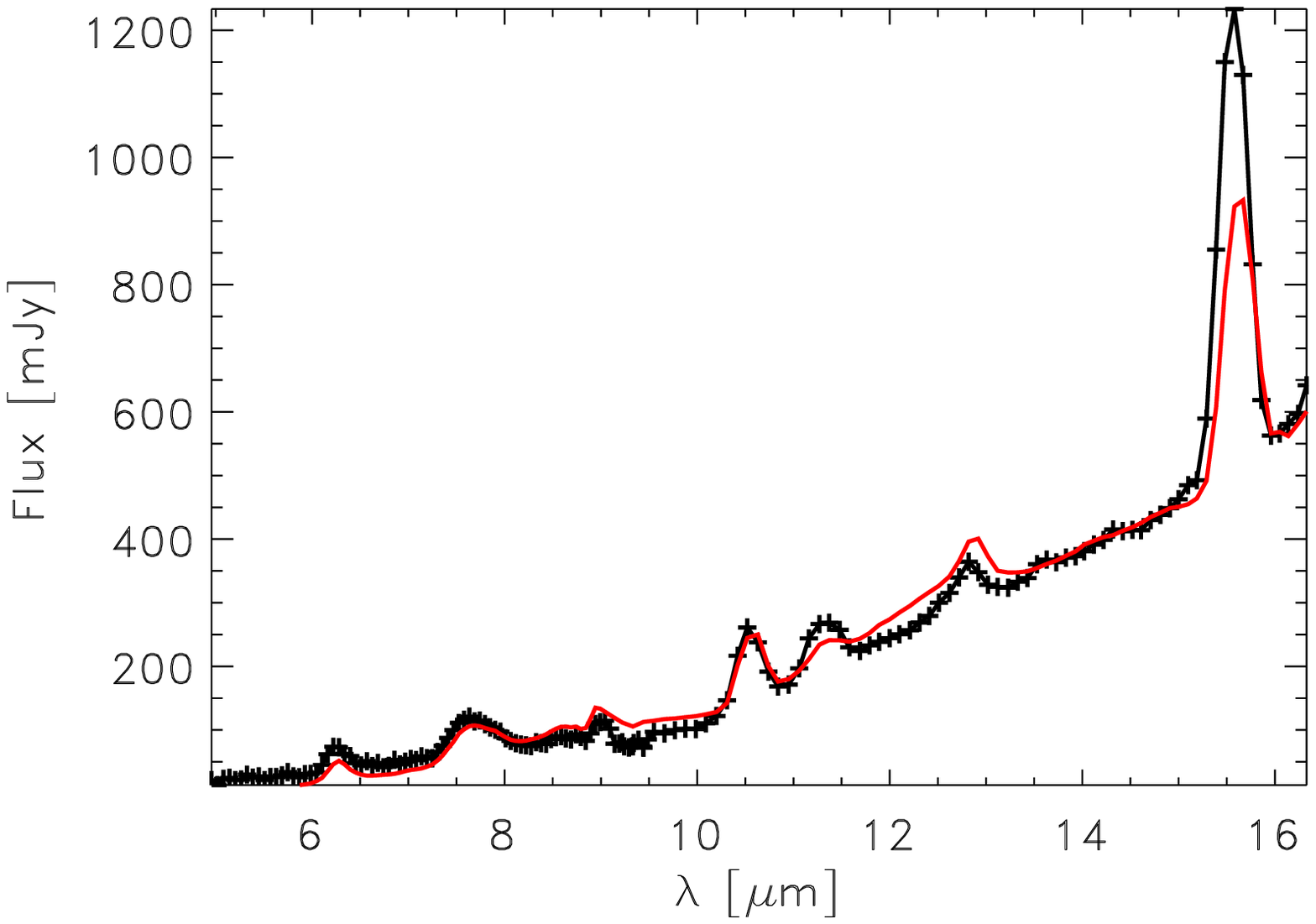} &
    \includegraphics[width=0.48\textwidth]{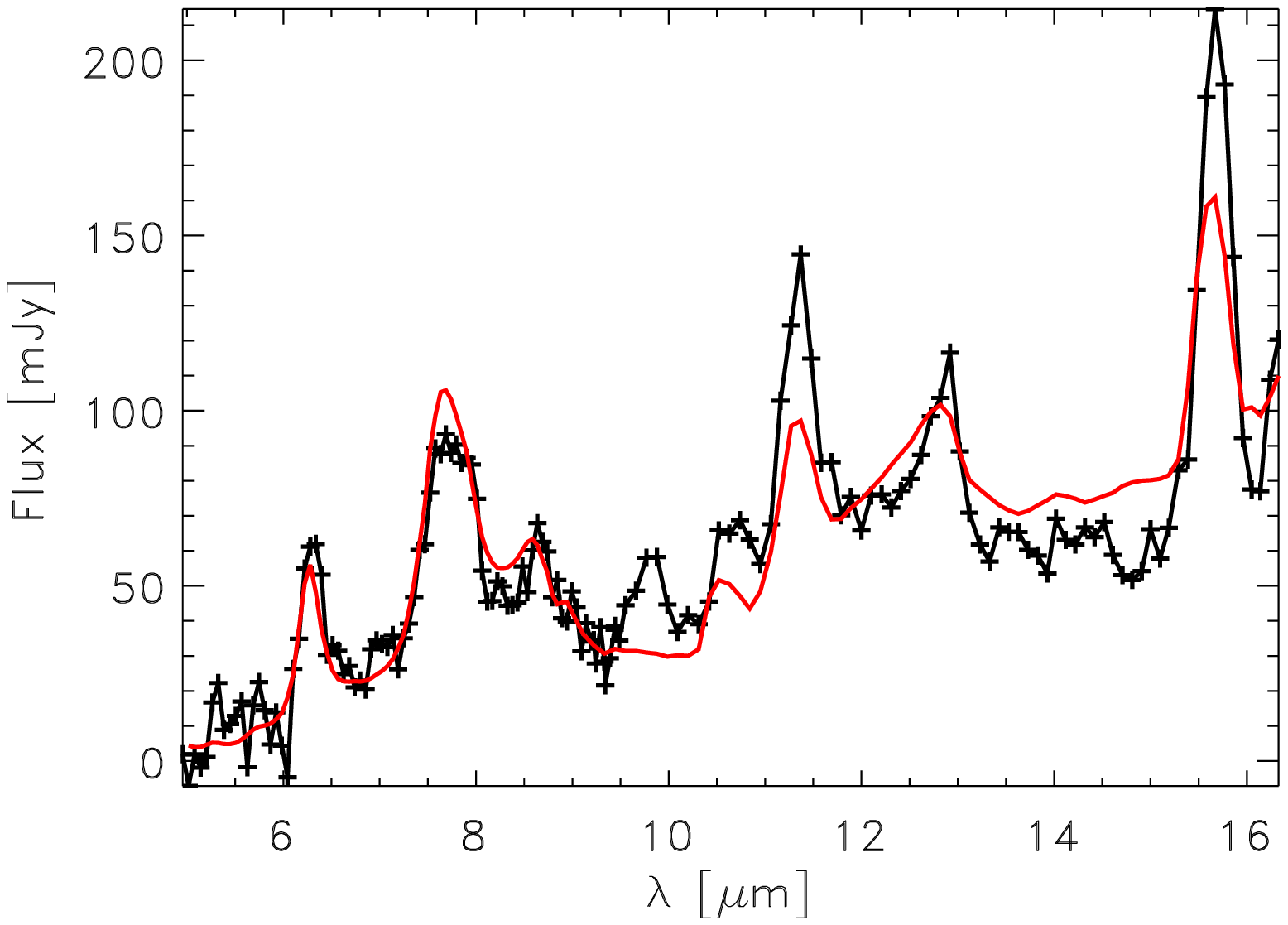}  \\
  \end{tabular}
  \caption{Template \hii\ region (blue) and PDR spectra (red) from M17  
           (upper left), used as examples to interprete the global MIR spectra
           of 3 dwarf galaxies in our sample: \iizw\ (top right), \ngc{1569} 
           (lower left) and \ngc{1140} (lower right). 
           The galaxy spectra are global averages with the observations 
           represented by black profiles and crosses and the red profiles are 
           the total spectra modeled with combinations of the template \hii\ 
           region and PDR spectra.
           The \hii\ template was extracted in a $30''\times30''$ aperture
           centered on $(18^h20^m26\fs2,-16\degr11\arcmin25.4\arcsec)$, and
           the PDR template, in a $30''\times30''$ aperture
           centered on $(18^h20^m19\fs9,-16\degr12\arcmin17.5\arcsec)$.}
  \label{fig:m17dwarfs}
\end{figure*}

\M{17} has been studied in great detail \citep{cesarskyd+96}: it contains an 
\hii\ region and PDR interface adjacent to a molecular cloud, illuminated by 
an O3 star and is viewed almost edge-on. 
With the spatial resolution of ISOCAM, individual physical regions can be 
isolated and used as templates to interpret the dwarf galaxy spectra. 
ISOCAM spectra of a typical \hii\ region and a PDR region, extracted from the 
\M{17} ISOCAM image \citep{cesarskyd+96} are shown in Fig.~\ref{fig:m17dwarfs}.
PAH emission peaks in the PDR interfacing the \hii\ region and molecular 
cloud, also extending somewhat into both regions. 
Inside the \hii\ region, the continuum emission from the hot VSGs is the 
principal dust component while the PAH emission is supressed, presumably 
destroyed by the hard, intense radiation field within \hii\ regions 
\citep[e.g.][]{leach87,voit92,siebenmorgen+04}. 
The template spectra were normalised by the total energy in the 
wavelength range and best combinations of these characteristic template 
spectra were then determined, using a least squares fit to infer the 
proportion of ionised and PDR material in the MIR spectra. 
When observing galaxies, an ensemble of physical components are in the beam, 
making it difficult to isolate the physical processes.

The global spectrum of \iizw\ is clearly dominated by contributions from \hii\
regions  as evidenced by the steeply-rising continuum and the prominent 
\neiii\ and \siv\ lines. 
Our crude model finds the MIR emission of \iizw\ is made up of emission from 
a factor of 30 more characteristic \hii\ regions compared to PDR-type material
when matching the shape of the continuum and the PAH bands on the global scale
(Fig.~\ref{fig:m17dwarfs}). 
Modeling the spectrum toward the central peak only, the ratio of \hii\ regions
to PDRs in the beam reduces to 20, since in this case, we are focusing on most
of the PDR gas as well as the ionised gas, all peaking toward the unresolved 
center. 
There appears to be a non-negligible component of {\it extended} ionised gas 
throughout the galaxy, when comparing the global and peaked spectra. 
Only traces of PDRs can be observed in the spectra in the form of very low 
level PAH bands in \iizw. 

The global MIR spectrum of \ngc{1569} is also dominated by \hii\ regions, 
with 20\%\ PDR origin and 80\%\ \hii\ region origin when matching the PAH 
emission and the VSG continuum in the global spectrum 
(Fig.~\ref{fig:m17dwarfs}). 
Note that when the peak spectrum is modeled, the contribution from \hii\ 
regions increases to a factor of 10 above that of PDRs.  
The \neiii\ and \siv\ lines are particularly intense in the dwarf galaxies, 
and are not reproduced well by simply using the template spectrum of the 
nebular emission surrounding the single O3 star from \M{17}. 
While the overall shape of the \iizw, \ngc{1569} and \M{17} \hii\ region 
spectra are similar, the differences in metallicity can explain the effects of
the ionic lines. 
For example, simply multiplying up the \hii\ region template spectrum of 
\M{17} to match those of the continuum and ionic lines of the dwarf galaxies, 
is not sufficient. 
The ionic lines require further modeling to be able to draw any quantitative 
information here. 
The nebular line characteristics are discussed further in 
Sect.~\ref{sec:neiii_neii}. 

The MIR spectrum of \ngc{1140}, which has the most prominent PAH emission and 
obviously flatter VSG continuum, is composed of equal contributions of a PDR 
component and \hii\ regions (Fig.~\ref{fig:m17dwarfs}). 
Since the extended emission is not very well resolved here, there is little 
difference in the MIR characteristic when inspecting the global spectrum or 
the peak spectrum.

\subsection{MIR component separation}
\label{sec:components}

\begin{figure*}[htbp]
  \centering
  \begin{tabular}{cc}
  \vspace*{-0.5cm}
    \includegraphics[width=0.48\textwidth]{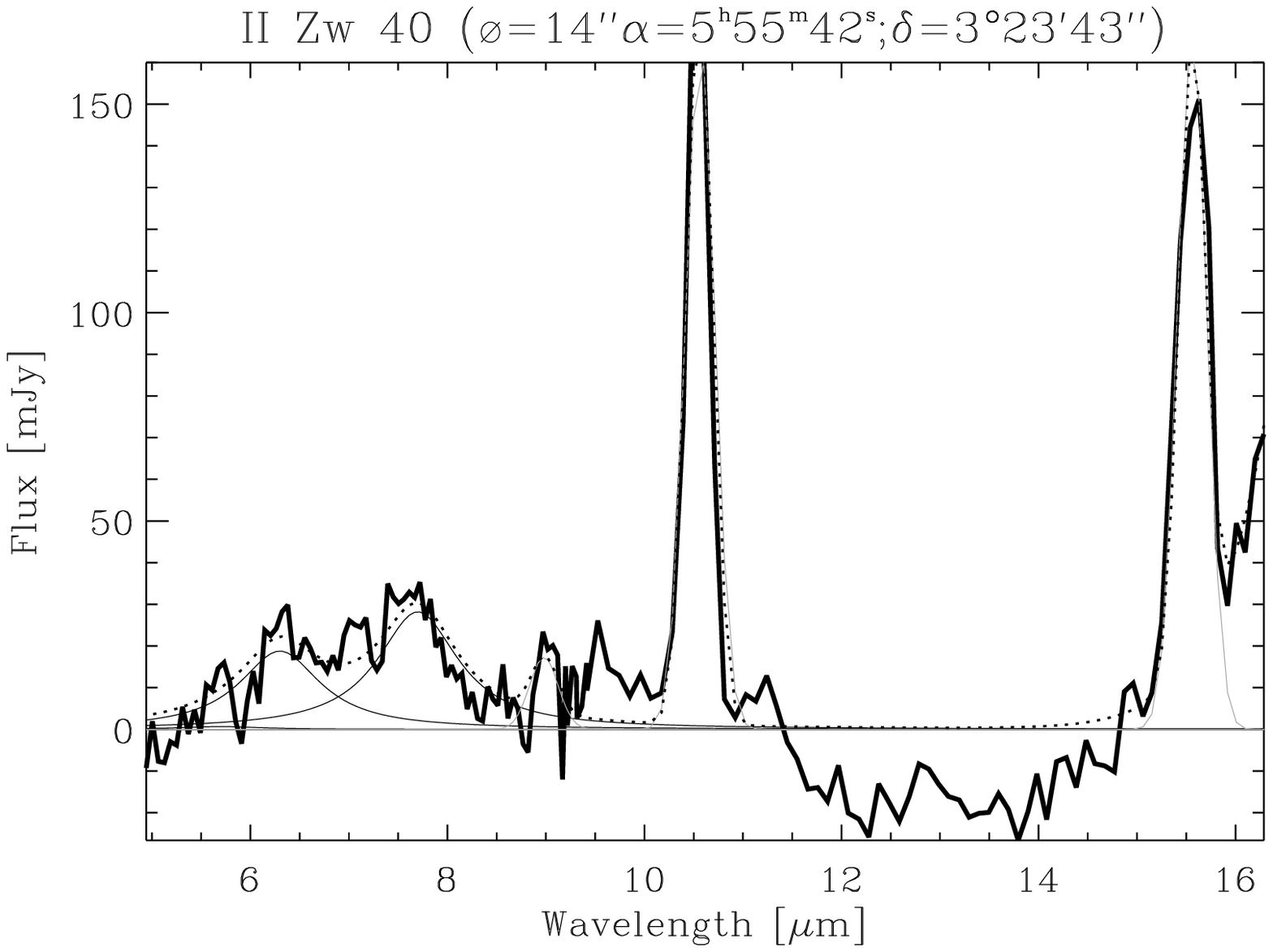}   &
    \includegraphics[width=0.48\textwidth]{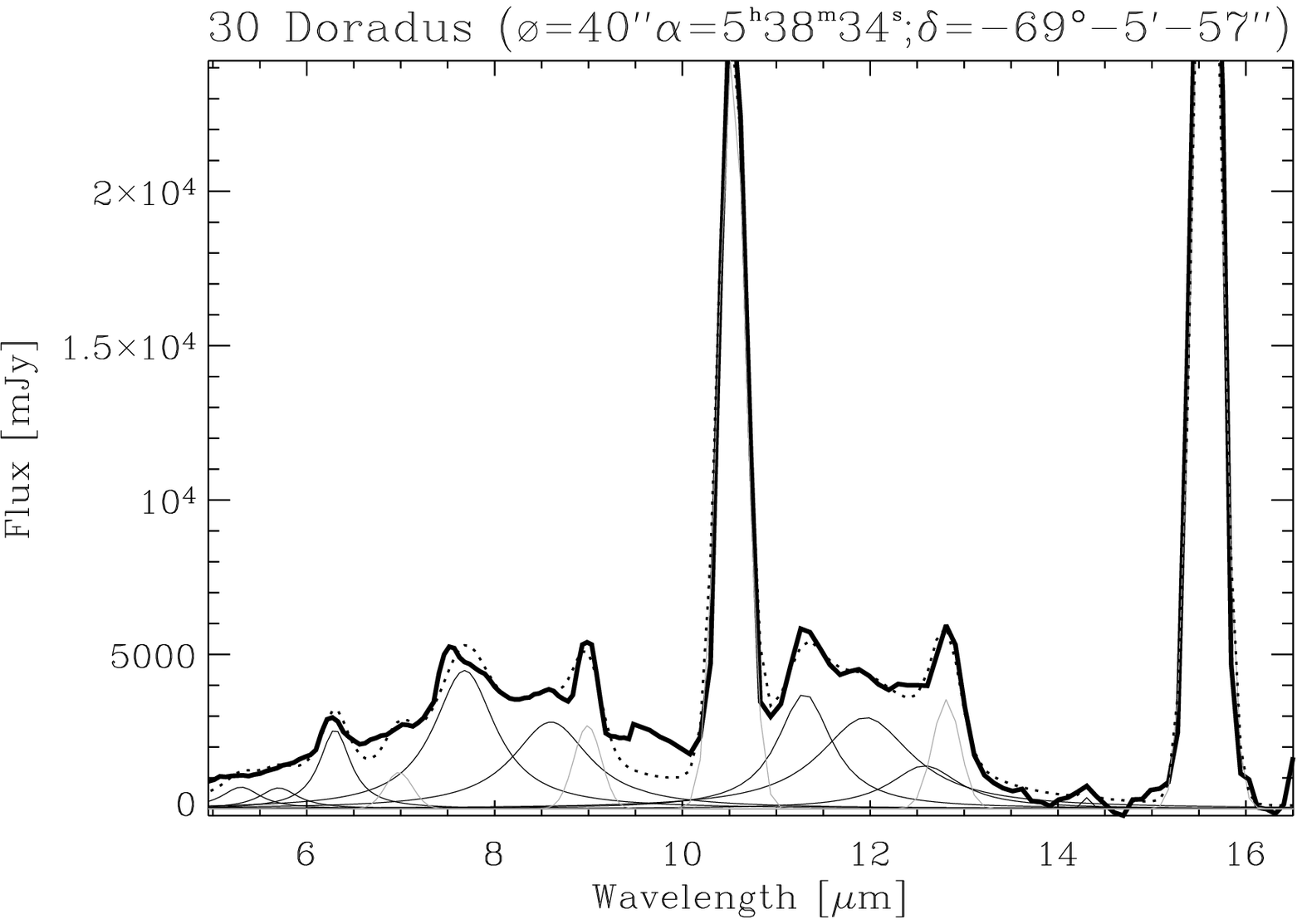}    \\
  \vspace*{0.2cm}
    \includegraphics[width=0.48\textwidth]{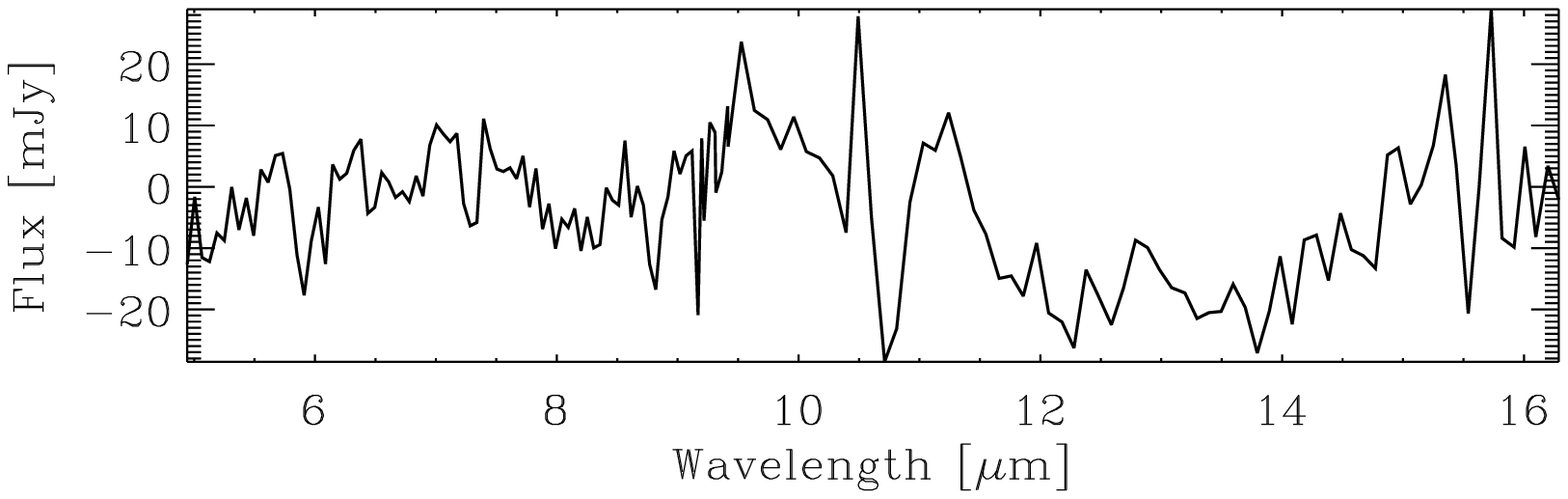}   &
    \includegraphics[width=0.48\textwidth]{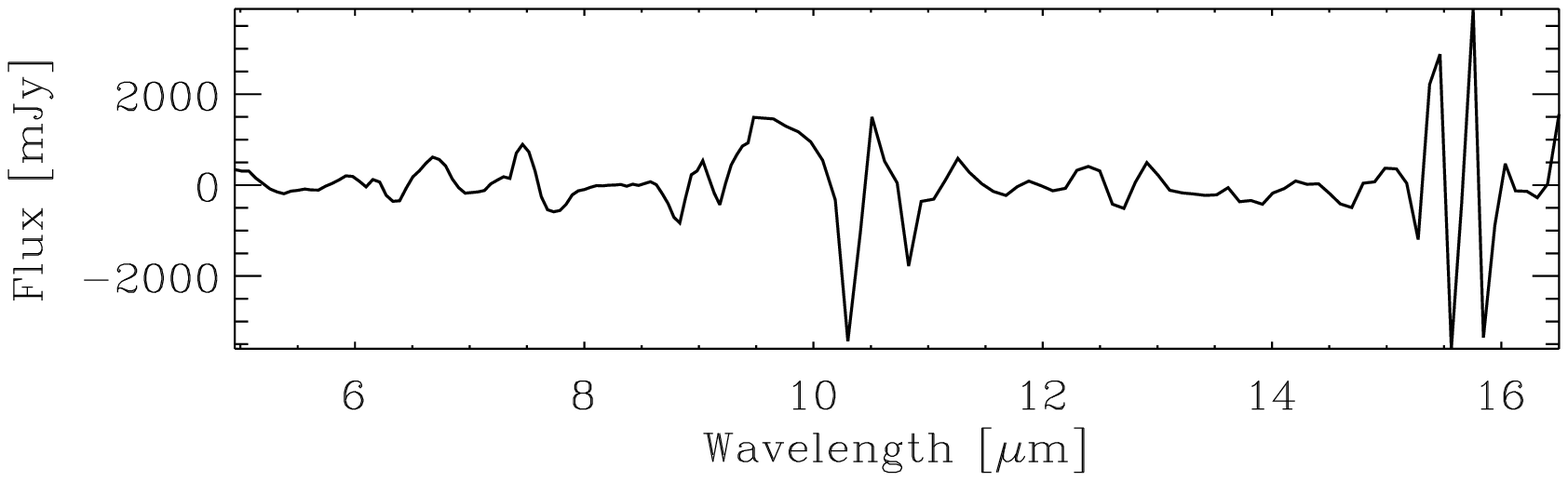}    \\
  \vspace*{-0.5cm}
    \includegraphics[width=0.48\textwidth]{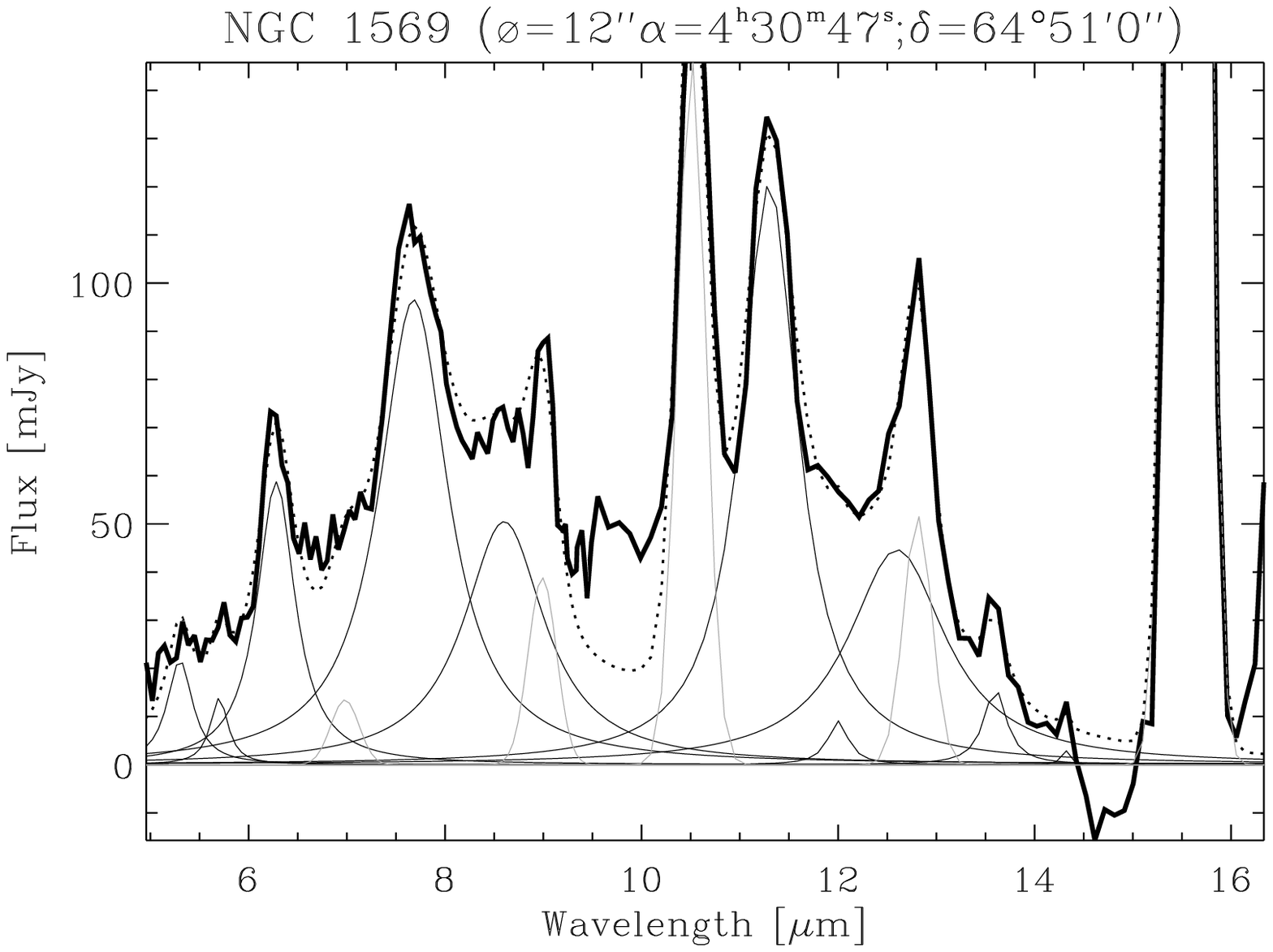}  &
    \includegraphics[width=0.48\textwidth]{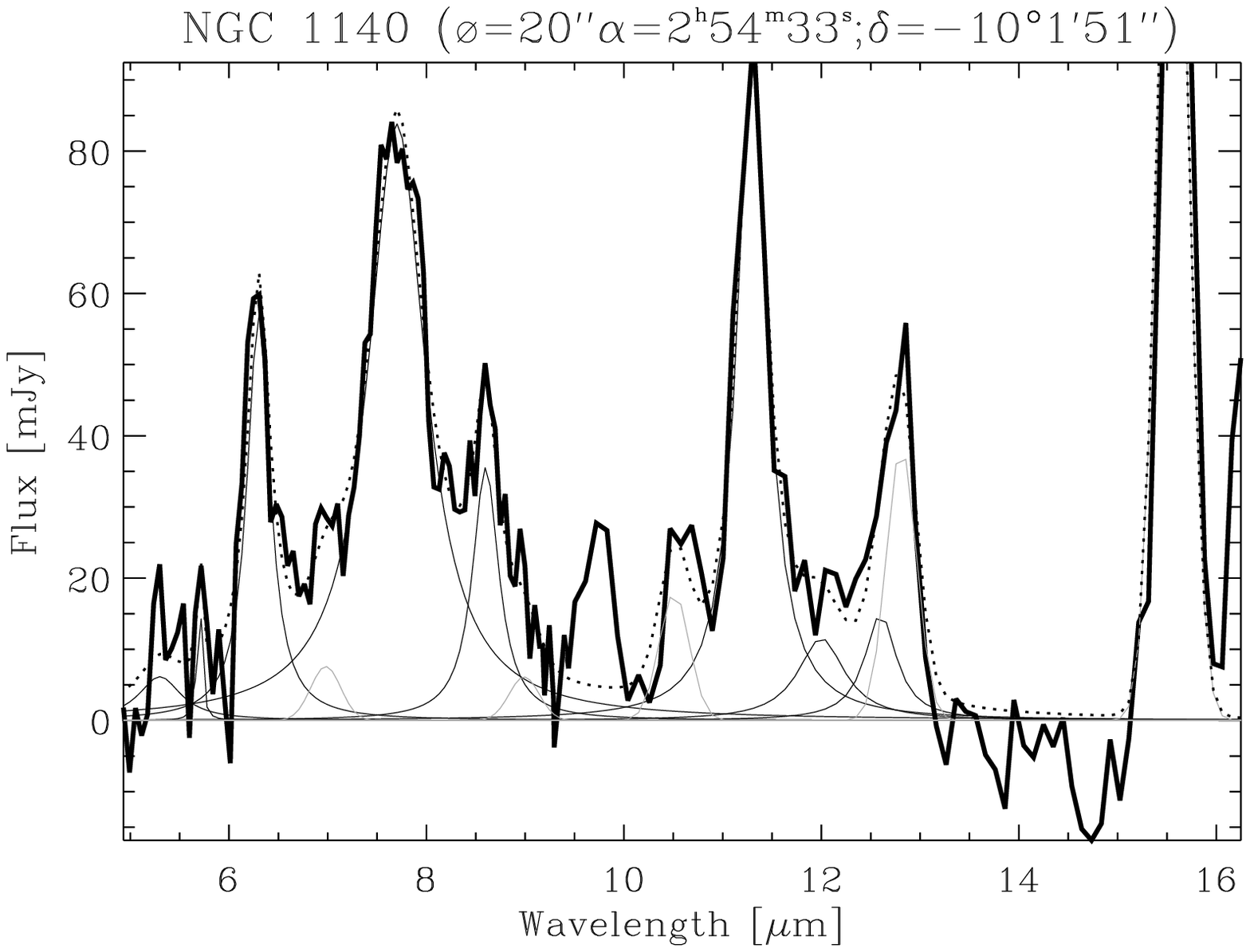}  \\
  \vspace*{0.2cm}
    \includegraphics[width=0.48\textwidth]{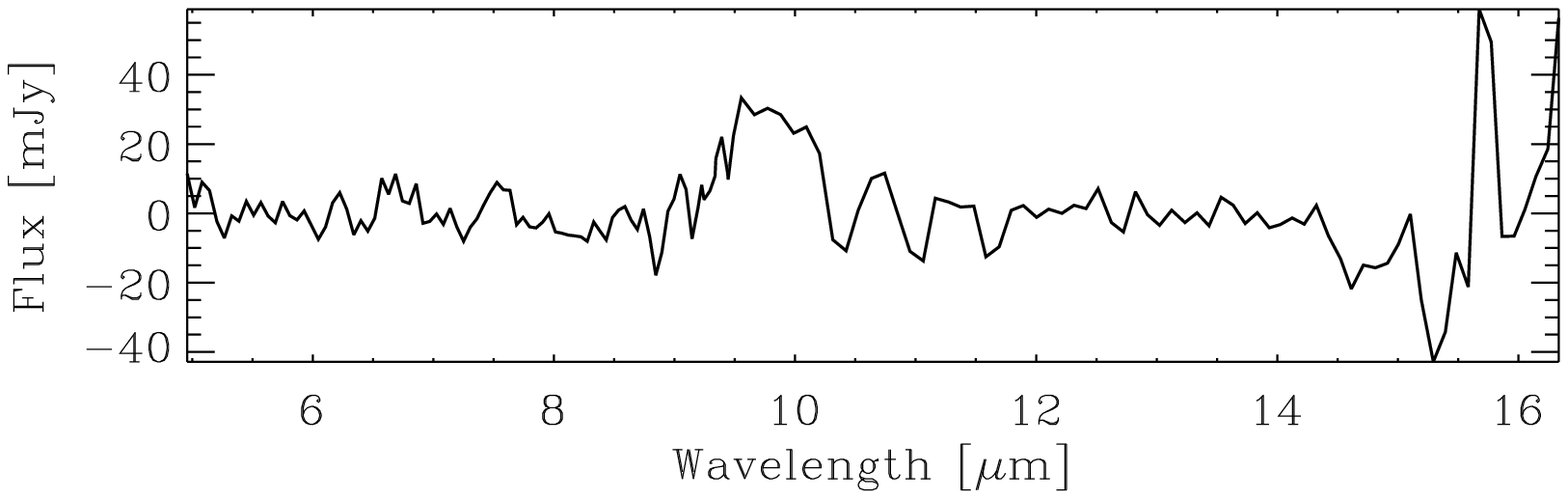}  &
    \includegraphics[width=0.48\textwidth]{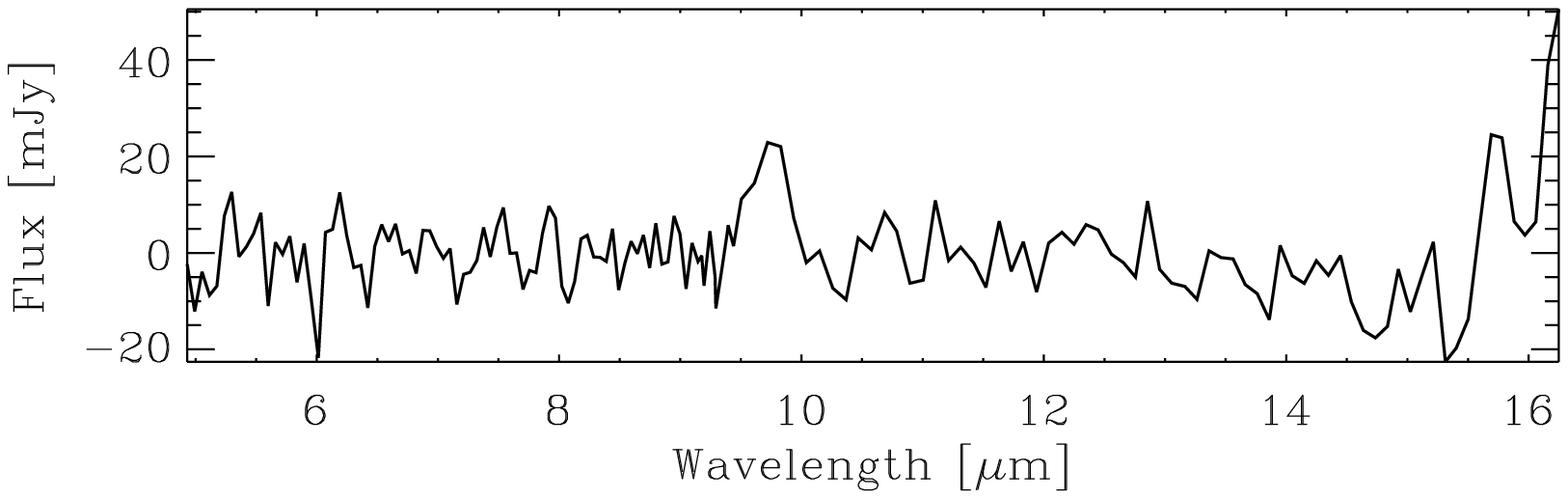}  \\
  \end{tabular}
  \caption{Examples of the fits of the ionic lines and the PAH bands. 
           The thick black line is the original spectrum after subtraction of 
           the continuum (see Sect.~\ref{sec:components} for line and 
           continuum fitting procedure); thin black lines are the Lorentzian 
           PAH band fits; thin grey lines are the Gaussian ionic line fits. 
           The sum of the model components are the dotted lines. 
           The residuals after the line and continuum fitting procedures 
           are also shown in the smaller panels beneath the complete spectra.}
  \label{fig:fitlines}
\end{figure*}

To analyze these spectra and separate the various MIR components, we followed 
the fitting procedure detailed in \citet{gallianophd, galliano+05}. 
We proceeded as follows:

\begin{enumerate}
  \item {\it small grain continuum: VSGs} A modified black body is used to model 
        the very small grains (VSGs). 
        We use an emissivity, $\beta = 1$, assuming that these very small 
        grains are mainly carbonaceous.
        The fit is not very sensitive to the value of $\beta$.
        We have also tried a solution with a power-law instead of a black body
        to model the very small grain continuum, however, this solution is
        mathematically less stable for noisy spectra. 
        Throughout the paper, where we refer to the VSGs in images or 
        plots, we are referring to this fitted VSG component, integrated 
        between 10 and 16 $\mic$.
  \item {\it nebular lines:} Gaussian functions are applied to the fitting of the 
	ionic lines.  
        We fixed the central wavelength and computed the line widths and the 
        line intensites.
        The lines included for the dwarf galaxies are: \neiiiline, \neiiline, 
        \sivline, \ariiiline\ and \ariiline. 
        At the spectral resolution of ISOCAM, the \neiiline\ is blended with 
        the $12.6\mic$ PAH band. 
        The fitted width of the line with the best signal-to-noise, usually 
        the \neiii\ or \siv\ line, is presumed to be the fixed width of the 
        other gas lines.  
  \item {\it PAH bands:} PAH bands are fitted with Lorentzian functions by 
        fixing the central wavelengths and leaving the width and intensity as 
        free parameters. 
        The bands taken into account here are: $5.3\mic$, $5.7\mic$, 
        $6.2\mic$, $7.7\mic$, $8.6\mic$, $11.3\mic$, $12.0\mic$, $12.6\mic$, 
        $13.6\mic$ and $14.3\mic$. 
        Note that where we refer to PAH bands in the following images
        or plots, we are referring to the integrated intensity of this whole 
        fitted component - essentially all of the PAH bands.
  \item The sum of these components is reddened with a mid-IR extinction 
        curve of \citet{mathis90}, with the silicate profiles of 
        \citet{dudley+97}. 
        This accounts for the intrinsic extinction of the galaxies. 
        The magnitude of the extinction is a free parameter.
\end{enumerate}
This fit is done after redshift correction, where necessary.
Each component is fitted simultaneously to obtain the best $\chi^2$. 
Examples of the line fits and their residuals are shown in 
Fig.~\ref{fig:fitlines}.

The residuals, shown in the small panels in Fig.~\ref{fig:fitlines}, appear 
relatively uniform, except for a noisy potential feature near $10\mic$ 
coinciding in wavelength with the SiO stretching silicate band, which is not 
included in the model for the emission spectra. 
It also appears close in wavelength to the position where the 2 grating 
spectrometers of the ISOCAM overlap (one CVF: $\lambda=5\mic$ to $9.5\mic$ 
and the other CVF: $\lambda=9\mic$ to $16\mic$). 
The residual feature near $10.5\mic$ seen \iizw\ (Fig.~\ref{fig:fitlines}), 
is from the fitted \siv\ line.

The ionic lines are very prominent in the spectra of dwarf galaxies 
(Fig.~\ref{fig:fitlines}; Table~\ref{tab:lines}), while, for the most part, 
there is a paucity of PAHs, on size scales traced by the ISOCAM beam as well 
as on global scales. 
\ngc{1140} seems to be somewhat of an exception (Fig.~\ref{fig:spec1}), as the
PAHs are rather conspicuous. 
Normally, the PAH bands figure prominently in the spectra of dust-rich 
starburst and spiral galaxies \citep[e.g.][]{laurent+00,sturm+00,forster+03}. 
While \ngc{1569} shows low level PAH emission with respect to the rising 
continuum, toward the central region(lower left of Fig.~\ref{fig:spec1}), the 
extended emission of \ngc{1569}, shown toward the lower right of 
Fig.~\ref{fig:spec1}, demonstrates that PAHs are rather conspicuous, while the
continuum is flatter. 
Globally, \ngc{1569} resembles the spectrum toward the central peak.
The contrast between the MIR spectra of the low-metallicity 
starbursts and the nearby metal-rich starburst galaxy, \M{82}, for example, 
can be observed in Fig.~\ref{fig:spec2}. 
The MIR spectrum toward the central 24" of \M{82}\ is clearly 
dominated by material originating from the disk, which shows dominant 
PAH emission 
\citep[e.g.][]{laurent+00,roussel+01b,dale+02,forster+03,vogler+05}. 
On a global scale, the \M{82} MIR spectra also looks like the central 24" with
similarly prominant PAH bands, but with somewhat lower rising continuum 
longward of  $14\mic$.
When viewed on {\it global} scales, the dwarf galaxies are not dominated by 
PAHs, but by the VSG continua characteristic of \hii\ regions 
\citep[e.g.][]{peeters+02,martin-hernandez+02a}.  

In resolved regions of the closest low-metallicity galaxies, the SMC and 
the LMC, PAHs are indeed present at low levels: 
\smcb\ region \citep[ and our Fig.~\ref{fig:spec2}]{reach+00}
and \smcn\ \citep[ and our Fig.~\ref{fig:spec2}]{contursi+00} 
and \xxxdor\ in the LMC (Fig.~\ref{fig:spec1}) and other regions in the 
LMC \citep{vermeij+02}. 
In the Local Group dwarf galaxy, \ngc{1569}, PAH band emission is likewise 
evident, yet relatively supressed.  
It may not be surprising to see more conspicuous local PAH band emission from 
the PDRs illuminated by \hii\ regions in low metallicity regions at very high 
spatial resolution.

The prominent ionic lines and significant continua observed on large scales in
these galaxies compel us to compare the observed MIR spectra with those of 
Galactic \hii\ regions. 
The presence of PAHs in some galaxies also implies some contribution from the 
molecular clouds/PDRs in the vicinity. 
In order to have a better idea as to which ISM components are traced by the 
MIR spectra in the beam, we compare some of these spectra with the Galactic 
region, \M{17} where the \hii\ region and PDR region are well resolved with 
ISOCAM \citep{cesarskyd+96}.

\begin{table*}[htbp]
\begin{center}
\caption{Ionic line intensities in $10^{-15}\; \rm W\,m^{-2}$
         for the low metallicity galaxies (global values).
         In the case of \xxxdor\ and \smcn, the fluxes are integrated over
         a $120''$ circular aperture.
         These line intensities are corrected for the internal extinction.
         This extinction correction is negligible in all the sources, except
         \iizw, and \M{82}.
         The upper limits are $3\sigma$ values.}
\label{tab:lines}
\begin{tabularx}{\textwidth}{l*{4}{X}}
\hline
\hline
\multicolumn{1}{c}{}
  & \neiii
  & \neii
  & \siv
  & \ariii     \\
\hline
\bf \ngc{1140}
  & $0.8 \pm 0.13$
  & $0.36 \pm 0.55$
  & $0.19 \pm 0.05$
  & $\lesssim 8.2 \times 10^{-3}$  \\
\bf \ngc{1569}
  & $14.7 \pm 1.0$
  & $1.90 \pm 0.28$
  & $6.7 \pm 1.2$
  & $1.4 \pm 0.6$ \\
\bf \ngc{5253}
  & $8.7 \pm  0.6$
  & $\lesssim 0.7$
  & $\lesssim 2.2$
  & $\lesssim 0.4$ \\
\bf \iizw
  & $1.48 \pm 0.12$
  & $\lesssim 0.26$
  & $5.24 \pm 0.53$
  & $\lesssim 2.2$ \\
\bf \xxxdor
  & $(1.04 \pm  0.04) \times 10^{3}$
  & $(1.87 \pm 0.07) \times 10^{2}$
  & $(1.35 \pm 0.05) \times 10^{3}$
  & $(2.56 \pm 0.12) \times 10^{2}$ \\
\bf \smcn
  & $42.0 \pm 2.4$
  & $\lesssim 3.5$
  & $66.5 \pm 4.1$
  & $15.0 \pm 2.9$ \\
\bf \smcb
  & $\lesssim 3.6$
  & $6.8 \pm 1.0$
  & $6.0 \pm 0.9$
  & $\lesssim  2.3$ \\
\bf \M{82}
  & $54.0 \pm 2.2$
  & $368 \pm 14$
  & $58.9 \pm 3.2$
  & $\lesssim 24$ \\
\hline
\end{tabularx}
\end{center}
\end{table*}

\subsection{Spatial distribution of various MIR components}
\label{sec:spatial}

For some of the nearby, well-resolved sources, such as \ngc{1569}, \iizw\ 
and \xxxdor, we study the spatial distribution of the various physical 
components in more detail. 
There was not sufficient signal to deconvolve each individual image of the CVF 
cubes, so the spatial resolution of the images were degraded, as specified 
below.
The MIR spectrum of each pixel was then computed (as described in section 
\ref{sec:components}) and the spectral analysis was performed to extract the 
various physical components for each pixel. 
The maps of the PAHs or ionic lines consist of the integrated intensities of 
those complete components. 
The VSG maps are constructed from the extracted VSG continuum component, 
integrated from 10 to $16 \mic$, to characterise the slope of the VSG 
continuum.
To generate the ratio maps, we computed the ratios of components only for the 
pixels where the flux of the denominator was greater than $3\;\%$ of the 
maximum to avoid non relevant values at the edge of the maps due to low flux 
values.

Inspection of the individual spectra shows that considering the low signals 
at short wavelengths and the uncertainty values, there is negligible local 
contribution at the shortest wavelengths, $5\mic$, from any possible cooler, 
evolved stellar population. 
The physical components which we are displaying as images here, are all 
emitting at wavelengths $> 6 \mic$. 
Thus, if our assumption concerning negligible local stellar contribution at 
short wavelengths was incorrect, these maps would not be effected.

\subsubsection{\ngc{1569}}
\label{sec:ngc1569}

\begin{figure*}[htbp]
  \centering
  \begin{tabular}{cc}
    \includegraphics[width=0.48\textwidth]{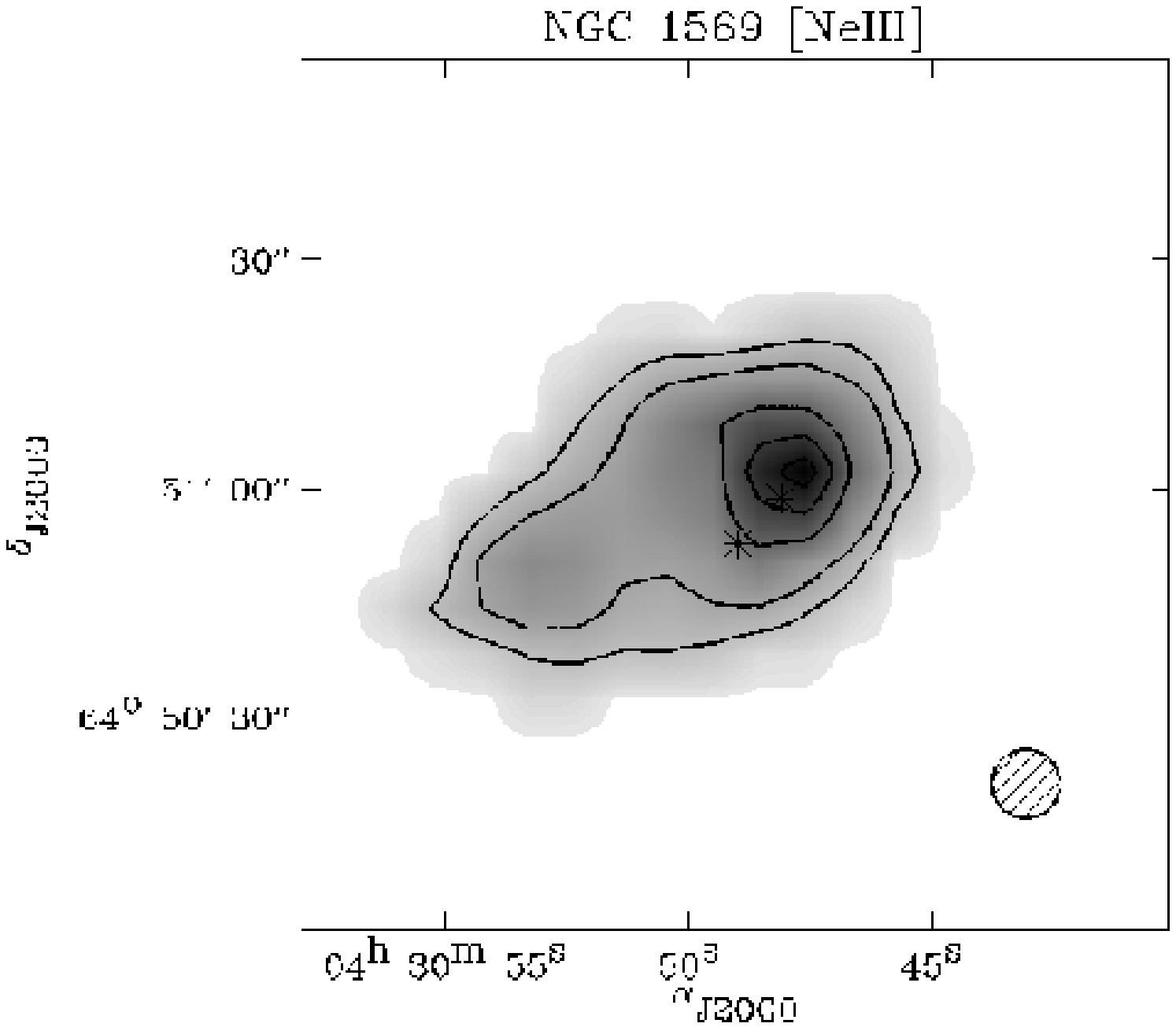}   &
    \includegraphics[width=0.48\textwidth]{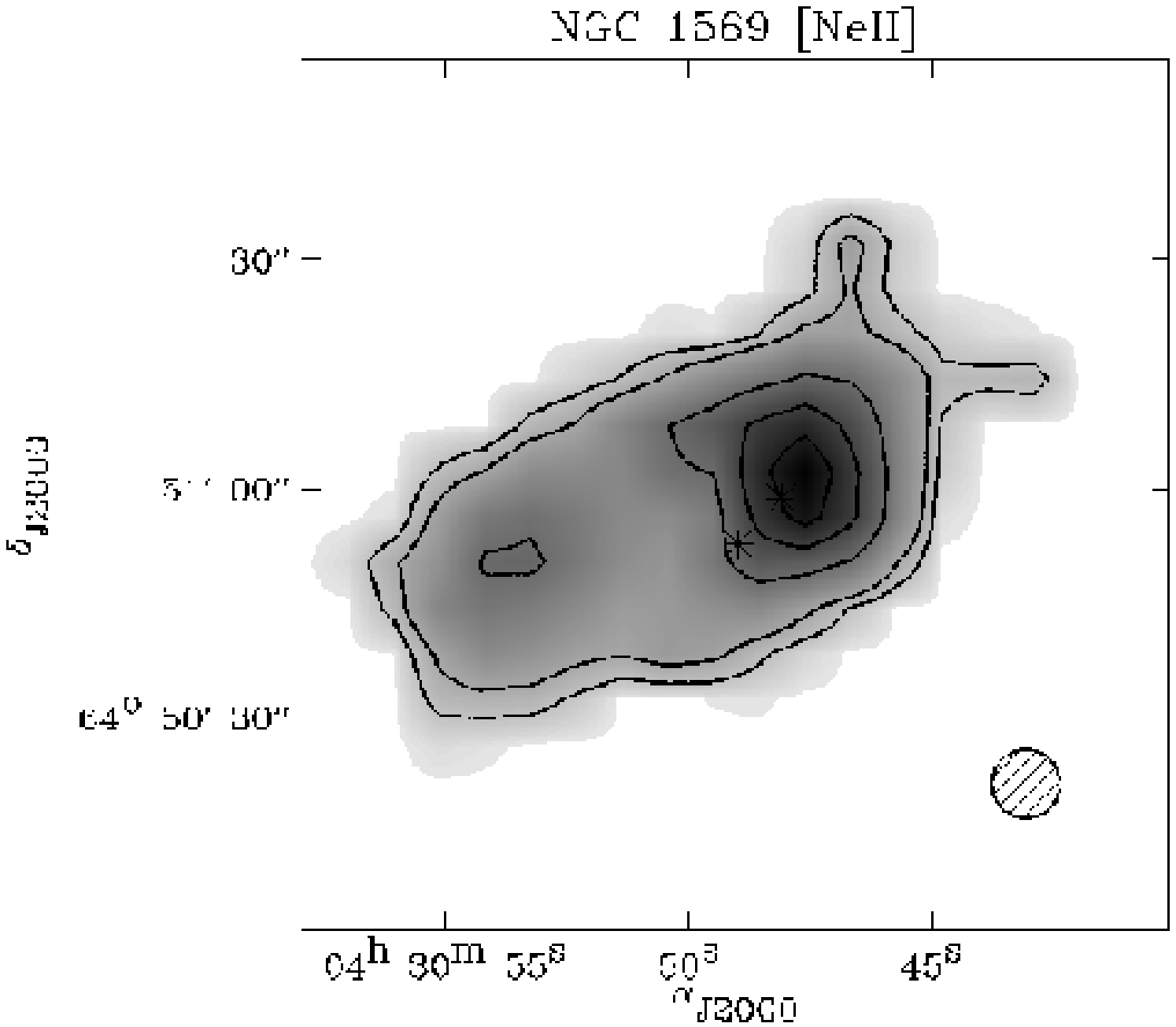}    \\
    \includegraphics[width=0.48\textwidth]{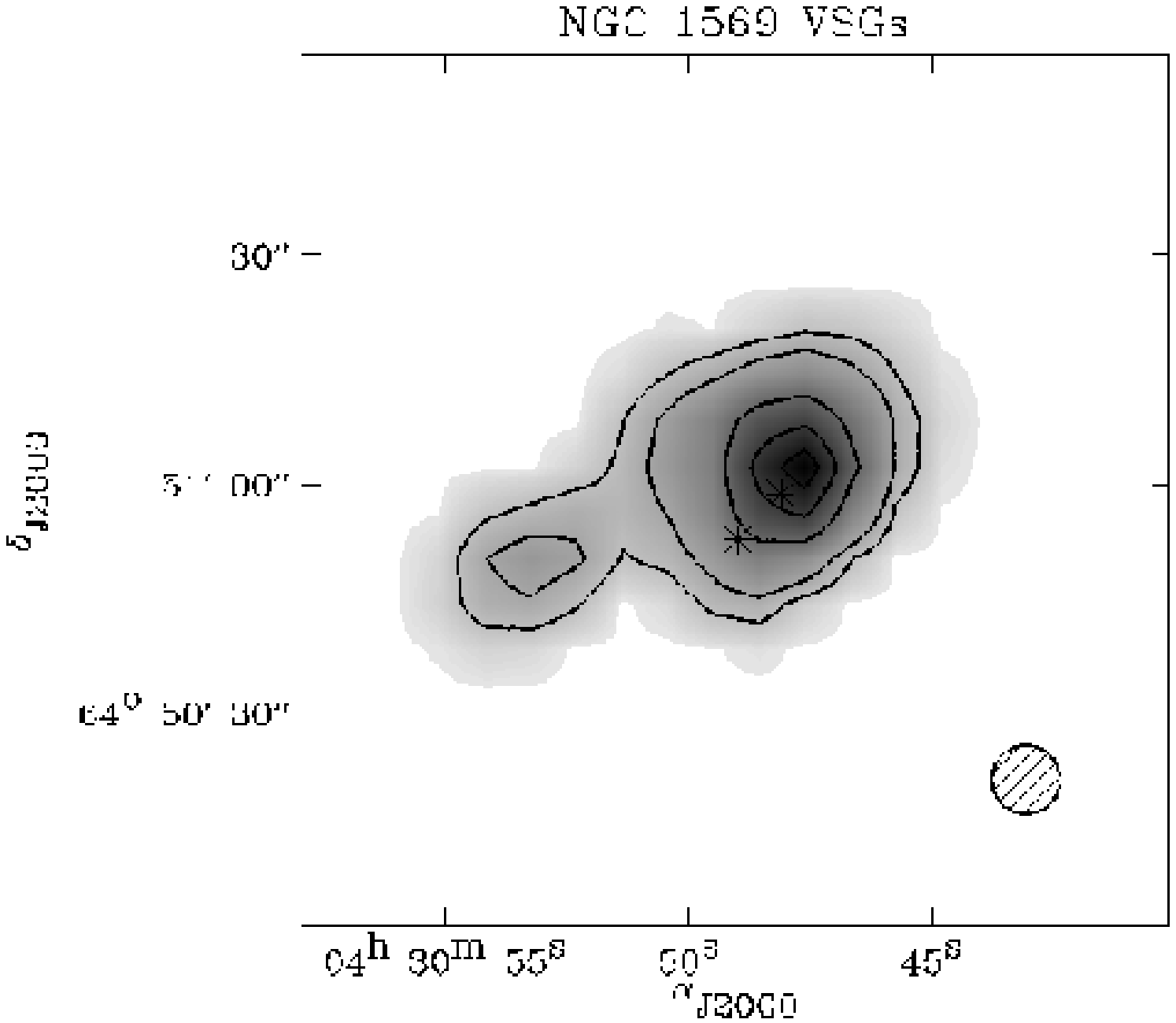} & 
    \includegraphics[width=0.48\textwidth]{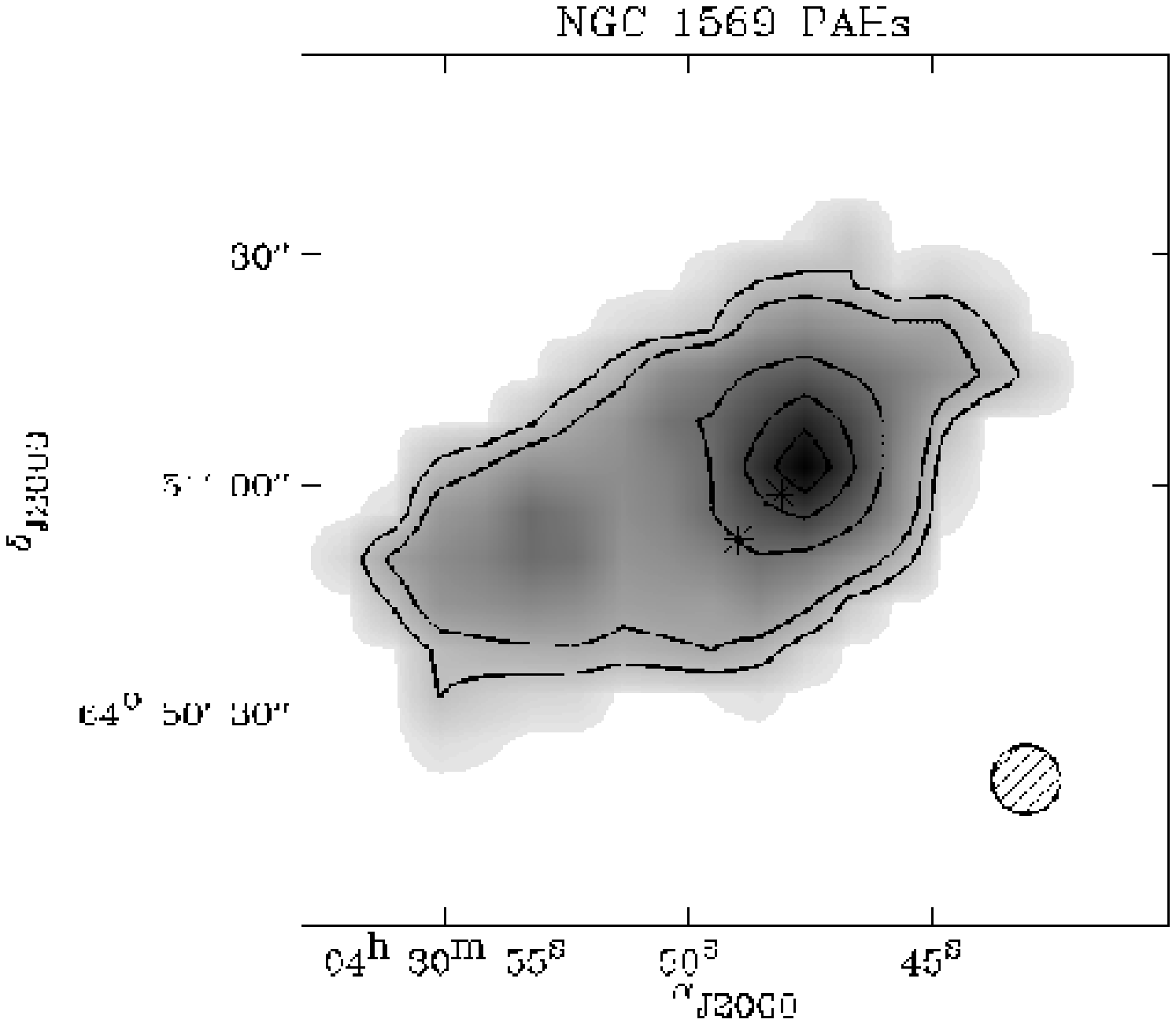}     \\
    \includegraphics[width=0.48\textwidth]{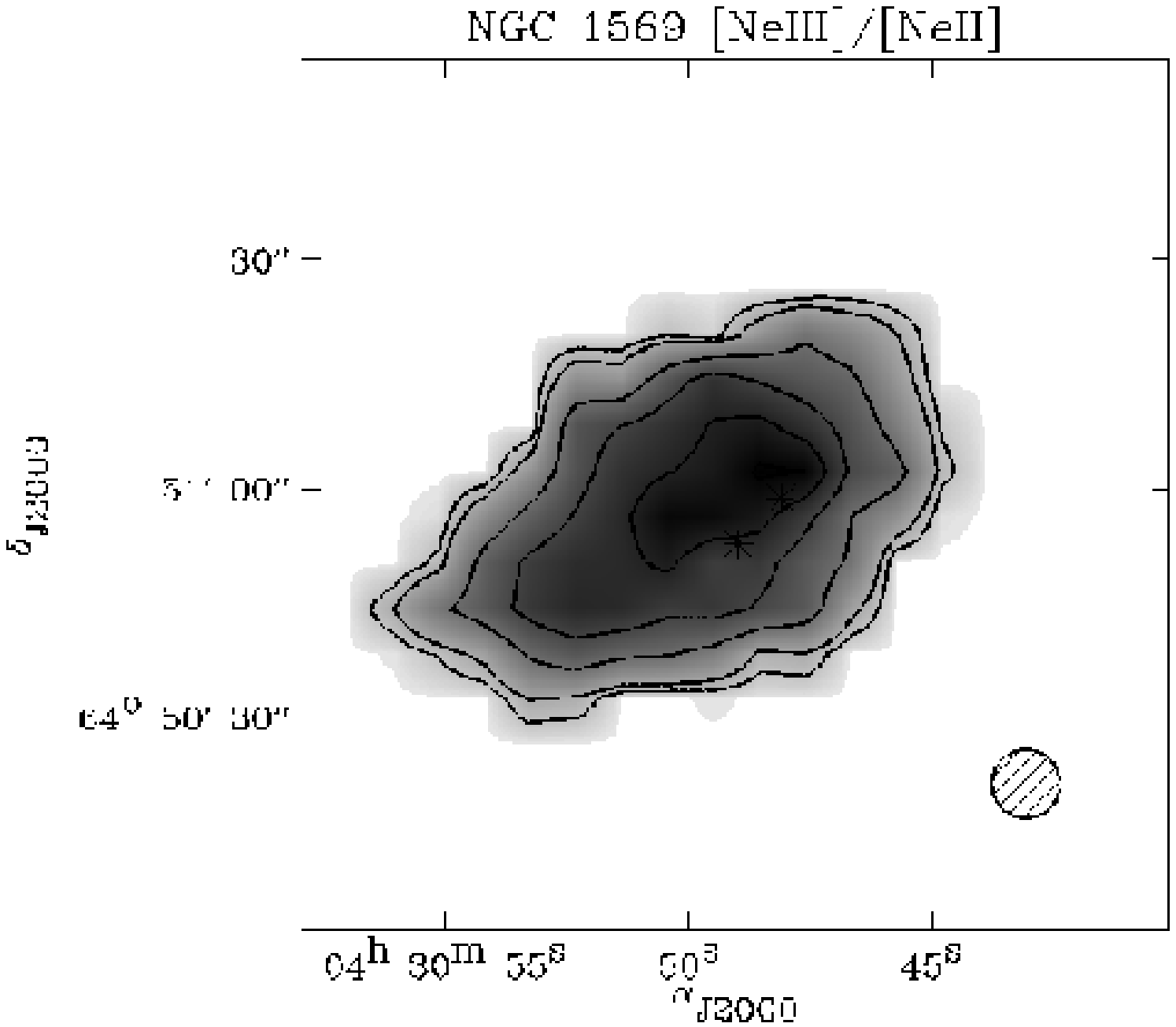}    &
    \includegraphics[width=0.48\textwidth]{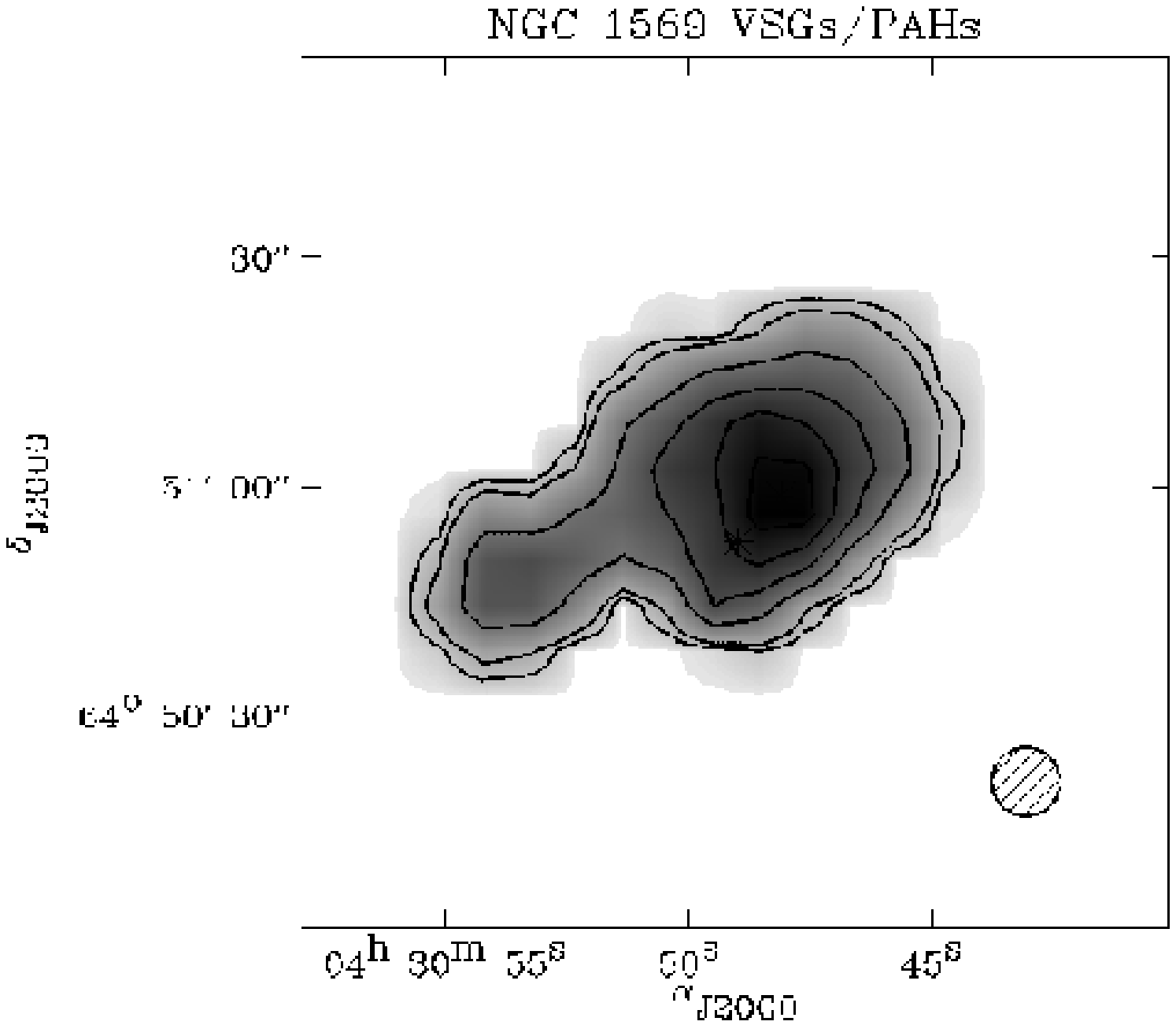}   \\
  \end{tabular}
  \caption{Maps of some physical components for \ngc{1569}.
           Up-left: \neiii; up-right: \neii;
           middle-left: VSG continuum; middle-right: PAHs;
           lower-left: \neiii/\neii; lower-right: VSGs/PAHs;
           The contour levels are 5, 10, 30, 50, 70 and 90\% of the peak 
           intensity.
           The peak intensities are: 
        $\mbox{\neiii}=2.2\times10^{-15}\;\rm W\,m^{-2}\,beam^{-1}$,
        $\mbox{\neii}=2.3\times10^{-16}\;\rm W\,m^{-2}\,beam^{-1}$,
        $I(\mbox{VSG})=5.2\times10^{-14}\;\rm W\,m^{-2}\,beam^{-1}$,
        $I(\mbox{PAH})=1.4\times10^{-14}\;\rm W\,m^{-2}\,beam^{-1}$,
        $\mbox{\neiii/\neii}=9.6$,
        $I(\mbox{VSG})/I(\mbox{PAH})=3.6$.
           The resolution has been degraded to the beam size at $16\mic$
           which is roughly $10''$ (100 pc).  
           The asterisks are the position of the super star clusters SSC-A
           and SSC-B.}
  \label{fig:n1569_compos}
\end{figure*}

The study of the physical components in the nearby galaxy, \ngc{1569}, 
provides a unique view of the physical components within a starburst dwarf 
galaxy in some detail. 
The resolution of our maps was degraded to $10''$ ($\sim$100 pc), which is 
roughly equivalent to the beam size at $16\mic$.  
Final maps for each of these components are shown in 
(Fig.~\ref{fig:n1569_compos}).

At first glance, with the ISOCAM spatial resolution, the ionic, dust and PAH 
components all peak toward the massive clusters of HII regions to the west 
\citep{waller91} and follow the starforming ridge, with a secondary peak 
toward the east (Fig.~\ref{fig:n1569_compos}).  
Super star clusters reside within the peak PAH emission zone, but only much 
higher resolution data would isolate the precise nature of the MIR component 
peaks. 
A variety of ISM components would be sampled toward \ngc{1569}\ within a beam 
size of $10''$, which is 106 pc. While the MIR peaks are all located toward 
the same locations, the extent of the emission and the rate of fall-off varies
from component to component. 
The PAHs have a relatively large extended component, with the low level 
emission (5\% of peak value) ranging out to 80\arcsec to 40\arcsec. 
The VSGs, normally confined to within \hii\ regions, appear relatively more 
compact. 
The \neiiiline\ and the \neiiline\ line emission, however, is very extended --
on the same scale as the PAH emission.
There is a substantial pervasive, diffuse ionised medium throughout the 
galaxy.
Notice the western filament extensions seen in the \neii\ and PAH images. 
These are spatially coincident with the superwinds impresively exhibited in 
the H$\alpha$ image of \citet{waller91}.  

Maps of the ratios of these components can shed some light on the behavior of 
the MIR tracers (Fig.~\ref{fig:n1569_compos}). 
The \neiii/\neii\ ratio looks relatively flat along the star formation ridge. 
As this ratio is related to the hardness of the ISRF (see 
Sect.~\ref{sec:neiii_neii}), the hard ISRF is not very well confined to only 
the most compact sources, but is more extended. 
This effect can be noted in the image of the VSGs/PAHs. 
The central region of the galaxy has a high ratio of VSGs/PAHs, i.e., 
relatively lower level of PAHs in the central, most active area, where the 
VSG hot grain continuum is also elevated.
The photons responsible for the  high level VSG continuum observed in the low 
metallicity galaxies would also excite the MIR PAH modes, if the PAHs are 
sufficiently abundant.
Further from the star formation sites, the PAH component begins to dominate 
over the VSGs. 
This effect could also be interpreted in light of PAH destruction due to the 
hardness of the radiation field. 
This possibility is discussed further in Section \ref{sec:pah_destruct}. 
The VSG distribution seems to be more sharply peaked with a smaller FWHM than 
that of the \neiii\ line (Fig.~\ref{fig:n1569_compos}).

\subsubsection{\xxxdor}
\label{sec:30dor}
\begin{figure*}[htbp]
  \centering
  \begin{tabular}{cc}
    \includegraphics[width=0.48\textwidth]{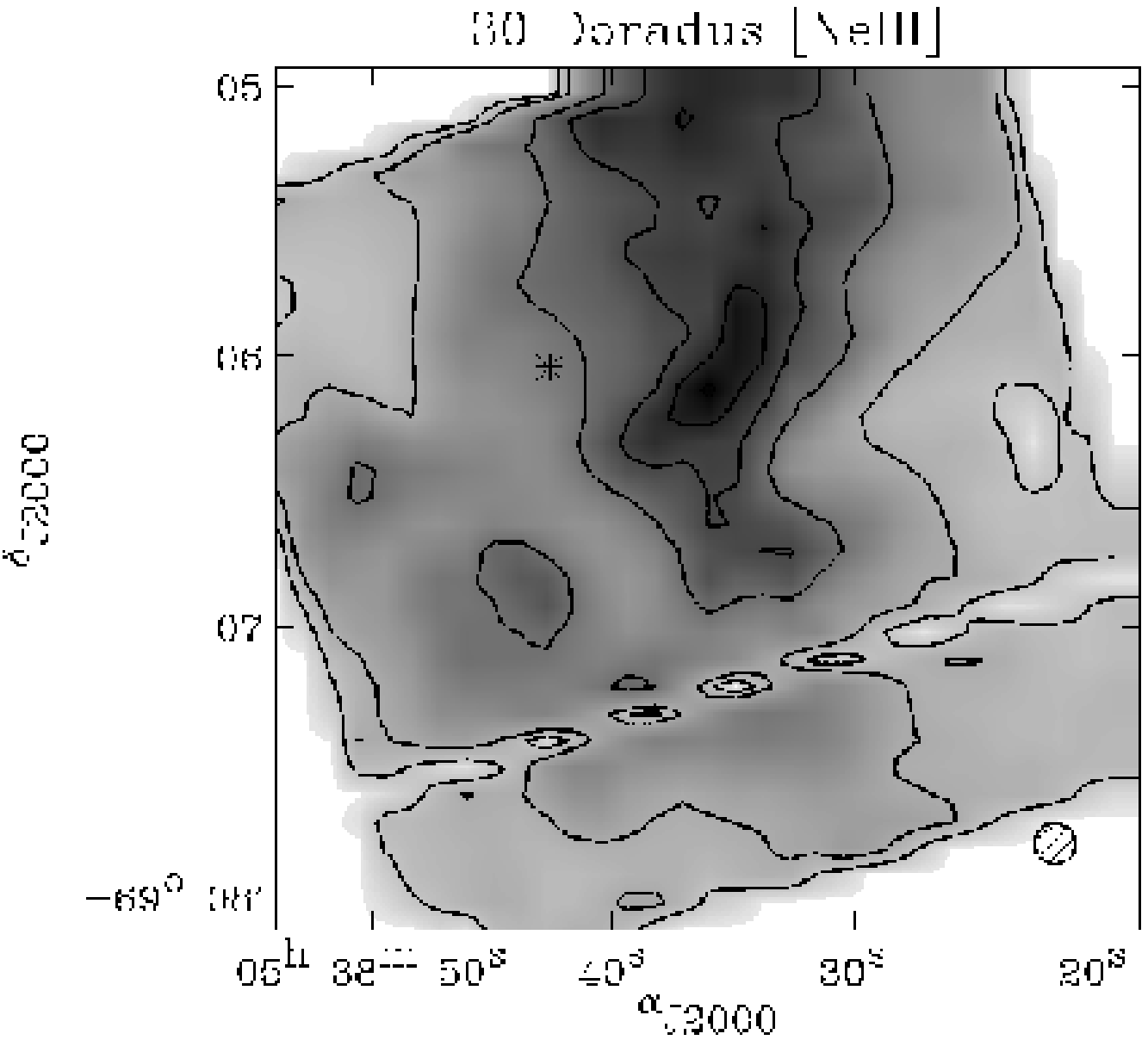}   &
    \includegraphics[width=0.48\textwidth]{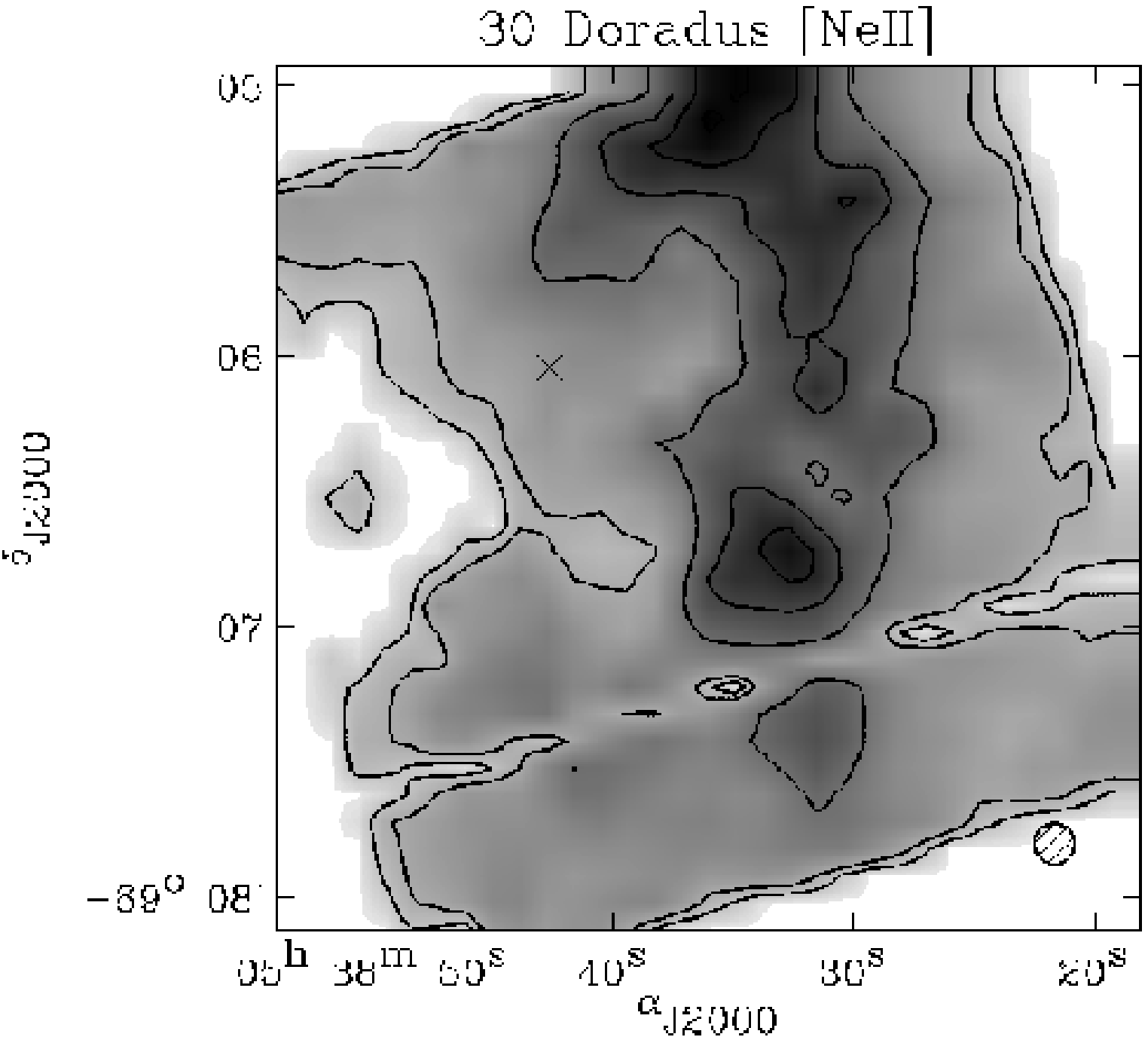}    \\
    \includegraphics[width=0.48\textwidth]{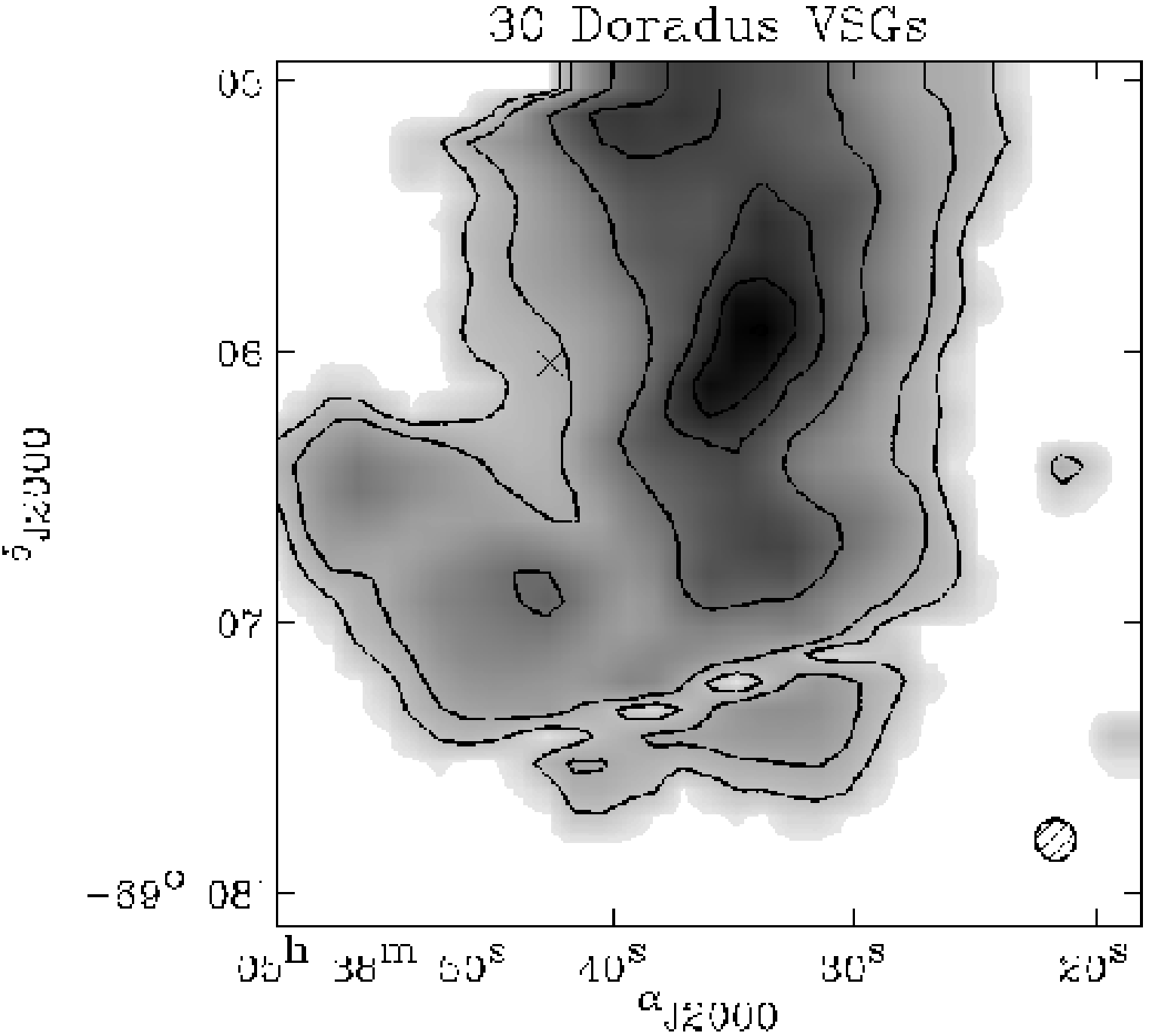} & 
    \includegraphics[width=0.48\textwidth]{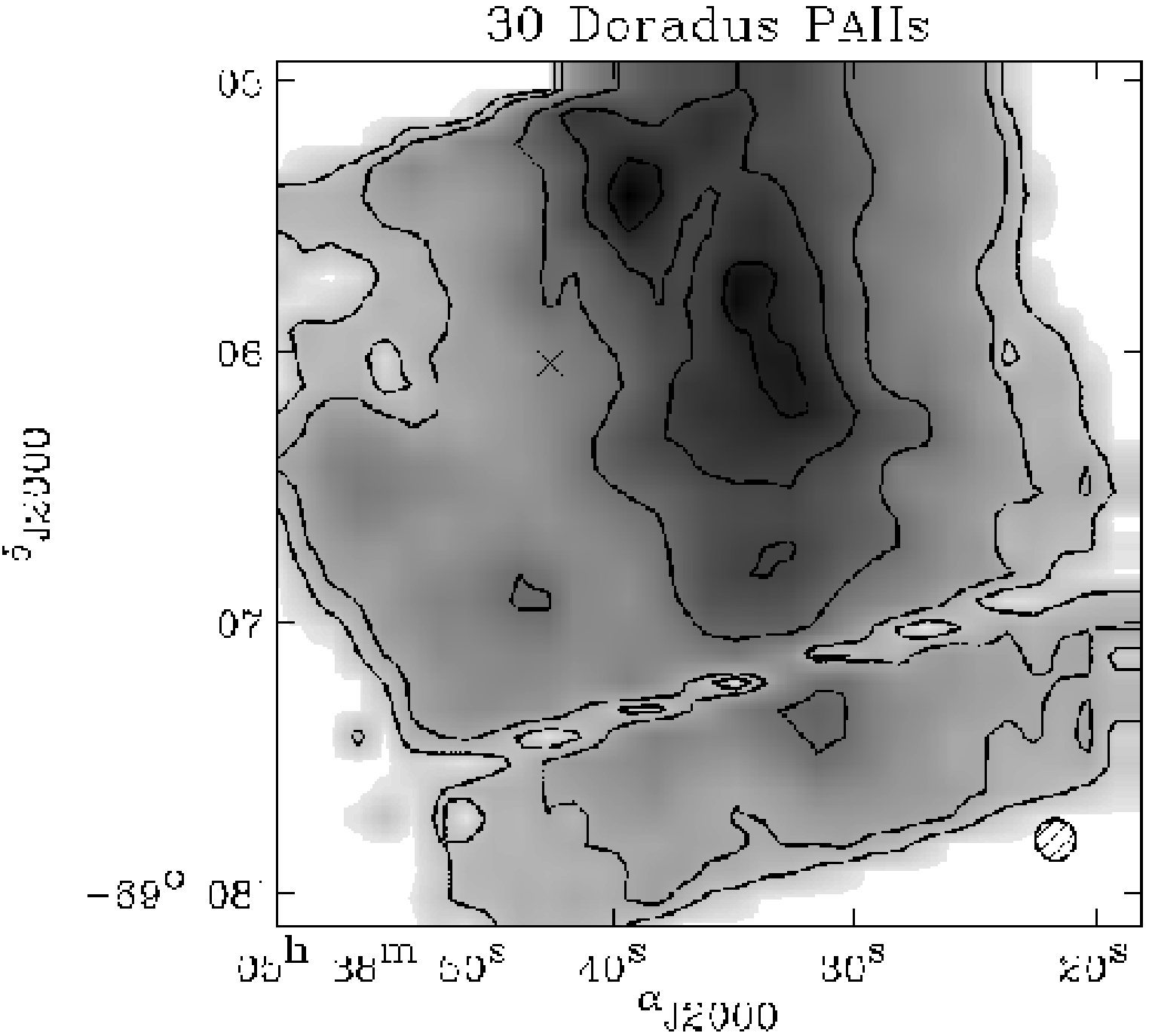}     \\
    \includegraphics[width=0.48\textwidth]{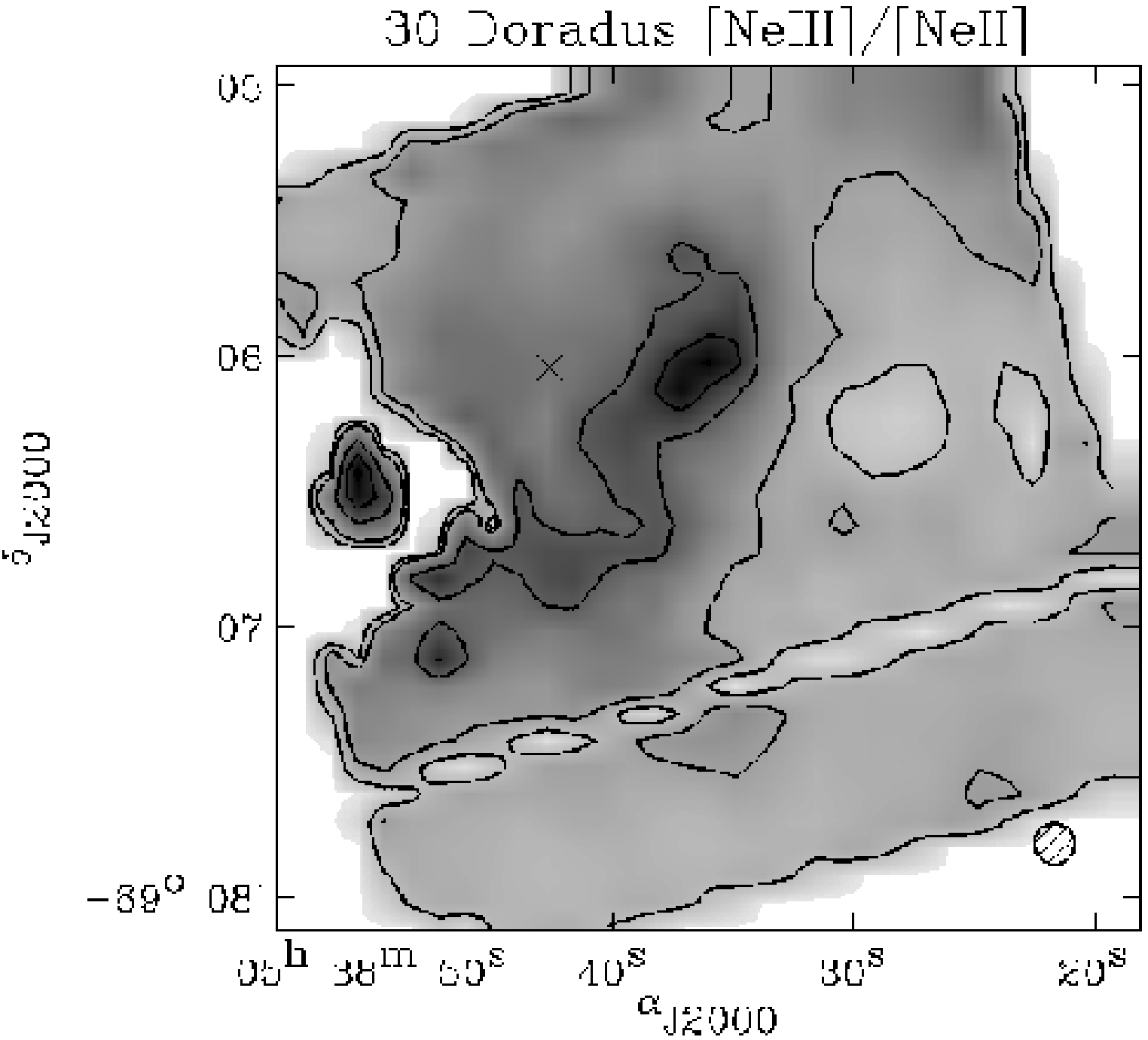}    &
    \includegraphics[width=0.48\textwidth]{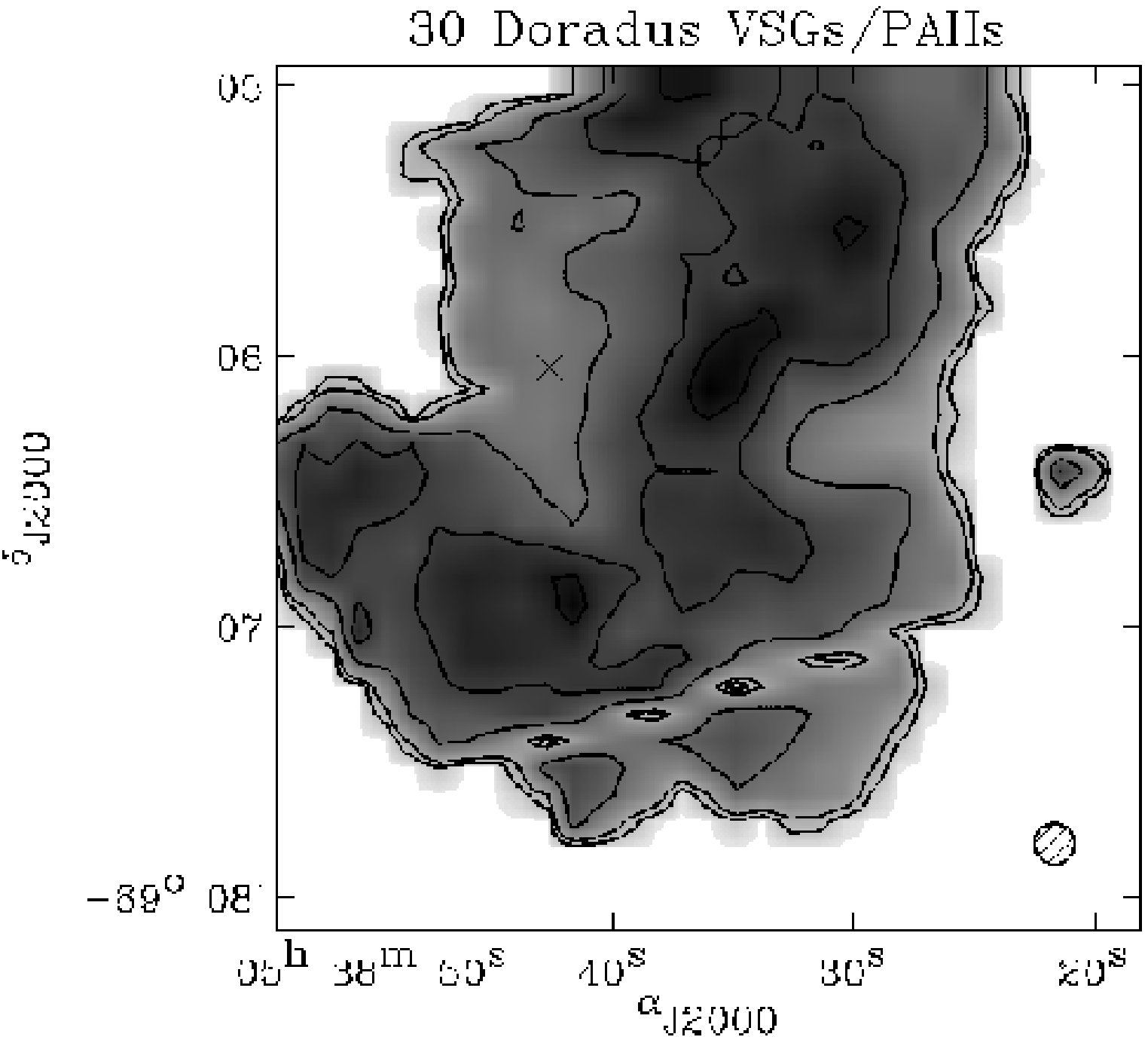}   \\
  \end{tabular}
  \caption{Maps of some physical MIR components for \xxxdor.
           Up-left: \neiii; up-right: \neii;
           middleleft: $15\mic$ VSG continuum; middle-right: PAHs;
           lower-left: \neiii/\neii; lower-right: VSGs/PAHs;
           The contour levels are 5, 10, 30, 50, 70 and 80\% of the peak 
           intensity.
           The peak intensities are: 
           $\mbox{\neiii}=1.6\times10^{-14}\;\rm W\,m^{-2}\,beam^{-1}$,
           $\mbox{\neii}=3.5\times10^{-15}\;\rm W\,m^{-2}\,beam^{-1}$,
           $I(\mbox{VSG})=7.2\times10^{-13}\;\rm W\,m^{-2}\,beam^{-1}$,
           $I(\mbox{PAH})=1.4\times10^{-13}\;\rm W\,m^{-2}\,beam^{-1}$,
           $\mbox{\neiii/\neii}=34$,
           $I(\mbox{VSG})/I(\mbox{PAH})=8.6$.
           The resolution has been degraded to the beam size at $16\mic$
           which is roughly $10''$ (2.4 pc). 
           The strip across the lower part of the image is the inactive 
           column 24 of ISOCAM. 
           The asterisk marks the position of the stellar cluster R136.}
  \label{fig:30dor_compos}
\end{figure*}

The images of the MIR components of \xxxdor\ were constructed in the manner 
described above for \ngc{1569}. 
The angular scale of $10"$ corresponds to a linear size of 2.4 pc. 
This angual scale should allow us to study the details in the region around 
the massive star cluster R136 and the surrounding PDR/molecular cloud region 
(Fig.~\ref{fig:30dor_compos}).  
The region directly toward the R136 star formation, has a minimum in emission
in all of the tracers, illustrating the effects of the massive star formation 
activity with winds efficiently blowing out the ISM in the immediate 
surroundings (Fig.~\ref{fig:30dor_compos}). 
The distribution of the \neiii\ and that of the \neii\ emission differ close 
to the R136 cluster, where the \neiii\ emission is more prominent. 
The difference in excitation potentials of the \neiii\ and \neii\ lines, is 
illustrated nicely by the partial shell of \neii\ emission around the cluster, 
which has a peak displaced further from the exciting cluster than that of the 
\neiii. 
The PAHs have peaks  beyond the peak of the VSG emission. 
They have a relatively broad distribution which appears to be flatter than 
that of the VSGs. 
The hardest form of the radiation field, traced by the \neiii/\neii\ ratio, 
is confined to a relatively limited region, about 10 pc from the R136 exciting 
source. 
The peaks of the PAH emission avoid the peaks of the \neiii/\neii\ ratio. 
This can also be seen in the distribution of the VSGs/PAH, which shows a peak 
coincident with the peak of the \neiii/\neii\ ratio.

The distribution of the various tracers of the ISM appear to overlap 
spatially. 
Even at this spatial resolution, there is not a very distinct separation 
between the various ISM components at such spatial scales. 
A very clumpy ISM at smaller spatial scales could be one reason for this. 
High resolution, ground-based MIR observations could reveal the geometry more 
accurately. 

\begin{figure}[htbp]
\centering
	\includegraphics[width=0.45\textwidth]{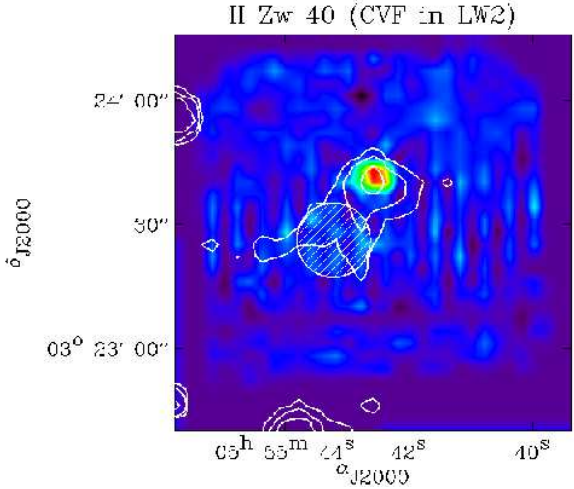}     
\caption{The image is the result of the CVF spectral image of \iizw\ 
integrated in the equivalent LW2 band; contours are the R band image of 
\citet{gildepaz+03} which shows the prominent merger tail. 
The hatched 18\arcsec circle is the region of the CVF image that was 
integrated, resulting in the CVF spectrum in Fig.~\ref{fig:spec1}(upper right 
spectrum). 
This region shows evidence for some faint PAH emission. 
The upper left spectra of Fig.~\ref{fig:spec1} is integrated in a 14" circle 
toward the central MIR peak.}
  \label{fig:iizw40_tail}
\end{figure}

\subsubsection{\iizw}
\label{sec:iizw40}
To visualise structure of  the MIR emission of \iizw, the CVF was integrated 
to simulate the LW2 broad band (5 to 8.5$\mic$) to create a map optimised in 
signal-to-noise. 
The overlaping tails of the interacting galaxies could then be readily 
identified using the R band image of \citet{gildepaz+03}. 
The smoothed CVF spectrum shown in Fig.~\ref{fig:spec1} integrated in the 
overlapping tail region, has a S/N of 3.4 at 7.7$\mic$ and indicates that the 
presence of PAHs is more apparent in the overlap region, where the continuum 
is not very striking, than toward the central MIR peak of the galaxy 
(Fig.~\ref{fig:iizw40_tail}). 
The contrast between the spectra in these 2 regions is seen in 
Fig.~\ref{fig:spec1} (upper left). 
The spectrum integrated within a 14" aperture toward the central MIR peak of 
the galaxy, shows little evidence for PAH emission, and very strong MIR 
continuum, and is dominated by emission from the starburst activity. 
The spectra of the merging tails of the interacting galaxies, on the other 
hand, shown in Fig.~\ref{fig:spec1} (upper right), contains more ISM material, 
perhaps indicative of the original disk material of the 2 merging galaxies. 
Still, globally, the MIR emission is dominated by the ionic lines and hot 
dust continuum.


\subsection{Contributions of various components in ISOCAM broad bands}

The most popular broad bands that were used to image galaxies with ISOCAM were
the $6.7\mic$ LW2 (5.0 to 8.5$\mic$) and the $15\mic$ LW3 (12 to 18$\mic$) 
bands. 
Often it was assumed that the $6.7\mic$ band traces the PAHs, emitting at 6.3, 
7.7 and $8.6\mic$, and that the $15\mic$ band traces the VSG continuum. 
Large extragalactic studies were carried out using these 2 bands 
\citep[e.g.][]{laurent+00,roussel+01a,dale+00}. Here we quantify the 
contribution of the different components in some useful 
ISOCAM broad bands, integrating the spectra over the appropriate wavelength 
ranges, and convolve them with the corresponding broad band filters. 
This exercise can serve as a useful aide in the interpretation of broad band 
data from low metallicity regions when MIR spectra are not available.

To evaluate the contribution for any component, integrated intensities of the 
features and continua were computed, following their decomposition as 
describled in Section~\ref{sec:components}.  
Table~\ref{tab:contrib} lists the percentage of the flux of the various MIR 
components, PAH bands, ionic lines and the VSG continua, which are contained 
within the ISOCAM broad bands LW2 (centered at $6.7\mic$, from 5 to 
8.5$\mic$), LW3 (centered at $15\mic$, from 12 to 18$\mic$) and the $12\mic$ 
LW10 band which resembles the IRAS band and is centered at $12\mic$, from 8 
to 15$\mic$), for the 7 low metallicity sources along with M82 for comparison. 
The LW2 band is dominated by the PAH band emission in all but 2 of the 
galaxies: the global spectra of the more active galaxies, \ngc{5253} and 
\iizw\ have about 1/2 of their total $6.7\mic$ (LW2) band flux originating 
from the wings of the Lorentzians of the fitted PAHs, while 1/2 arises from 
the small grain continua. 

However, most of the $15\mic$ (LW3) band flux originates from the 
hot VSG continua, except for the case of \smcb, which traces the diffuse ISM, 
thus accounting for the significant contribution from PAHs 
\citep[e.g.][]{mattila+96}. 
Even in \ngc{1140}, where the continuum is relatively flat 
(see Fig.~\ref{fig:spec1}) the VSGs still make up $75\;\%$ of the total LW3 
broad band flux. 
In the active starburst galaxies, \ngc{5253} and \iizw, as well as the 
starforming reigons, \smcn\ and \xxxdor, $85\;\%$ to $97\;\%$ of the $15\mic$ 
LW3 flux is originating from the VSG component in the $12\mic$ broad band.

In the $15\mic$ LW3 broad band, the \neiii\ line alone can contribute between 
6\% to 10\% of the total flux for the starbursting dwarf galaxies/\hii\ 
regions, such as \ngc{1569}, \ngc{1140}, \xxxdor\ and \smcb. 
The \neiiiline\ line is one of the brightest ionic lines in the MIR spectra, 
along with the \sivline\ line (see Table~\ref{tab:lines}).

The broader, IRAS-type $12\mic$ band, LW10, has at least 37\% of its flux 
originating from VSGs in all sources studied here, except \smcb. 
In the active starburst galaxies, which globally resemble \hii\ regions, 
\ngc{5253} and \iizw, almost all of the LW10 IRAS $12\mic$ flux is due to the 
VSG emission. 
In the sources dominated by the disk or diffuse ISM, such as \M{82} and 
\smcb, the PAHs are the more important contributors with almost 90\% of the 
$12\mic$ broad band flux originating from PAHs.  
The \sivline\ is 5\% to 9\% of the LW10 band flux in the \hii\ regions, 
\xxxdor\ and \smcn\ and in the low-metallicity \hii\ region-like galaxies, 
\ngc{1569} and \iizw. With these beam sizes, the ionic lines in \M{82} are 
certainly small contributors to any of the broad bands --~they are dwarfed by 
the prominent PAH bands and the VSGs. 

In the nearby (4.5 Mpc) spiral galaxy, M83, which was well-mapped in many 
ISOCAM broad bands as well as the CVF, it was possible to isolate the spectra 
of the nuclear region, the spiral arms, the interarm region and the diffuse, 
extended ISM \citep{vogler+05}. 
While PAHs are the dominant contribution to the $15\mic$ LW3 broad band flux 
in the diffuse ISM and the galactic disk, they are responsible for no more 
than 60\% of the nuclear flux. 
Toward the nucleus, the PAH bands comprise most of the LW2 broad band flux in 
M83, since the starburst region itself is not resolved by the ISOCAM beam, 
which is, instead, tracing the circumnuclear material. 
{\it Globally}, the MIR emmission in M83, as well as other spiral galaxies, is 
dominated by the disk material (e.g., PAH bands), not the nucleus 
\citep[e.g.][]{roussel+01b,dale+01, vogler+05}. 
In contrast, the component contribution to the global broad bands of the 
dwarf galaxies sampled here typically have MIR characteristics similar to 
those of the nuclei of spiral and starburst galaxies, except for the fact 
that the abundance of PAH bands is much lower in these low metallicity sources.
The dearth of dominant disk-emitting material is evident in the MIR spectra.

  \begin{table*}[htbp]
\begin{center}
\caption{Contribution of the different physical components
         for the global galaxies in various ISOCAM broadbands.
         The percentage is the fraction of the component compared
         to the total flux in the band (Table~\ref{tab:lines}).}
\label{tab:contrib}
\begin{tabularx}{\textwidth}{l|*{2}{X}|*{4}{X}|*{5}{X}}
\hline
\hline
\multicolumn{1}{c|}{}         &
\multicolumn{2}{c|}{LW2}  &
\multicolumn{4}{c|}{LW3}    &
\multicolumn{5}{c}{LW10}    \\
  &
 PAHs    &  VSGs    &
 \neiii &  \neii   &  PAHs     &  VSGs    &
 \neii   &  \siv    &  \ariii  &  PAHs    &  VSGs   \\
\hline
\bf \ngc{1140} &
$        75 \;\%$ &
$        24 \;\%$ &
$       8.6 \;\%$ &
$       4.0 \;\%$ &
$       5.5 \;\%$ &
$        75 \;\%$ &
$       3.0 \;\%$ &
$       1.3 \;\%$ &
               \ldots &
$        23 \;\%$ &
$        73 \;\%$ \\
\bf \ngc{1569} &
$      95 \;\%$ &
$       3 \;\%$ &
$      11 \;\%$ &
$     1.4 \;\%$ &
$     6.3 \;\%$ &
$      80 \;\%$ &
$     1.6 \;\%$ &
$     4.6 \;\%$ &
$     0.6 \;\%$ &
$      29 \;\%$ &
$      63 \;\%$ \\
\bf \ngc{5253} &
$        42 \;\%$ &
$        56 \;\%$ &
$       2.9 \;\%$ &
               \ldots &
               \ldots &
$        95 \;\%$ &
               \ldots &
               \ldots &
               \ldots &
               \ldots &
$        95 \;\%$ \\
\bf \iizw &
$        50 \;\%$ &
$        49 \;\%$ &
$         3 \;\%$ &
               \ldots &
               \ldots &
$        96 \;\%$ &
               \ldots &
$       6.3 \;\%$ &
$       0.9 \;\%$ &
$         1 \;\%$ &
$        94 \;\%$ \\
\bf \xxxdor &
$        88 \;\%$ &
$        11 \;\%$ &
$       5.9 \;\%$ &
$       1.1 \;\%$ &
$       3.5 \;\%$ &
$        89 \;\%$ &
$       1.3 \;\%$ &
$       7.7 \;\%$ &
$       0.9 \;\%$ &
$        14 \;\%$ &
$        75 \;\%$ \\
\bf \smcn &
$        85 \;\%$ &
$         9 \;\%$ &
$       7.8 \;\%$ &
               \ldots &
$       2.6 \;\%$ &
$        85 \;\%$ &
               \ldots &
$        11 \;\%$ &
$       1.8 \;\%$ &
$        18 \;\%$ &
$        68 \;\%$ \\
\bf \smcb &
$        97 \;\%$ &
               \ldots &
               \ldots &
$       4.2 \;\%$ &
$        49 \;\%$ &
$        43 \;\%$ &
$       6.5 \;\%$ &
$       4.6 \;\%$ &
               \ldots &
$        88 \;\%$ &
$       0.9 \;\%$ \\
\bf \M{82} &
$        97 \;\%$ &
$       0.4 \;\%$ &
$       0.7 \;\%$ &
$       4.6 \;\%$ &
$        26 \;\%$ &
$        68 \;\%$ &
$       4.0 \;\%$ &
$       0.5 \;\%$ &
               \ldots &
$        59 \;\%$ &
$        36 \;\%$ \\
\hline
\end{tabularx}
\end{center}
\end{table*}

\subsection{Very hot $5.5\mic$ dust}

\begin{figure}[htbp]
  \begin{center}
    \includegraphics[width=\linewidth]{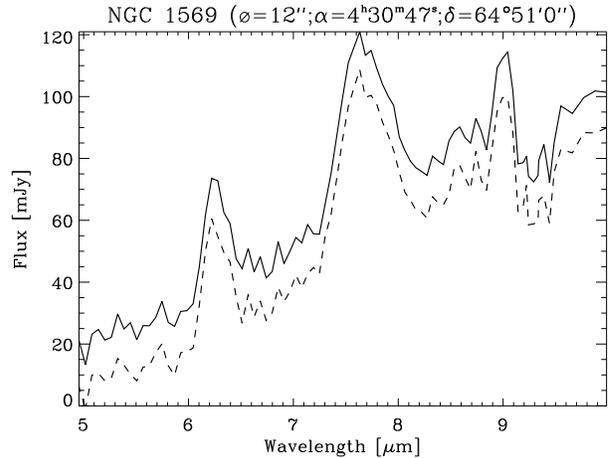}
    \caption{A zoom into the short wavelength part of the CVF spectrum of 
             \ngc{1569} toward the 
             central 12$"$ of the peak (Fig.~\ref{fig:ngc1569_veryhot}).
             (solid line).
	     The dashed line represents the lower limit of this spectra,
             taking into account the errors ($F_\nu - \Delta F_\nu/2$) 
             (Sec.~\ref{sec:obs}).}
    \label{fig:zoomngc1569}
  \end{center}
\end{figure}

\begin{figure}[htbp]
  \begin{center}
    \includegraphics[width=\linewidth]{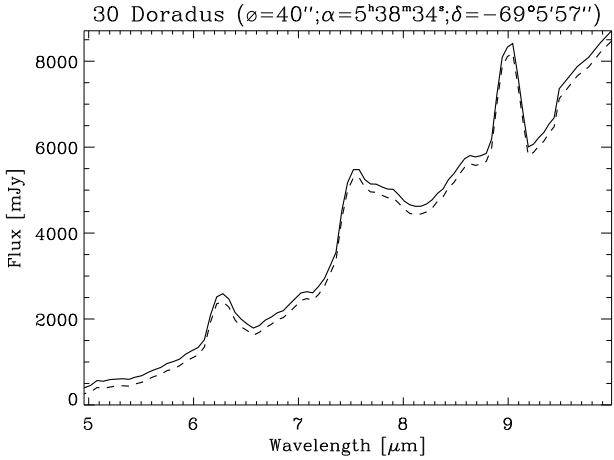}
    \caption{Zoom on the short wavelength part of the global CVF of \xxxdor\
             (solid line).
             The dashed line represents the lower limit of this spectra,
             taking into account the uncertainty ($F_\nu - \Delta F_\nu/2$) 
             Sec.~\ref{sec:obs}).}
    \label{fig:zoom30dor}
  \end{center}
\end{figure}

Inspection of individual spectra throughout the galaxies, zooming into the 
short wavelength end of the ISOCAM spectra, reveals that the continua usually 
decreases to near zero value at the beginning of the ISOCAM CVF spectra at 
$5.5 \mic$, considering the uncertainty of the measurements. 
Isolated regions of \ngc{1569}\ (Fig.~\ref{fig:zoomngc1569}) and \xxxdor\ 
(Fig.~\ref{fig:zoom30dor}), however, are the exceptions. 
These regions show excess at 4.9 to $5.6 \mic$ which could either be due to 
very hot dust or the stellar population. 
While the detailed modelling of the stellar population and gas and dust of 
low metallicity galaxies demonstrates that the stellar spectra are dominated 
by younger stellar populations on global scales, usually less than 10~Myr 
\citep[e.g.][]{galliano+03, galliano+05}, there can also be some contribution, 
on global scales to the stellar continuum toward shorter wavelengths in low 
metallicity dwarf galaxies, but usually more prominant at wavelengths shorter 
than those traced by the ISOCAM CVF \citep[e.g.][]{engelbracht+05}.

\begin{figure}[htbp]
  \centering
  \includegraphics[width=\linewidth]{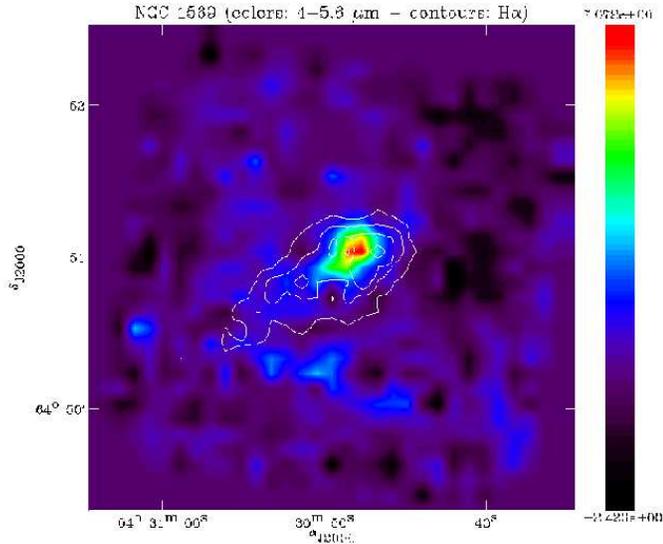}
  \caption{Very hot dust emission.
           The image is the integration of the CVF spectrum of \ngc{1569} from 
           4.9 to $5.6\mic$. Contours are from the H$\alpha$ map of 
           \citet{waller91}.}
  \label{fig:ngc1569_veryhot}
\end{figure}

\begin{figure}[htbp]
  \centering
  \includegraphics[width=\linewidth]{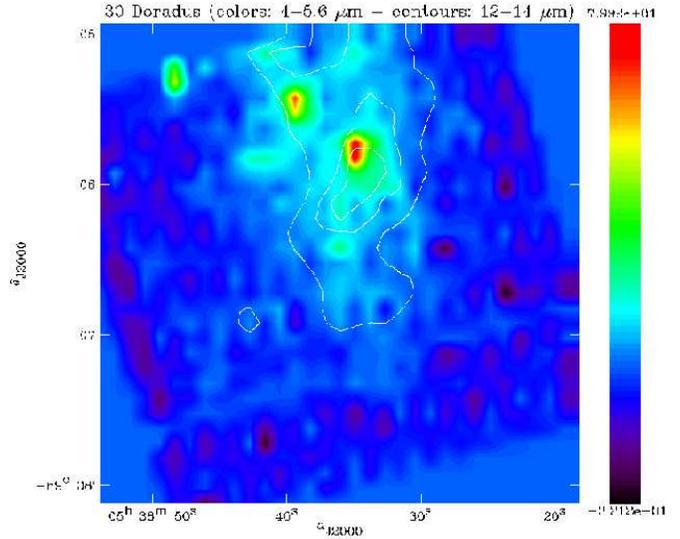}
  \caption{Very hot dust emission.
           This map is the integration of the CVF spectrum of \xxxdor\ from 
           4.9 to $5.6\mic$. Contours are the VSG distribution, which also 
	  resembles that of the \neiiiline\ (see Fig.~\ref{fig:30dor_compos}).}
  \label{fig:30dor_veryhot}
\end{figure}

Figure~\ref{fig:ngc1569_veryhot} shows the spatial distribution of the 
continuum 
between 4.9 and 5.6$\mic$ in \ngc{1569} overlayed on the H$\alpha$ map by 
\citet{waller91}. 
The 4.9 to 5.6$\mic$ emission is very localised and peaks at the site of the 
brightest H$\alpha$ peak, which is in close proximity to a super star cluster. 
The $5.5\mic$ emission falls off rapidly, suggesting an origin of very 
localised, hot dust emission. 
It is also spatially associated with the \neiii\ and VSG peaks, and not 
necessarily related to the PAHs (Fig.~\ref{fig:n1569_compos}).
Likewise, toward the \xxxdor\ region, the 4.9 to 5.6$\mic$ continuum is very 
localised (Fig.~\ref{fig:30dor_veryhot}), peaking toward the \neiii\ and VSG 
peak,
about 10 pc from the R136 cluster (Fig.~\ref{fig:30dor_veryhot}).
Hot dust emission at these wavelengths has been inferred in other starburst 
galaxies \citep{hunt+02}.
If dust is emitting at these short wavelengths, the dominant particles must be
small (nanometer sizes) and stochastically heated with temperature 
fluctuations as high as $\simeq 1000$~K \citep{guhathakurta+89}.

\section{Interpretation}
\label{sec:interp}

Using the spectral fitting procedure described above 
(Sect.~\ref{sec:components}), the global ratios 
of the physical MIR components, the nebular lines, the PAHs and the VSGs have 
been computed. 
In this section we present and study relationships between these components 
in our various sources and provide interpretation to help elucidate the 
nature of the ISM of low metallicity environments. 

\subsection{The PAH bands}

One of the most striking aspects of the spectra of these low metallicity 
starburst systems is the dearth of PAH emission along with the steeply-rising 
VSG continuum beyond 10 $\mic$. 
This is a very unusual circumstance, since the photons which produce the 
elevated temperatures of the VSG continuum would also illuminate the PAHs, if 
they were present.

This consistent discrepency of the paucity of bright PAH bands along with 
elevated continuum in the MIR spectra of a sample of low metallicity galaxies 
was demonstrated with ISOCAM spectro-imaging in 
\citet{thuan+99,madden00,madden05}. 
Ground-based observations from 1991, already showed that PAHs were not prominent in 2 dwarf galaxies \citep{roche+91}.  
Recent Spitzer observations also confirm this 
\citep{houck+04,engelbracht+05,wu+05}. 
Dustier starbursts or spiral galaxies, on the other hand, do show prominent 
PAHs in their MIR spectra, even on global scales 
\citep[e.g.][]{genzel+98, laurent+00, forster+03, vogler+05}. 
There can be several reasons for the relatively low PAH intensity seen here: 
a) the PAHs are destroyed globally in mow-metallicity systems via evaporation 
and photodissociation processes, due to radiation field properties, such as 
hardness and intensity \citep[e.g.][]{madden00, madden05};  b) the PAHs are 
present but only on small spatial scales, which may be a consequence of (a). 
In this case, their emission could be suffering from dilution due to the 
telescope beam, and could be more apparant on smaller spatial scales; 
c) PAHs were never a chemically abundant species in these galaxies, due to 
their low metalliciy nature.

\begin{figure}[htbp]
  \centering
  \includegraphics[width=\linewidth]{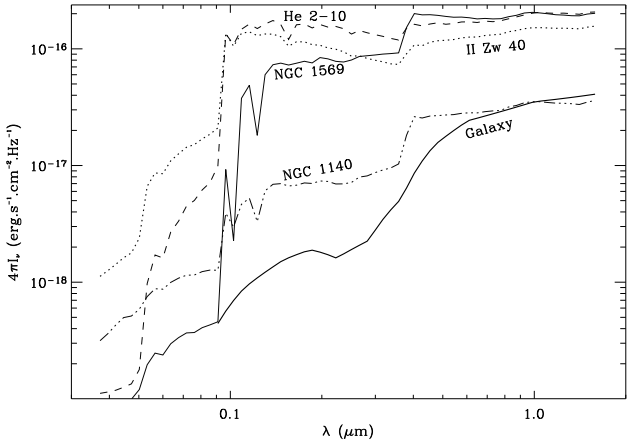}
  \caption{The variations of the modeled ISRFs for 4 of the dwarf galaxies 
        compared to the Galaxy \citep[taken from ][]{galliano+03,galliano+05}.}
  \label{fig:isrfs}
\end{figure}

We begin to investigate effects due to the properties of the radiation field 
by first comparing the intrinsic ISRFs of 4 dwarf galaxies, as modeled by 
\citet{galliano+03, galliano+05}. 
As can been seen in Figure~\ref{fig:isrfs}, the global intrinsic ISRFs of the 
dwarf galaxies are all harder and more intense relative to the global ISRF of 
the Galaxy. 
For example, the global ISRF of NGC1140, where the MIR spectra do indeed show 
evidence for relatively prominent PAH bands, is obviously softer, and less 
intense than those of \iizw\ and \ngc{1569}, where the PAH bands are less 
prominent. 
We look more closely into this quantitatively in the following sections.

\subsection{\neiii/\neii\ line ratios: effects on the survival of PAHs?}
\label{sec:neiii_neii}

The diagnostic value of the MIR nebular line ratios, particularly that of the 
\neiiiline\ and the \neiiline\ lines, has received a great deal of attention 
lately, initiated by the abundant ISO spectra in a wide variety of sources.  
The ionisation potentials of neutral Ne and Ne$^+$ are 21.5 and 40.9~eV, 
respectively. 
The ratio of the \neiii/\neii\ fine structure lines are independent of the 
neon abundances and are relatively independent of extinction, since their 
wavelengths are not very different. 
They are sensitive to the ionisation structure of the nebulae and to the 
spectral form of the UV radiation field, thus, they trace 
populations of massive hot stars, characterised by their hard radiation 
fields. 
Deciphering the precise ionisation structure of nebulae from ionic line 
ratios involves untangling the intertwined effects of metallicity, morphology, 
spectral type and temperature to finally arrive at the stellar composition and
age - important characteristics for studying the evolution of starbursts 
within galaxies and from galaxy to galaxy. 
This is not completely straightforward. 

\begin{figure}[htbp]
  \centering
	\includegraphics[width=\linewidth]{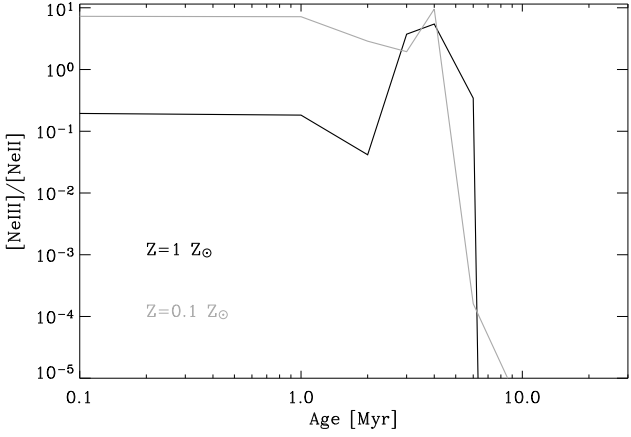}
  \caption{The behavior of the \neiii/\neii\ line ratio as a function of age 
           for clusters of solar metallicity (black curve) and 0.1 solar 
           metallicity (grey curve). 
           A single instantaneous starburst was assumed with a Salpeter IMF.}
  \label{fig:neiiineii}
\end{figure}

The diagnostic value of the \neiii/\neii\ line ratios can be seen in 
Fig.~\ref{fig:neiiineii}, for solar and 0.1 solar metallicities. 
These examples are shown for a star cluster with a Salpeter IMF (upper mass 
cut-off, m$_{up}$, of 120 M$_{\odot}$; lower mass cut-off, m$_{low}$, of 
0.1$M_{\odot}$). 
The line ratio is boosted to ratios of the order of 10 to 100 between 1 to 
3~Myr, due to contribution from the O stars. After this time, the O stars have
moved off the main sequence and the \neiii/\neii\ line ratios plummet to very 
low values.

Decreasing the metallicity has effects on the \neiii/\neii\ line ratios due to
reduced line blanketing and blocking in the stellar atmospheres. 
For a given stellar type, the main sequence temperatures are hotter and the 
form of the stellar SED is harder. 
Metallicity effects on ionic line ratios have been studied by a number of 
authors \citep[e.g.][]{thornley+00,martin-hernandez+02b,giveon+02,rigby+04}. 
The \neiii/\neii\ line ratios in low metallicity regions are strikingly high, 
often 1 to 2 orders of magnitude greater than more metal-rich starburst 
galaxies as demonstrated in \citep{madden05}. 
The outward increase of Galactic \neiii/\neii\ line ratios has also been 
attributed to the decrease in metallicity \citep{giveon+02}.

\citet{thornley+00} studied a large sample of starburst galaxies, and  
found relatively low \neiii/\neii\ line ratios observed in solar metallicity 
starburst galaxies ( $\leq$ 0.8). One interpretation could be a lack of 
massive stars ($\gtrsim 50M_{\odot}$), but this would be difficult to 
reconcile given other evidence for massive stars. 
Their favoured explanation for low \neiii/\neii\ line ratios was the aging of 
relatively short-lived (10$^{6}$ to 10$^{7}$yr) starbursts 
\citep[see also ][]{rieke+93,engelbracht+98}. 
In this scenario, the hottest stars are not necessarily the dominant source of
the ionisation in the starburst galaxy, accounting for the low line ratio 
values.  
Another scenario proposed for the low \neiii/\neii\ line ratio observed in 
metal-rich starburst galaxies is that the massive stars could spend most of 
their main sequence lifetimes concealed in very dusty, extincted regions, 
undetected even out to MIR wavelengths \citep{rigby+04}.

The nebular line intensities for our low metallicity sources are listed in 
Table~\ref{tab:lines} while the line ratios are shown in 
Table~\ref{tab:ratios}. 
We also include M~82 for comparison with a metal-rich starburst galaxy.  
Except for M~82 and the diffuse, low metallicity region, \smcn, all of the 
other sources show \neiii/\neii\ line ratios $> 2$ and as high as 15 or more 
averaged over full galaxy sizes. Note that the lower limits on the 
\neiii/\neii\ ratios are due to the low signal-to-noise of the [NeII] line. 
Many of these starburst regions are known to be harbouring super star clusters
(SSCs) and the modeled ISRFs for some of these sources is obviously much 
harder than that of the Galaxy, for example (Fig.~\ref{fig:isrfs}). 

Not only are the \neiii/\neii\ line ratios very high within the resolved 
galaxy \ngc{1569} (Sec.~\ref{sec:ngc1569}; Fig.~\ref{fig:correl}), but high 
averaged \neiii/\neii\ line ratios are also observed on averaged full galaxy 
scales for the more distant, unresolved galaxies (Table \ref{tab:ratios}). 
To account for an averaged high \neiii/\neii\ line ratio in the unresolved 
galaxies, the ionising photons would need to be present in a relatively large 
area of the galaxy \citep{madden00, madden05}. One possibility would be an 
ISM that is very porous, allowing the ionising radiation to penetrate the 
galaxy on large scales (more on this  hypothesis in 
Sec.~\ref{sec:pah_destruct}).

\begin{table*}[htbp]
\caption{Ionic line intensity ratios for the low metallicity galaxies
         (global values).}
\label{tab:ratios}
\begin{center}
\begin{tabularx}{\textwidth}{*{5}{X}}
\hline
\hline
\multicolumn{1}{c}{}  & \neiii/\neii
  & \siv/[\neiii
  & \ariii/\neiii
  & I(PAH)/I(VSG) \\
\hline
\bf \ngc{1140}
  & $2.2 \pm 0.7$
  & $0.24 \pm 0.11$
  & $\lesssim 0.01$
  & $1.33 \pm 0.37$ \\
\bf \ngc{1569}
  & $7.7 \pm 1.7$
  & $0.45 \pm 0.11$
  & $0.09 \pm 0.05$
  & $1.20 \pm 0.22$ \\
\bf \ngc{5253}
  & $\gtrsim 11$
  & $\lesssim 0.25$
  & $\lesssim 0.05$
  & $\lesssim 0.30$ \\
\bf \iizw
  & $\gtrsim 5.7$
  & $3.53 \pm 0.65$
  & $\lesssim 1.5$
  & $\lesssim 0.88$ \\
\bf \xxxdor
  & $5.59 \pm 0.41$
  & $1.29 \pm 0.10$
  & $0.25 \pm 0.02$
  & $0.36 \pm 0.01$ \\
\bf \smcn
  & $\gtrsim 10.3$
  & $1.59 \pm  0.18$
  & $0.36 \pm 0.09$
  & $1.4 \pm 0.6$ \\
\bf \smcb
  & $\lesssim 0.5$
  & $\gtrsim 1.7$
  & \ldots
  & $\gtrsim 56$ \\
\bf \M{82}
  & $0.147 \pm 0.012$
  & $1.09 \pm 0.10$
  & $\lesssim  0.4$
  & $3.78 \pm 0.11$ \\
\hline
\end{tabularx}
\end{center}
\end{table*}

\begin{figure}[htbp]
  \centering
  \includegraphics[width=\linewidth]{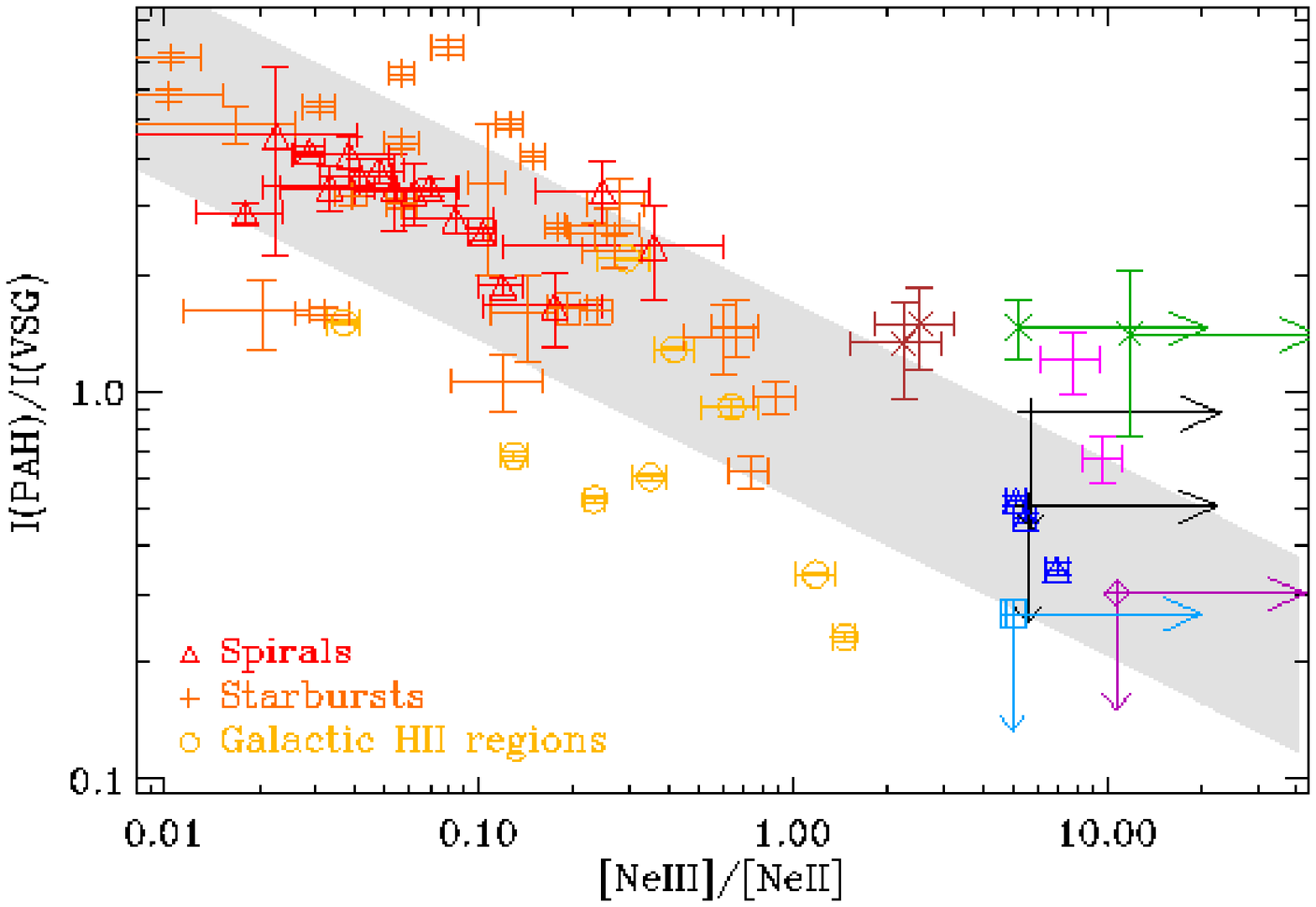}
  \caption{Correlation between \neiii/\neii\ and $15\mic$/PAH,
           for \hii\ regions \citep{peeters+02}, spiral, starburst and dwarf 
           galaxies \citep{madden05}. 
           The dwarf galaxies are located on the right of the plot (see 
           Fig.~\ref{fig:correlglobzoom} for the color code).
           The grey strip is the linear fit to the spread of points 
           $\pm 1\sigma$. 
           The arrows on the dwarf galaxy values indicate measurement limits 
           which are limited by the low detection levels of the \neiiline\ and
           the PAHs. 
           See the electronic version of the Journal for color version of the 
           plots.}
  \label{fig:correlglob}
\end{figure}

\begin{figure}[htbp]
  \centering
  \includegraphics[width=\linewidth]{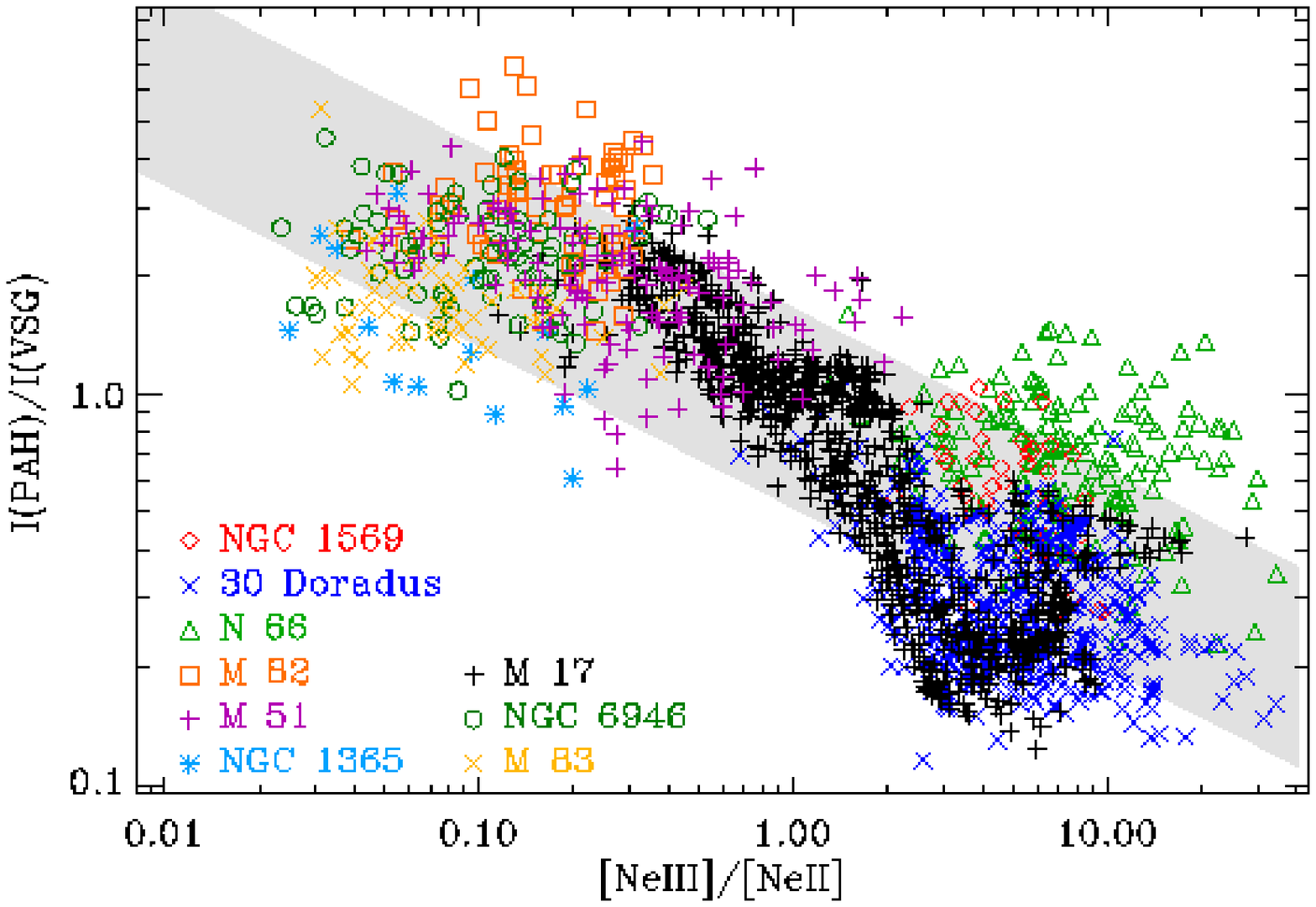}
  \caption{Correlation between \neiii/\neii\ and PAH/VSG
           for regions within resolved sources, such as the low metallicity 
           sources \ngc{1569}, \xxxdor\ and \smcn. 
           Also included for comparision are regions within the more 
           metal-rich galaxies, M~51, M~82, NGC~6946, M~83 and NGC~1365. 
           Note that the metal-poor objects occupy a spread of values located 
           to the lower right of the plot, while the more metal-rich objects 
           have lower \neiii/\neii\ ratios and higher PAH/VSG line ratios. 
           The grey strip is from Fig.~\ref{fig:correlglob}. 
           See the electronic version of the Journal for color version of the 
           plots.
           }
  \label{fig:correl}
\end{figure}

To investigate the effect of the hard radiation fields on the PAHs, we inspect 
the behavior of the \neiii/\neii\ and PAH/VSGs for our dwarf galaxies 
(Fig.~\ref{fig:correlglob}), first demonstrated in \citep{madden05}. 
For comparison, we also include many other sources such as the \hii\ region 
sample of \citet{peeters+02}, starburst galaxies from the ISOCAM sample of 
\citet{laurent+00} and spiral galaxies from the sample of \citet{roussel+01a}.
The MIR spectra were modeled and the various components were extracted as 
described in Section~\ref{sec:components}. 
The PAH quantity presented here is the sum of the fluxes the 5 PAH 
bands\footnote{PAH bands refer to the features extracted in the way described 
in Sec.~\ref{sec:components}, without the continuum from the VSGs} 
($\lambda=6.2$, 7.7, 8.6, 11.3, $12.6\mic$) in our CVF spectra. 
Using all of the features to measure the behavior of the PAHs as opposed to a 
single feature, avoids the known effects of feature-to-feature ratio 
variations, which can be very noticable under widely varying UV environments 
\citep[e.g.][]{vermeij+02, gallianophd}. 
The quantity characterising the VSG emission is the fitted, feature-free 
continuum of the MIR spectra (Sec.~\ref{sec:components}), integrated between 
10 and 16$\mic$. 
Notice the variations of 3 orders of magnitude of the observed \neiii/\neii\ 
line ratios. 
Metal-rich starburst regions typically show a range of low excitation values 
of about 0.01 to 0.15 \citep[see also][]{thornley+00}. 
The ratio of the \neiii/\neii\ lines in \hii\ regions range from 0.1 to 1.0 
and the low-metallicity sources show the highest values up to and beyond 
ratios of 10.  
The large values of \neiii/\neii\ line ratios are indicative of the hard 
ISRFs within the low-metallicity regions.

\begin{figure}[htbp]
  \centering
  \includegraphics[width=\linewidth]{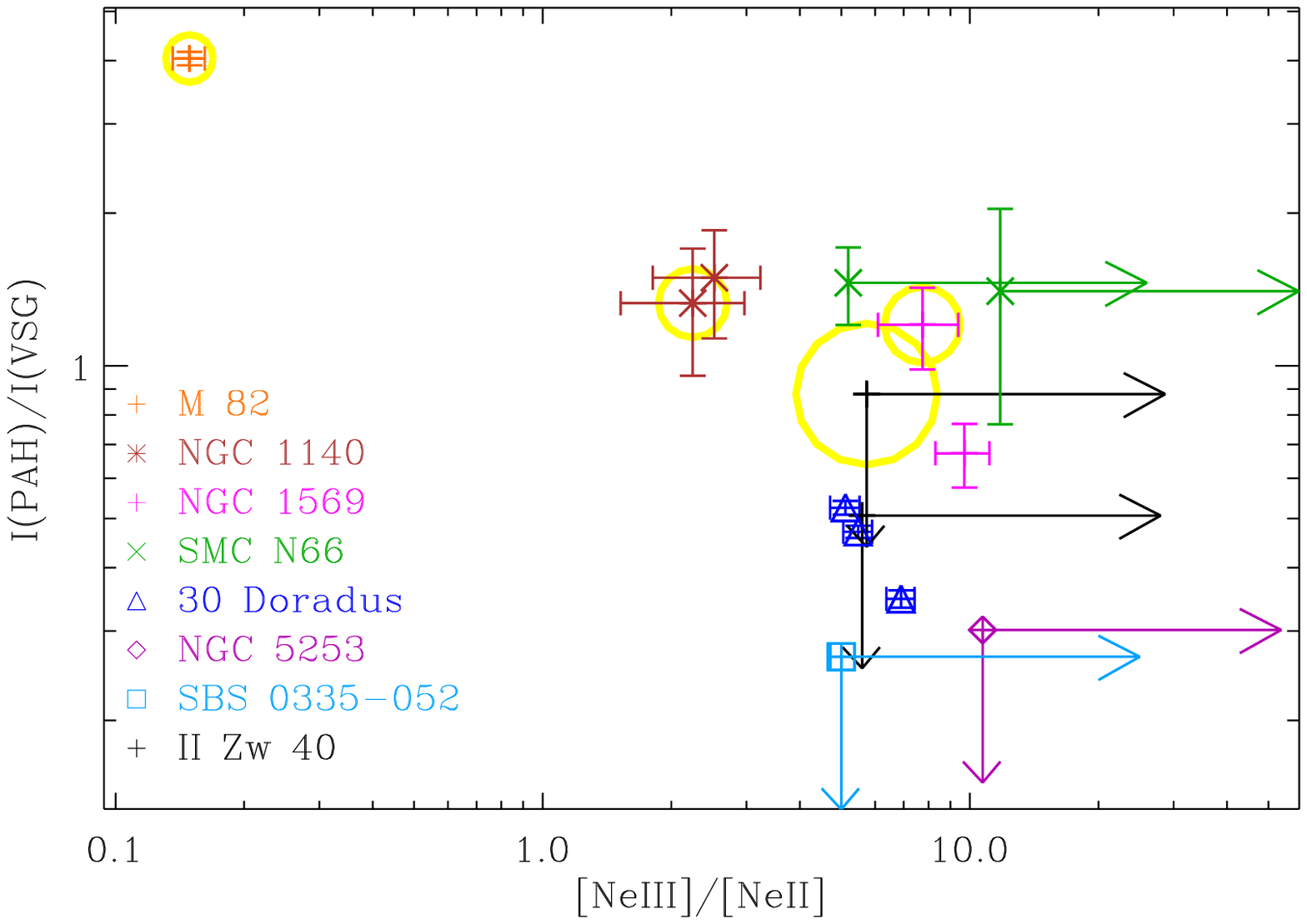}
  \caption{Correlation between \neiii/\neii\ and PAH/VSG,
           for global values of dwarf galaxies, and \M{82} for contrast (a 
           zoom into Fig.~\ref{fig:correlglob}). 
           The \neiii/\neii\ value for \sbs\ is from \citet{houck+04}. 
           For some well-resolved galaxies we also include a central, smaller 
           region. 
           The size of the circles around some sources represents the 
           relative hardness of the modeled ISRFs (Fig.~\ref{fig:isrfs}), as 
           described by equation 2. 
           For example, the larger the circle, the harder the intrinsic ISRF. 
           The arrows indicate measurement limits which are limited by the 
           low detection levels of the \neiiline\ and the PAHs. 
           See the electronic version of the Journal for color version of the 
           plots.}
  \label{fig:correlglobzoom}
\end{figure}

We find a correlation which indicates that as the hardness of the radiation 
field increases, as traced by the high \neiii/\neii\ 
line ratios, the ratio of PAH/VSG decreases.

Fig.~\ref{fig:correlglobzoom} is a zoom into Fig.~\ref{fig:correlglob}, 
showing the low metallicity sources, which occupy the higher values of the 
\neiii/\neii\ line ratios. 
As another check, we quantify the hardness of the ISRFs directly in 
Fig.~\ref{fig:isrfs} by:
\begin{equation}
\frac{\displaystyle\int_0^\infty I_\nu \ddiff\nu}
    {\displaystyle\int_0^\infty \frac{I_\nu}{h\nu} \ddiff\nu},
\end{equation}

This ratio is schematically represent the relative hardness of the ISRF of the
galaxies by the size of the circles in Fig.~\ref{fig:correlglobzoom}.  
We see that the relative hardness of the ISRF increases as the ratio of 
PAH/VSG decreases, consistent with the increasing \neiii/\neii\ line ratios.  

Thus, both the increasing \neiii/\neii\ line ratios, measured from the MIR 
spectra, as well as the hardness of the intrinsic ISRF, measured from the 
modeled SEDs, show a striking increase as the PAH/VSG decreases.  
Thus, this is consistent with the the hard radiation field being responsible 
for the destruction of the PAHs, and the steeply rising continuum emission 
from the hot, small grains.

In Fig.~\ref{fig:correl}, we plot the values of the \neiii/\neii\ line ratios
and PAH/VSG for individual pixels within \ngc{1569}, \xxxdor, \smcn, as well 
as the metal-rich starburst and spiral galaxies, M82 and M51, NGC~6946, M~83 
and NGC~1365. 
A spread of the \neiii/\neii\ line ratios of about 1 and up to 2 orders of 
magnitude exists within these sources, while the values of the ratios of the 
PAH/VSG span over a range of about 5 to 10 within sources. 
This plot also demonstrates that the 3 well-resolved low metallicity sources 
have high \neiii/\neii\ line ratios even from point-to-point within the 
galaxies, confirming the galaxy-wide spatial extent of the hard radiation 
field see throughout \ngc{1569} (Sec~\ref{sec:spatial}). 
The more metal-rich galaxies never have exceptionally high values of 
\neiii/\neii\ line ratios nor low PAH/VSG values. 
While metallicity gradients can be an explanation for some of the spread in 
the dustier galaxies, it can not be the explanation for \ngc{1569}, \xxxdor\ 
and \smcn. 
\smcn\ and \xxxdor\ are confined \hii\ and PDR regions that can not vary 
significantly in metallicity and \ngc{1569}, like other low metallicity 
galaxies, is not known to have a metallicity gradient \citep{kobulnicky+97}.  
We also confirm that within these sources the largest values of the 
\neiii/\neii\ line ratios coincide with the prominent \hii\ regions/SSCs 
within the object, and decrease outward. Thus the hardness of the stellar 
radiation does appear to be traced here by the \neiii/\neii\ line ratios, and 
may control the destruction of the PAHs within galaxies.

\subsection{PAH Destruction Mechanisms}
\label{sec:pah_destruct}

PAHs are known to be mostly absent from \hii\ regions. 
The MIR spectra of AGNs in contrast to those of starburst galaxies, was 
already noted to be void of PAH features in pioneering work 20 years ago 
\citep{roche+85, aitken+85} and also demonstrated more recently in 
\citet{siebenmorgen+04}. 
Most explanations follow on the suggestion of the destruction of PAHs being 
due to the hard radiation field of the central source. 
The heating by energetic photons can basically, either ionise or dissociate 
PAHs. 
However, the precise processes and environmental factors involved, and how 
this varies as a function of the chemical properties such as the state of 
ionisation and the size of the molecule, continues to be a subject of 
investigation \citep[e.g.][]{voit92,jochims+94,joblin+95,allain+96a,allain+96b,allamandola+99,lepage+03}. 
The apparant lack of small PAHs ($< 20$ atoms) in the ISM 
\citep[e.g.][]{clayton+03} suggests that such small molecules are readily 
destroyed under widely varying ISM conditions. 
Depending on photon energies, critical sizes for dehydrogenating PAHs can be 
20-30 atoms \citep{lepage+03} and even larger ones can be ionised. 
The derrived threshold size of the PAH molecule depends on the model used. 

Recent high spatial resolution MIR observations of 2 starburst galaxies by 
\citet{tacconi-garman+04} also show the PAH and continuum emission peaking at 
star formation sites, while an obvious suppression of the PAH feature to 
continuum occurs specifically toward the star formation peaks. 
Their investigatigation suggests photoionisiation and/or photodissociation of 
PAHs as plausible mechanisms to explain the observations. 

The fact that PAHs are not present in Galactic \hii\ regions and the PAH 
features to continuum ratio decreases toward the star formation peaks 
(Fig.~\ref{fig:n1569_compos}), and the effect of the overall tendency for the 
PAH to continuum ratios to decrease as the \neiii/\neii\ line ratios increase, 
are consistent with the conclusion that the lack of PAHs observed in dwarf 
galaxies is due to the hardness of the ISRF. 

\subsection{Other factors possibly playing a role in the observed low PAH 
            intensities}

The possibility for low-metallicity systems to {\it chemically}, have low 
abundances of PAHs must also be explored. Taking into account possible ISM 
enrichment sources for carbonaceous particles, such as AGB stars and/or 
supernovae Type II (SN II), and not considering PAH destruction processes, we 
can eliminate SN II as large contributors of PAHs, due to their relatively 
short life times. 
The galaxies considered would already have experienced mass enrichment from 
SN II. 
\citet{dwek05, galliano+05b} have considered this issue and note that AGB 
stars begin to contribute to the carbon reservoir only when 4 M$_{\odot}$ 
stars have entered the AGB stage ($~ 5\times10^{8}$ yr). 
Assuming AGB stars are the dominant source of PAHs, this could explain why 
PAHs are not abundant in extremely low metallicity systems ($<0.1$ solar). 
For the more moderate metallicity systems considered here, PAHs should indeed 
be present, under this hypothesis. 
If, on the other hand, SN II were to be a dominant source of PAHs, then PAHs 
should likewise be present in these systems, if destruction were not taken 
into account. 
Note that model results can depend strongly on the adopted yields 
\citep[e.g.][]{dwek05}. 
Photodestruction of PAHs as well as chemistry effects are probably both 
playing important roles in low-metallicity systems.  

\begin{figure}[htbp]
  \centering
  \includegraphics[width=\linewidth]{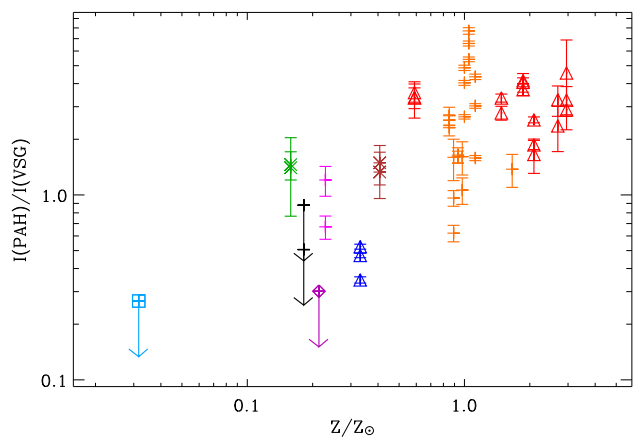}
  \caption{Correlation between PAH/VSG and metallicity (Z/Z$_\odot$). 
           Symbols for the dwarf galaxies, starburst galaxies and Galactic 
           \hii\ region are those used in Fig.~\ref{fig:correlglob} and 
           Fig.~\ref{fig:correlglobzoom}.}
  \label{fig:correlpahmet}
\end{figure}

Fig.~\ref{fig:correlpahmet} shows the observed PAH/VSG as a function of the 
metallicity for the sources considered here. There does appear to be a 
correlation related to the relative intensity of PAH bands and the 
metallicity of the object, such that as the metallicity decreases, the 
PAH/VSG ratio decrease. 
The range of metallicities that we consider here is limited. 
However, \citet{engelbracht+05} find such a correlation using broad band 
observation for a wide range of low metallicity galaxies. 

What are the chances of us missing compact photodissociation regions, where 
PAHs could be residing in very small clumps and which could be suffering from 
significant beam dillution? 
The ISOCAM beam at the shorter wavelengths, is about 6", which is about 500~pc
for the most distant source in our study (\ngc{1140} $\sim$ 23 Mpc). 
The fact that we see quite the pervasive ionic gas in the ISM over galaxy-wide
scales in resolved galaxies such as \ngc{1569}, as well as on the global 
scales of unresolved galaxies, supports the idea of a very porous ISM where 
the mean free path length of the ionising photons can be relatively large due 
to the decrease in dust attenuation. 
The fact that the FIR fine structure lines, such as the important cooling line 
of 158$\mic$ [CII], are relatively bright in all of these galaxies, infers a 
substantial amount of dense clumpy, small filling-factor PDR material 
\citep{madden00}. 
In support of this hypothesis, to account for the observed FIR to submm SEDs 
in dwarf galaxies, \cite{galliano+05} invoked a scenario of small dense 
clumps, with size scales on the order of a few tens of pc, having a small 
volume factor of 10$^{-4}$ to 10$^{-3}$. 
If this hypothesis were correct, then PAHs could indeed be hidden from our 
view in the relatively large ISOCAM beam.  
One example of this point is in \ngc{5253}, where the $15\times 30\arcsec$ 
ISOSWS beam barely produced evidence for low level PAH bands 
(Fig.\ref{fig:spec1}), while more recent ground-based observations by 
\citet{alonso-herrero+04} use a beam size comparable to 5pc in \ngc{5253} with
which they are able to detect the weaker PAH band at $3.3 \mic$. 
While this could be a possible explanation for the MIR spectra considered 
here, it would probably not support a significant mass of PAHs residing in 
these small clumps.

\section{Conclusion}
\label{sec:conc}
We present new ISOCAM MIR spectra of the low metallicity dwarf galaxies, 
\all, along with MIR spectra from the \xxxdor\ region in the low metallicity 
LMC.  
These are compared with other published spectra of low metallicity regions as 
well as several dustier starburst galaxies. 
The low metallicity nature of the ISM presents a different picture than that 
of dustier starburst galaxies or our own \mw. We summarise our findings:

\begin{itemize}

\item The characteristics of the ISM that dominate the MIR emission of low 
metallicity regions differ strikingly from those of dustier starburst or 
spiral galaxies. 
The most prominent aspect of the MIR spectra of low metallicity regions, at 
least down to the scale of the ISOCAM beam, is the dearth of PAH bands. 
Even in the {\it full galaxy averaged} spectra of these galaxies, the MIR 
characteristics are often dominated by steeply rising small grain continua 
and prominent ionic lines, most notably the \neiii\ and \siv\ lines, in 
contrast to dustier starburst galaxies, which globally show prominent emission 
arising from the extragalactic disk component (i.e. PAHs and relatively 
flatter MIR continuum).  
The distribution of various ISM components in low metallicity environments 
appears to have been altered by the decrease in dust abundance, which 
effectively translates into a larger mean free path length of the UV photons. 
This may explain why the low metallicity galaxies tend to resemble giant \hii\ 
regions viewed on large extragalactic scales, with smaller, clumpy 
PDRs interspersed within the ISM  \citep[e.g.][]{galliano+03, galliano+05}.

\item We quantify the contributions of ionic lines, continuua and PAH bands 
in various broad bands, which were often used to map galaxies. 
While the $12.7 \mic$ PAH band dominates the $15 \mic$ ISOCAM broad band 
emission on galaxy-wide scales spiral galaxies, this same broad band is 
typically dominated by the hot small grain continuum emission in low 
metallicity galaxies with up to 10\% of the $15 \mic$ broad band arising from 
the \neiiiline\ alone. 
In the $6.7 \mic$ broad band of ISOCAM, PAHs, even as faint as they are, 
dominate, in some cases. However, in the more active low metallicity galaxies 
such as \ngc{5253} and \iizw, for example, this band captures 1/2 PAH emission 
and 1/2 hot dust continuum emission. Thus caution must be used when 
interpreting broad band emission.

\item \ngc{1569} and \xxxdor\ are well resolved spatially and the distribution 
of the various tracers are inspected. 
\xxxdor\ images show an example of the higest spatial resolution in our sample 
with a resolution of 2.4 pc. 
The \neiiiline\ emission is distributed in a sharp ridge closer to the R136 
exciting star than the \neii\ line emission. 
The PAH peaks avoid the more confined \neiii/\neii\ peak, which is tracing the 
hard radiation field, while the VSG peak prefers the \neiii/\neii\ peak. 
In \ngc{1569}, with the spatial resolution of 106 pc, most of the species, in 
general, appear to peak toward the star formation sites. 
In \ngc{1569}, the extents of the \neiiline\ and the \neiiiline\ lines are 
remarkable, even tracing some of the H$\alpha$ ejecta. 

\item We find high local \neiii/\neii\ ratios throughout the resolved low 
metallicity galaxies, and likewise high values averaged over the unresolved 
galaxies, often as high as 10 --~much higher than dustier starburst galaxies. 
Due to the lower dust abundance, hard UV photons presumably can permeate over 
larger size scales within galaxies. 
We find a prominent correlation between the PAHs/VSGs ratios and the 
\neiii/\neii\ ratios for the low metallicity sources as well as Galactic \hii\ 
regions and more metal-rich starburst galaxies. 
They span 3 orders of magnitude in the sample presented here. 
The correlation of the \neiii/\neii\ ratios and the PAHs/VSGs is also seen 
within sources which have sufficient spatial resolution. 
The hard, permeating radiation field seems to play an important role in the 
destruction of PAHs in low metallicity regions.
It does not appear to be simply a matter of metallicity controlling the 
spectral characteristics of dwarf galaxies,
but rather a combined complicated effect of the intrinsic radiation field 
plus the low metallicity and, undoubtedly, the geometry of the sources and 
the structure end evolution of the ISM. MIR spectra from a broader sample of 
lower metallicity galaxies with a variety of star formation activity plus 
higher spatial resolution observations would help to shed more light on these 
issues.

\end{itemize}


\acknowledgements
A special thanks to Els Peeters for providing the \hii\ region spectra. 
We thank Ren\'e Gastaud, Pierre Chanial and H\'el\'ene Roussel for expert 
advice on various stages of the ISOCAM data reduction. 
We thank the anonymous referee for helpful comments that improved the clarity 
of the paper. 
This work was supported in part by a National Research Council Research 
Associateship for F.~Galliano at NASA GSFC.

\bibliographystyle{aa}
\bibliography{article}


\end{document}